\documentclass[aps,prx,showpacs,twocolumn,amsmath,amssymb,superscriptaddress,letterpaper]{revtex4}
\usepackage{times}
\usepackage{amsfonts}
\usepackage{mathrsfs}
\usepackage{graphicx}
\usepackage{dcolumn}
\usepackage{bm}
\usepackage{color,xcolor}

\usepackage[colorlinks,bookmarks=false,citecolor=blue,linkcolor=red,urlcolor=blue]{hyperref}
\usepackage{ulem}

\usepackage{multirow}
\usepackage{makecell}
\usepackage{booktabs}
\usepackage{caption}

\def\be{\begin{equation}}       \def\ee{\end{equation}}
\def\bea{\begin{eqnarray}}      \def\eea{\end{eqnarray}}

\newcommand{\clr}{\color{red}}

\begin{document}
\title{Double and Quadruple Flat Bands tuned by Alternative magnetic Fluxes in Twisted Bilayer Graphene}

\author{Congcong Le}
\affiliation{RIKEN Interdisciplinary Theoretical and Mathematical Sciences (iTHEMS), Wako, Saitama 351-0198, Japan}

\author{Qiang Zhang}
\affiliation{ Institute of Physics, Chinese Academy of Sciences,
Beijing 100190, China}

\author{Fan Cui}
\affiliation{ Institute of Physics, Chinese Academy of Sciences,
Beijing 100190, China}
\affiliation{ University of Chinese Academy of Sciences, Beijing 100049, China}

\author{Xianxin Wu}\email{xxwu@itp.ac.cn}
\affiliation{CAS Key Laboratory of Theoretical Physics, Institute of Theoretical Physics, Chinese Academy of Sciences, Beijing 100190, China}

\author{Ching-Kai Chiu}\email{ching-kai.chiu@riken.jp}
\affiliation{RIKEN Interdisciplinary Theoretical and Mathematical Sciences (iTHEMS), Wako, Saitama 351-0198, Japan}

\date{\today}

\begin{abstract}
Twisted bilayer graphene (TBG) can host the moir\'{e} energy flat bands with two-fold degeneracy serving as a fruitful playground for strong correlations and topological phases. However, the number of degeneracy is not limited to two. Introducing a spatially alternative magnetic field, we report that the induced magnetic phase becomes an additional controllable parameter and leads to an undiscovered generation of four-fold degenerate flat bands. This emergence stems from the band inversion at $\Gamma$ point near the Fermi level with a variation of both twisted angle and magnetic phase. We present the conditions for the emergence of multi-fold degenerate flat bands, which are associated with the eigenvalue degeneracy of a Birman-Schwinger operator. Using holomorphic functions, which explain the origin of the double flat bands in the conventional TBG, we can generate analytical wave functions in the magnetic TBG to show absolute flatness with four-fold degeneracy. Moreover, we identify an orbital-related intervalley coherent state as the many-body ground state at charge neutrality. In contrast, the conventional TBG has only two moir\'{e} energy flat bands, and the highly degenerate flat bands with additional orbital channels in this magnetic platform might bring richer correlation physics.

\end{abstract}


\maketitle

\textit{Introduction} -- Moir\'{e} twistronics, focusing on the exotic properties of  moir\'{e} superlattcies from two-dimensional (2D) twisted van der Waals multilayers, has attracted enormous attention in contemporary research\cite{Bistritzer-pnas11,cao-nat18-insulator,cao-nat18,Ashvin-prx18-insulating,Lu-nat19-sc_orbital_ci,chen2019,chen2019Signatures,devakul2021magic,ghiotto2021quantum,Ledwith2022}. One well-known example is the twisted bilayer graphene~(TBG), where moir\'{e} interlayer coupling can significantly alter the low-energy physics, and its electronic structure is described by the Bistrizer-MacDonald model for a small twisted angle~\cite{Bistritzer-pnas11}. Most prominently,  when the twisted angle is adjusted to be magic values, the low-energy Dirac bands evolve to isolated two-fold flat bands~(FBs) with "fragile" topology~\cite{Ashvin-prx18-insulating}.
Thus, these FBs in the TBG offer an exciting playground to explore emergent correlated and topological physics, and a wide range of intriguing correlated phenomena have been experimentally observed, such as correlated insulating states\cite{cao-nat18-insulator,Lu-nat19-sc_orbital_ci,Shen20-nphy-corelated,zondiner2020,wong2020,Das2022,christos2022}, unconventional superconductivity\cite{Lu-nat19-sc_orbital_ci,christos2022,park2021tunable,cao-nat18,Yankowitz-sci19-tuning,stepanov2020} and quantum anomalous Hall effect\cite{Serlin-sci19-fm,Sharpe-sci19-fm,KTLaw-nc20-omagnetoelectric,tseng2022anomalous}.

When chiral symmetry is preserved, at magic angles the moir\'{e} FBs are absolutely flat in the entire moir\'{e} Brillouin zone~(BZ). This absolute flatness can be proved by the construction of the wave functions~(WFs) from holomorphic functions~\cite{Ashvin2019}. In addition, calculating the spectral of a compact Birman-Schwinger operator can precisely determine all real magic angles for the two-fold FBs (per valley/spin)~\cite{Simon2021,Simonv4}. Till now, however, most studies have focused on the double FBs in the TBG because this two-fold degeneracy originates from non-degenerate bands of the Dirac cones. An interesting and outstanding question is: is it possible to realize FBs with higher degeneracy? Such an exploration is highly desirable as this realization will not only provide new insights into the origin of magic angles and FBs but also offer a more fruitful platform hosting diversified strongly-correlated physics.

Beyond the two-fold degeneracy of the moir\'{e} FBs, in this work, we show that an undiscovered generation of FBs with four-fold degeneracy emerges in the presence of spatially alternating magnetic fields, motivated by the electronic tunability in 2D systems with external uniform magnetic fields~\cite{Phong2022, dong2022dirac,Arbeitman2022,Guan2022} modified by the hexagonal Haldane model~\cite{Haldane}. Then, the Dirac cone of the monolayer acquires an additional magnetic phase $\varphi$. Importantly, this magnetic phase expands the solutions of magic angles to the complex eigenvalues of the Birman-Schwinger operator, in contrast to the real eigenvalues of the conventional TBG.  Interestingly, some complex eigenvalues correspond to quadruple FBs (per valley/spin) with absolute flatness and non-trivial topology. Moreover, to reveal the origin of the quadruple bands, we show that the WFs can be generated by holomorphic functions with two real-space nodes, and the generating functions naturally lead to absolute flatness with the four-fold degeneracy. Finally, we discuss that the four-fold degeneracy can enrich correlated physics due to additional filling selections in these multi-fold FBs.


\begin{figure}
\centerline{\includegraphics[width=0.38\textwidth]{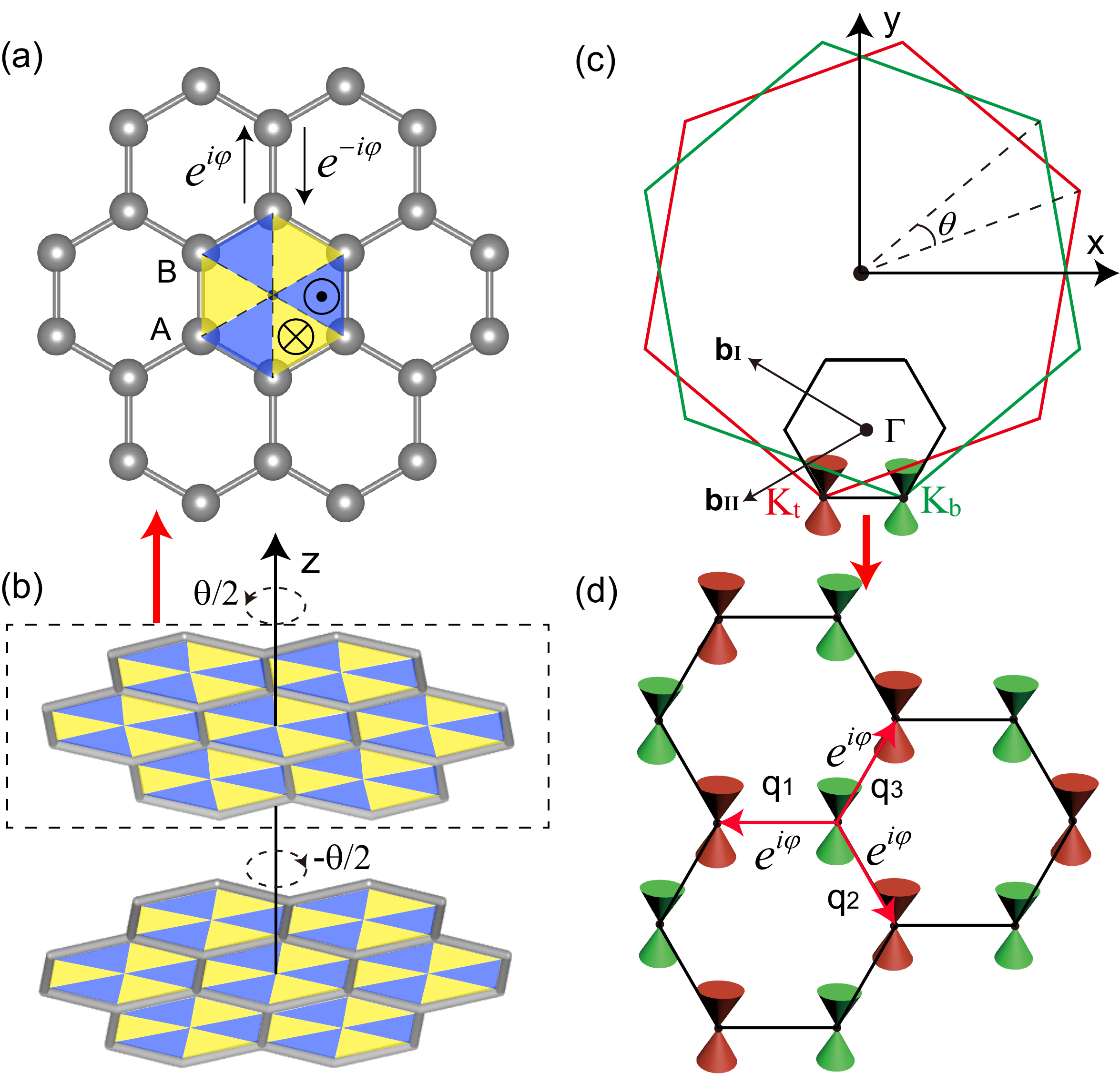}}
\caption{(color online) (a) Distribution of periodic local magnetic fluxes in hexagonal
plaquette of monolayer graphene with basis A and B.
(b) TBG with alternative fluxes. (c) The two monolayer BZs with the twist form a moir\'{e} BZ.
The red and green hexagons exhibit the BZs of the bottom and top layers, each rotated by an angle $\pm\theta/2$, and the black hexagon indicates the moir\'{e} BZ with momentum bases $\mathbf{b}_{\textbf{I}/\textbf{II}}=\sqrt{3}k_{\theta}(-\sqrt{3}/2,\pm1/2)$. (d) The reciprocal lattice structures. The interlayer coupling with the phase $\varphi$ couple the red and green Dirac cones separately located in the top and bottom layers.
\label{fig1} }
\end{figure}

\textit{Model}--To be specific, we consider monolayer graphene with alternative magnetic fluxes in each triangle inside a hexagonal plaquette (Fig.\ref{fig1}(a)), similar to the Haldane model\cite{Haldane}. The blue ($\bigodot$) and yellow ($\bigotimes$) triangles denote outward and inward to the plane fluxes, and the total flux through each hexagonal plaquette vanishes. With a Peierls substitution, the nearest-neighbor hopping acquires a magnetic phase $\varphi$. The Dirac Hamiltonian in the real space gains an additional phase and reads (See supplemental Material~(SM))
 \begin{eqnarray}
h^\varphi_{D}(\mathbf{r})=-v_0\left(\begin{array}{cc}
0 & e^{-i\varphi}2{\partial}     \\
e^{i\varphi}2\bar{\partial} &  0
\end{array}\right),
\end{eqnarray}
with $\bar{\partial}=\frac{1}{2 i}(\partial_{x}+i \partial_{y})$ and Fermi velocity $v_0$. The standard Dirac Hamiltonian is restored for monolayer graphene in the absence of magnetic flux. Then, we introduce a twist with angle $\theta$ between these two graphene layers (Fig.\ref{fig1}(b)). We use the basis of $\Phi(\mathbf{r})=\left(e^{i\frac{\varphi}{2}}\psi^{t}_{A}, e^{i\frac{\varphi}{2}}\psi^{b}_{A}, e^{-i\frac{\varphi}{2}}\chi^{t}_{B}, e^{-i\frac{\varphi}{2}}\chi^{b}_{B}\right)^{\top}$, which absorbs the magnetic phase, where $t/b$ denotes the top/bottom graphene layers and $A/B$ is the sublattice index. The chirally symmetric continuum model of TBG with alternating fluxes can be written as
\begin{eqnarray}
H^{\varphi}(\mathbf{r})&=&\left(\begin{array}{cc}
0 & D^{\varphi*}(-\mathbf{r})   \\
D^{\varphi}(\mathbf{r}) &  0  \\
\end{array}\right)\label{model1},
\\
D^{\varphi}(\mathbf{r})&=&\left(\begin{array}{cc}
2\bar{\partial} & e^{i(\varphi+\beta)}\alpha_0 U(\mathbf{r})     \\
 e^{i(\varphi-\beta)}\alpha_0 U(-\mathbf{r}) &  2\bar{\partial}
\end{array}\right)\label{model2},
\end{eqnarray}
where $U(\mathbf{r})=e^{-i \mathbf{q}_{1}\cdot \mathbf{r}}+e^{-i \mathbf{q}_{2}\cdot \mathbf{r}}e^{i\phi}+e^{-i \mathbf{q}_{3}\cdot \mathbf{r}}e^{-i\phi}$ with $\phi=2\pi/3$. $\mathbf{q}_1=k_{\theta}(-1,0)$, $\mathbf{q}_{2,3}=k_{\theta}(1/2,\mp\sqrt{3}/2)$ are shown in Fig.\ref{fig1}(d) with $k_{\theta}=2k_D\sin(\theta/2)$, and $k_D=|\mathbf{K}_{t/b}|$ is Dirac momentum in the monolayer graphene. The above Hamiltonian is characterized by a dimensionless real parameter $\alpha_0e^{i\beta}=\frac{\omega_{AB}}{v_0k_{\theta}}$, where $\omega_{AB}$ is the strength of the interlayer coupling with a constant phase $\beta$. Unlike the magnetic phase $\varphi$, the global phase $\beta$ can be gauged away. For simplicity, we define a complex parameter $\alpha (\varphi) \equiv \alpha_0 e^{i\varphi}$. In addition, $H^{\varphi}(\mathbf{r})$ is off-diagonal as we consider the chiral limit, where we neglect intra-sublattice (AA/BB) interlayer hopping. Since the twisted angle is small, we also neglect the rotation effect on the Dirac Hamiltonian $h^\varphi_{D}(\mathbf{r})$ \cite{Ashvin2019}. In the momentum space, the interlayer coupling $U(\mathbf{r})$ can be illustrated by the network of the two Dirac cones with the extension of the momentum hoppings, where the hopping between the nearest-neighbor momentum sites acquires a phase $e^{i\varphi}$ owing to the alternative fluxes (Fig.\ref{fig1}(c) and (d)). This non-zero phase plays a pivotal role in generating multi-fold FBs.

\begin{figure}
\centerline{\includegraphics[width=0.5\textwidth]{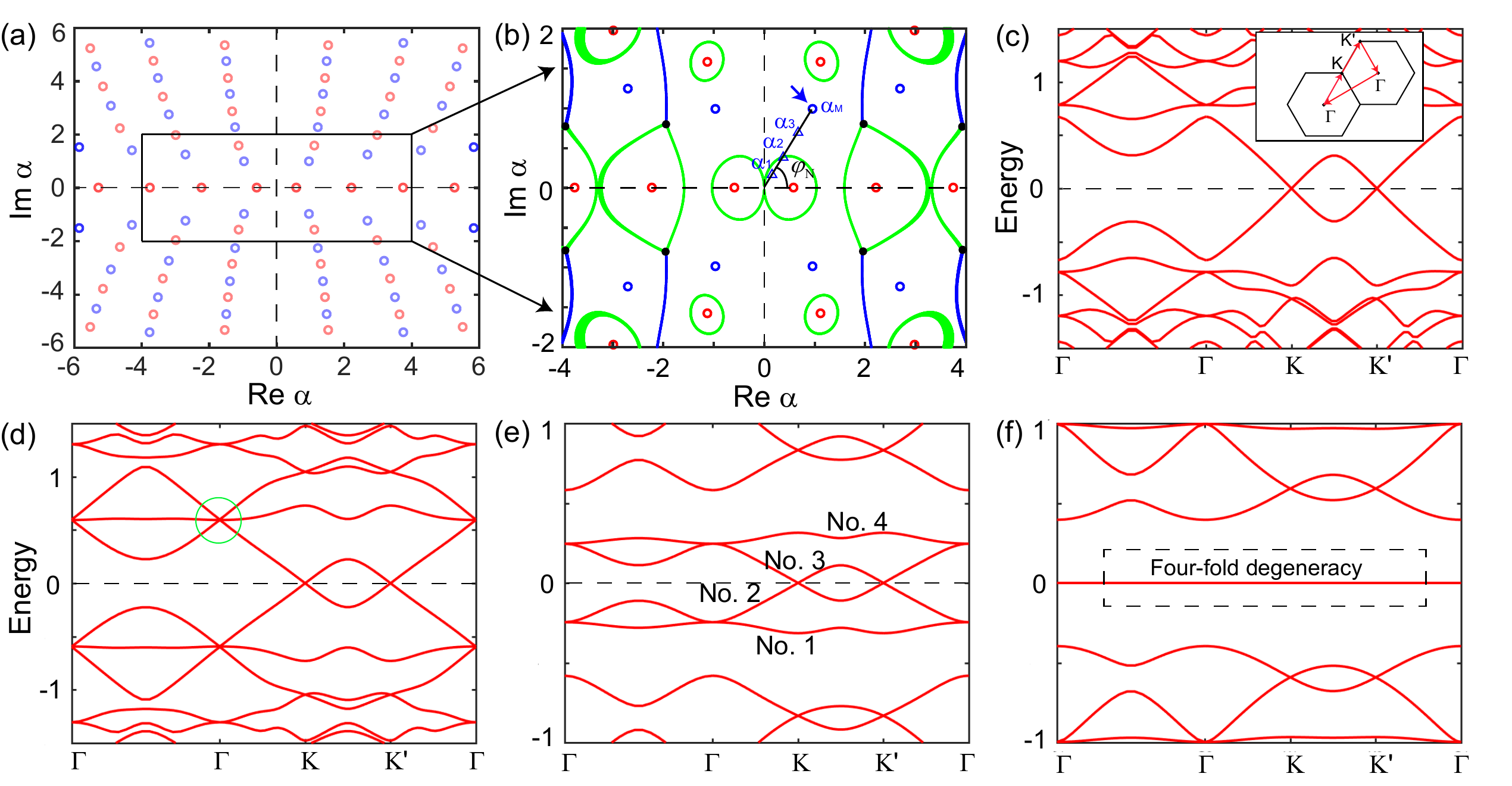}}
\caption{(color online)  (a) The spectra $\mathcal{A}$ shows the magic values of the dimensionless complex parameter $\alpha$ corresponding to double (red) and quadruple (blue) FBs.
(b) The distribution of lowest energy band touching at $\Gamma$ point in the spectra $\mathcal{A}$.
The band structures of the Hamiltonian $H^{\varphi}(\mathbf{r})$ with magnetic phase $\varphi_N=0.254\pi$ in the different $\alpha_0$ values marked by blue triangle in (b): (c) $\alpha_1=0.3e^{i\varphi_N}$, (d) $\alpha_2=0.55e^{i\varphi_N}$, (e) $\alpha_3=1.0e^{i\varphi_N}$ and (f) $\alpha_M=1.379e^{i\varphi_N}$. The inset in (c) indicates the spectrum path in the moir\'{e} BZ. The green circle in (d) denotes band inversion between double degenerate and non-degenerate points at $\Gamma$.
\label{fig2} }
\end{figure}

\textit{Multi-fold FBs}-- We further discuss the conditions of absolutely zero-energy FBs for the above continuum model. After a gauge transformation $V=diag(e^{-i\mathbf{k}\cdot \mathbf{r}},e^{-i\mathbf{k}\cdot \mathbf{r}})$, the Hamiltonian $H^{\varphi}_{\mathbf{k}}(\mathbf{r})=V^{\dagger}H^{\varphi}(\mathbf{r})V$  can be written as
\begin{eqnarray}
H^{\varphi}_{\mathbf{k}}(\mathbf{r})=\left(\begin{array}{cc}
0 & D^{\varphi*}_{\mathbf{k}}(-\mathbf{r})   \\
D^{\varphi}_{\mathbf{k}}(\mathbf{r}) &  0  \\
\end{array}\right),
\end{eqnarray}
where $D^{\varphi}_{\mathbf{k}}(\mathbf{r})\equiv D^{\varphi}(\mathbf{r})-\mathbf{k}$ and the momentum $\mathbf{k}$ is defined in moir\'{e} BZ. The explicit form of $D^{\varphi}_{\mathbf{k}}(\mathbf{r})$ is given by
\begin{eqnarray}
&D^{\varphi}_{\mathbf{k}}(\mathbf{r})=(2\bar{\partial}-\mathbf{k})(I+\alpha (\varphi)T_{\mathbf{k}}),\nonumber
\\
&T_{\mathbf{k}}={(2\bar{\partial}-\mathbf{k})^{-1}}\left(\begin{array}{cc}
0 & U(\mathbf{r}) \\
U(-\mathbf{r}) & 0
\end{array}\right),
\end{eqnarray}
where $T_{\mathbf{k}}$ ($\mathbf{k}\ne 0$) is known as the Birman-Schwinger operator \cite{Simon2021,Simonv4}. The appearance of zero-energy FBs implies that the determinant of Hamiltonian $H^{\varphi}_{\mathbf{k}}(\mathbf{r})$ vanishes for all $\mathbf{k}$, i.e. $\det(D^{\varphi}_{\mathbf{k}}(\mathbf{r}))=0$. Due to $\det(2 \bar{\partial}-\mathbf{k})\ne 0$, for all non-zero $\textbf{k}$ the matrix $[I-\alpha (\varphi) T_\mathbf{k}]$ should have at least one zero eigenvalue and for $\mathbf{k}=0$ two low-energy states are fixed at zero energy. According to Ref.\cite{Simon2021,Simonv4}, the eigenvalues of $T_{\mathbf{k}}$ are independent of $\textbf{k}$ and we can define a spectrum $\mathcal{A}=1 /\operatorname{Spec}\left(T_{\mathbf{k}}\right)$ and a corresponding two-component eigenstate $\psi^0_\mathbf{k}(\mathbf{r})$. Therefore, once the parameter $\alpha$ is tuned to be one of the complex eigenvalues in $\mathcal{A}$, $D^{\varphi}_{\mathbf{k}}(\mathbf{r})\psi^0_\mathbf{k}(\mathbf{r})=0$ so that zero-energy FBs emerge in $H_{\mathbf{k}}^{\varphi}(\mathbf{r})$ and $\alpha$ has a magic value.


In the absence of the alternative magnetic field ($\varphi=0$), the magic parameters $\alpha$ in the conventional TBG appear recursively on the real axis of spectrum $\mathcal{A}$\cite{Simon2021,Simonv4}, and indicate the presence of double FBs (Fig.\ref{fig2}(a)). When the alternative magnetic field is recovered, the magic values of $\alpha$ are expanded to the entire complex plane. The spectrum $\mathcal{A}$ exhibit no degeneracy (red circles) and two-fold degeneracy (blue circles), respectively (Fig.\ref{fig2}(a)). The non-degenerate and two-fold degenerate eigenvalues correspond to double and quadruple FBs in magnetic TBG (see SM). That is, the degeneracy of FBs is twice the degeneracy at magic values. Introducing the alternative fluxes extends magic parameters to more discrete values and leads to the emergence of quadruple FBs.

To investigate the evolution of the degeneracy in the spectra of the magnetic TBG, we take the complex eigenvalue closest to the origin in Fig.\ref{fig2}(b) as a example with magnetic phase $\varphi_{N}=0.254\pi$ to demonstrate the changes of the low energy bands. The degeneracy of FBs is intimately related to the degeneracy of the lowest energy states at $\Gamma$ in the TBG spectrum. To understand the transition of FBs with different degeneracies as $\alpha$ varies, we can track the band touch of the low-energy states at $\Gamma$ (Fig.\ref{fig2}(b)). The green (blue) curves represent a three-fold (four-fold) band touching at $\Gamma$, resulting in band inversions between one singlet and one doublet (between two doublets). Importantly, the three-fold degeneracy curves (Fig.\ref{fig2}(d)) separate the magic parameters $\alpha$ with  double and  quadruple FBs. Inside one of these degeneracy loops, two connected bands near zero energy in a direct gap (Fig.\ref{fig2}(c)) evolve to double FBs at magic $\alpha$ (red circles). As $\alpha$ crosses a three-fold degeneracy line, the band inversion changes two connected bands to four connected ones in the gap (Fig.\ref{fig2}(e)). Outside these degeneracy loops, the four connected bands evolve to quadruple FBs (Fig.\ref{fig2}(f)) at magic $\alpha$ (blue circles). Then, as discussed in SM, we consider the effect of intra-sublattice interlayer coupling, which can result in a slight dispersion of the quadruple FBs. In addition, within the quadruple FBs, two of them (No.2/3) that constitute Dirac points possess exhibit Chern numbers of $\pm1$ while the Chern numbers of the remaining two bands (No.1/4) vanish.

\begin{figure}
\centerline{\includegraphics[width=0.43\textwidth]{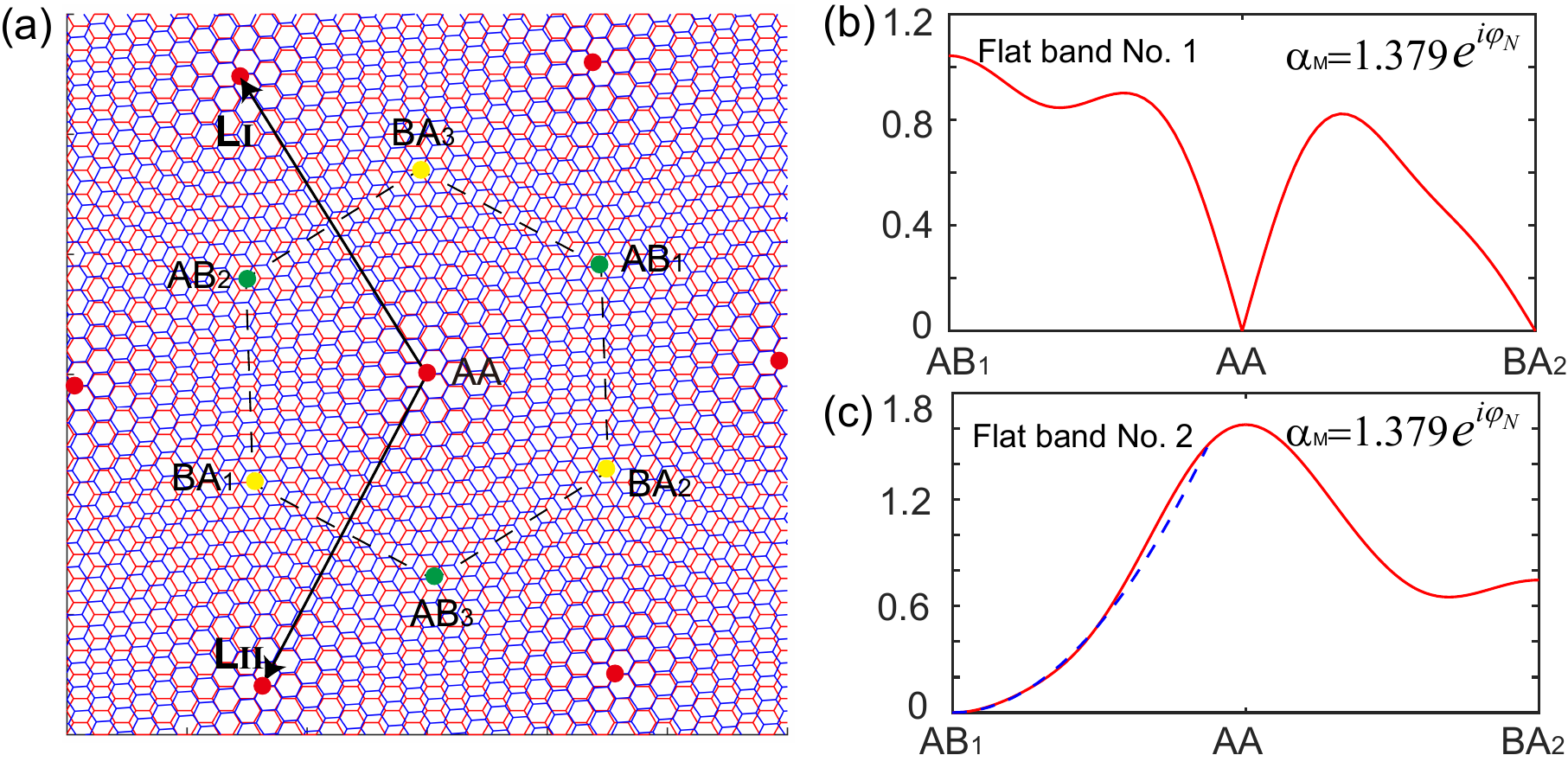}}
\caption{(color online) (a) Schematic moir\'{e} pattern. The red, green and yellow points indicate the $\mathbf{AA}$, $\mathbf{AB}$ and $\mathbf{BA}$ stacking points, and $\mathbf{L}_{\text{I}/\text{II}}=4\pi/3k_{\theta}(-1/2,\pm\sqrt{3}/2)$ are moir\'{e} lattice vector. (b) and (c) The norm of the 2-component WFs $\psi_{\mathbf{K},1/2}(\mathbf{r})$ at $\mathbf{K}$ point for FB No.$1/2$ at magic $\alpha_M$. The blue dashed line in (c) represents the fitting of quadratic function.
\label{fig3} }
\end{figure}

 \textit{Origin of  quadruple FBs}-- In the conventional TBG, the key to rigorously showing the emergence of the double absolutely FBs is that at the magic angles, the absolute values of the WFs with the zero energy at $\textbf{K}/\textbf{K}'$ have a linear-momentum-dispersion node at BA/AB stacking (Fig.\ref{fig3}(a)). Using this WF at $\textbf{K}$, one generates two eigenfunctions with zero energy at any momentum point ~\cite{Ashvin2019}. For the magnetic TBG, we can use a similar approach to show the absolute flatness of the quadruple bands at magic $\alpha$ by identifying the nodes of the WFs at $\textbf{K}$ and extending this function to the entire Moir\'{e} BZ.

	To distinguish the quadruple FBs, consider $\alpha$ slightly away from the magic value (Fig.\ref{fig2}(e)). The two lowest bands are labeled by No.1/2, and the remaining two bands can be generated by chiral symmetry operator $S=\tau_z\sigma_0$. In addition, due to $C_2T$ symmetry, the non-degenerate 4-component WF $\Psi_{\mathbf{k},i}(\mathbf{r})$ can be expressed by a 2-component vector $\psi_{\mathbf{k},i}(\mathbf{r})$ with band index $i=1,2$. At the magic $\alpha_M=1.379e^{i\varphi_N}$, the norm $[\psi^{\dagger}_{\mathbf{K},1/2}(\mathbf{r})\psi_{\mathbf{K},1/2}(\mathbf{r})]^{1/2}$ shows that for FB No.$1$ at $\textbf{K}$, two nodes appear at the AA and BA stacking with a dominant linear real-space dependence (Fig.\ref{fig3}(b)). Meanwhile, for FB No.$2$ at $\textbf{K}$, there is only one node at the AB stacking, and the norm exhibits real-space quadratic dependence near it, implying a second order node (Fig.\ref{fig3}(c)). These nodal features in sharp contrast to the conventional TBG are the important ingredients for the absolute flatness of the quadruple bands.

To simplify the absolute flatness' problem, we focus on the solution of $D^{\varphi}(\mathbf{r}) \psi_{\mathbf{k}}(\mathbf{r})=0$ for all $\mathbf{k}$.  Previously, we numerically obtain the two-component WFs $\psi_{\mathbf{K}}(\mathbf{r})$ pinned at zero energy at $\mathbf{K}$, which is a good starting point to construct WFs of the FBs at other $\textbf{k}$. 
As $\mathbf{k}\neq\mathbf{K}$, the conjectural WF can be written as $\psi_{\mathbf{k}}(\mathbf{r})\equiv f_{\mathbf{k}}(z) \psi_{\mathbf{K}}(\mathbf{r})$ with $z=x+iy$ since $D^{\varphi}(\mathbf{r})\psi_{\mathbf{k}}(\mathbf{k})=f_{\mathbf{k}}(z)[D^{\varphi}(\mathbf{r})\psi_{\mathbf{K}}(\mathbf{r})]=0$. The holomorphic function $f_{\mathbf{k}}(z)$ is either a constant or unbounded by Liouville's theorem, indicating that $f_{\mathbf{k}}(z)$ is meromorphic and thus has poles as $\mathbf{k}\neq\mathbf{K}$. The eigenfunctions $\psi_{\mathbf{k}}(\mathbf{r})$ are valid when the poles of $f_{\mathbf{k}}(z)$ are smoothed out by the nodes of $\psi_{\mathbf{K}}(\mathbf{r})$ in the Moir\'{e} unit cell. For the double FBs in the conventional TBG, only one node appears in $\psi_{\mathbf{K}}(\mathbf{r})$. Differently, for the quadruple FBs, a second-order node at AB stacking point emerges in $\psi_{\mathbf{K},2}(\mathbf{r})$ of the band No.$2$ at the magic $\alpha$, while there are already two linear nodes in the $\psi_{\mathbf{K},1}(\mathbf{r})$ of the band No.$1$. We note that $\psi_{\mathbf{K},2}(\mathbf{r})$ is always pinned at the zero-energy and possesses the node only at the magic $\alpha$. In contrast, $\psi_{\mathbf{K},1}(\mathbf{r})$ always has the two nodes and drops to zero energy at the magic $\alpha$.  The zero energy and the presence of the nodes at the magic $\alpha$ lead to the solution of the zero-energy absolutely FBs in the entire Moir\'{e} BZ.


We turn to the construction of WFs for the quadruple FBs. To cancel the poles of $\Psi_{\mathbf{K}}(\mathbf{r})$, we choose two theta functions in the denominator of $f_{\textbf{k}}(z)$ and adjust $f_{\textbf{k}}(z)$ to render the two-component WF $\psi_{\mathbf{k}}(\mathbf{r})$ to obey the Bloch boundary conditions on moir\'{e} lattice vectors $\mathbf{L}_{\text{I/II}}$, namely $\psi_{\mathbf{k}}(\mathbf{r}+\mathbf{L}_{\text{I/II}})=e^{i \mathbf{k}\cdot\mathbf{L}_{\text{I/II}}}U \psi_{\mathbf{k}}(\mathbf{r})$ with $U=\operatorname{diag}(e^{-i\phi}, e^{i\phi})$~\cite{Ashvin2019}. The analytical expressions for the quadruple FB WFs $\Psi_{\mathbf{k},\text{1/2}}(\mathbf{r})$ read,
\begin{eqnarray}
\Psi_{\mathbf{k},\text{1}}(\mathbf{r})=\frac{\vartheta_{a_1, b_1}(\nu | \omega)\vartheta_{a_1^{\prime}, b_1^{\prime}}(\nu | \omega)}{\vartheta_{\frac{1}{2}, \frac{1}{2}}(\nu | \omega)\vartheta_{\frac{5}{6}, \frac{7}{6}}(\nu | \omega)} \Psi_{\mathbf{K},\text{1}}(\mathbf{r}), \label{wave function1maintext}
\\
\Psi_{\mathbf{k},\text{2}}(\mathbf{r})=\frac{\vartheta_{a_2, b_2}(\nu | \omega)\vartheta_{a_2^{\prime}, b_2^{\prime}}(\nu | \omega)}{[\vartheta_{\frac{7}{6}, \frac{5}{6}}(\nu | \omega)]^2}\Psi_{\mathbf{K},\text{2}}(\mathbf{r}),\label{wave function2maintext}
\end{eqnarray}
with $\nu={z}/{{L}_\text{I}}$, $\omega= \frac{{L}_\text{II}}{{L}_\text{I}}=e^{i\phi}$ and ${L}_{\text{I/II}}=(\mathbf{L}_{\text{I/II}})_x+i(\mathbf{L}_{\text{I/II}})_y$. To satisfy the Bloch boundary conditions, the rational characteristics $a$ and $b$ obey
\begin{align}
a_i +a_i' =& \frac{1}{3} +\frac{(\mathbf{k}-\mathbf{K}) \cdot \mathbf{L}_{\text{I}}}{2 \pi}+ n_{\text{I}}, \label{a eq}\\
b_i + b_i' =& \frac{2}{3} -\frac{(\mathbf{k}-\mathbf{K}) \cdot \mathbf{L}_{\text{II}}}{2 \pi}+ n_{\text{II}}, \label{b eq}
\end{align}
where $n_{\text{I}}$ and $n_{\text{II}}$ are arbitrary integers due to the lattice transnational symmetry. The definition of the theta function $\vartheta_{a, b}(\nu | \omega)$ and the detailed derivations of Eq.\ref{wave function1maintext}/\ref{wave function2maintext} are provided in the SM. The equations above cannot determine the explicit expressions of the rational characteristics, which are functions of $\mathbf{k}$, unless the node locations are given.


\begin{figure}
\centerline{\includegraphics[width=0.4\textwidth]{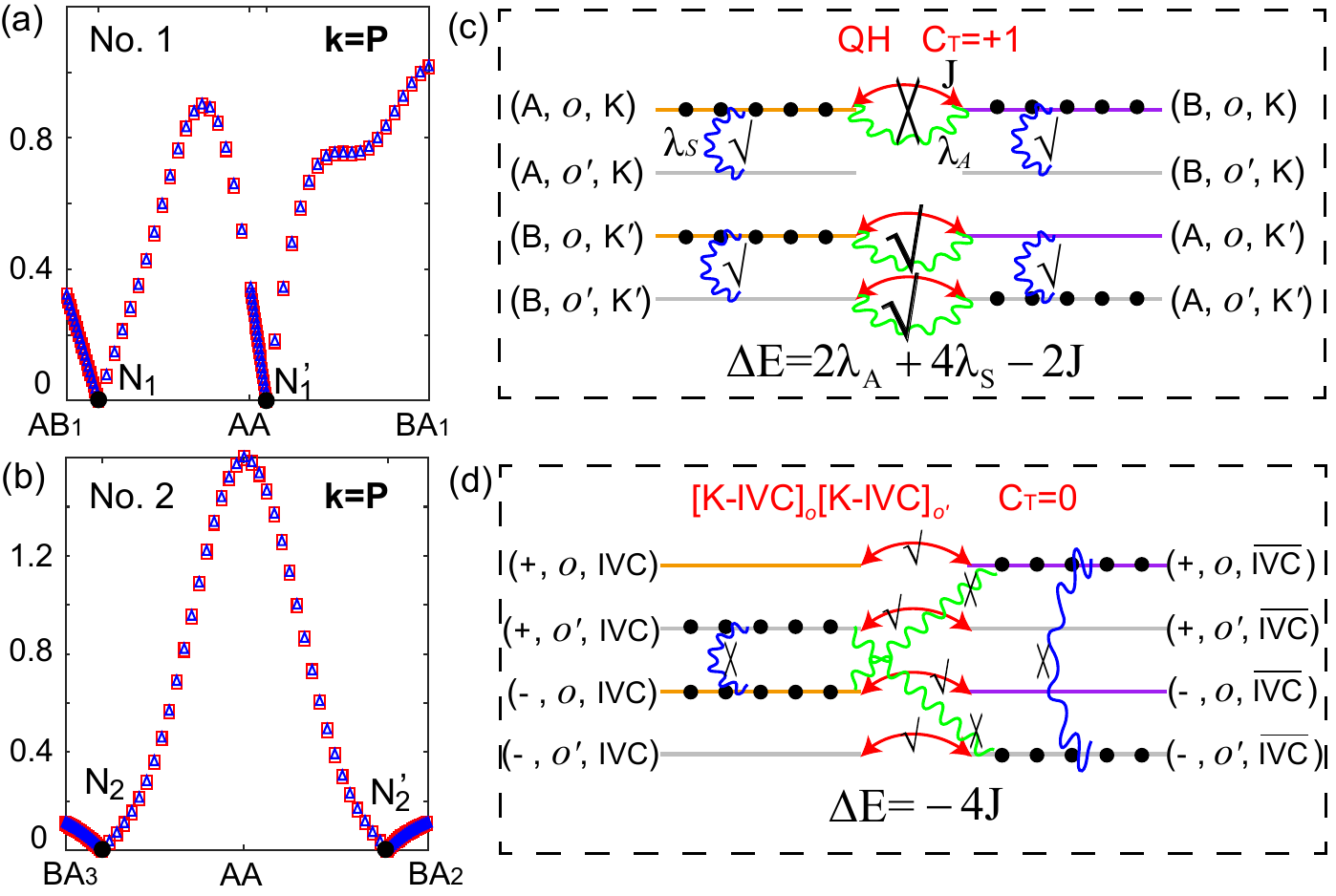}}
\caption{(color online) (a) and (b) The norm of WFs $\psi_{\mathbf{k},1/2}(\mathbf{r})$ for FBs No.$1/2$ at $\mathbf{k}=\mathbf{P}$ with $\mathbf{P}=0.5\mathbf{b}_\textbf{I}-0.3\mathbf{b}_\textbf{II}$. $N_{1/2}$ and $N^{\prime}_{1/2}$ are corresponding nodal points. (c) and (d) The configurations for QH and [K-IVC]$_o$[K-IVC]$_{o^{\prime}}$ states with $C_T=+1$ and $C_T=0$, and the IVC bases are a coherent superposition of two valley FB bases and the relations can be found in SM. The FBs labelled by gray, orange and pink lines possess Chern number $C=0$, $C=+1$, and $C=-1$. The symbols $\surd$ or $\times$ indicate whether the processes are permitted or forbidden.
\label{fig6}}
\end{figure}



 To prove the absolutely FBs, we numerically calculate the flat-band WFs from Hamiltonian $H^{\varphi}(\mathbf{r})$ and determine the node locations. The node locations fix the values of $a,\ b$ in theta functions Eq.\ref{wave function1maintext}/\ref{wave function2maintext}. To confirm the validity of the WF expressions in Eq.\ref{wave function1maintext}/\ref{wave function2maintext}, we first select a general point $\mathbf{P}$ as an example. Fig.\ref{fig6} (a) and (b) show the norm of WFs $\psi_{\mathbf{k},1/2}$ for FB No.$1/2$ at $\mathbf{k}=\mathbf{P}$, where the WFs from Hamiltonian $H^{\varphi}(\mathbf{r})$ and Eq.\ref{wave function1maintext}/\ref{wave function2maintext} are labelled in red squares and blue triangles, respectively. We find a good agreement between them, and the agreement can be extended to other general $\textbf{k}$ points (see SM). Therefore, through the evolution of nodes and real-space WF, we have verified the validity of WF in Eq.\ref{wave function1maintext}/\ref{wave function2maintext}, and quadruple FBs are absolutely flat. In SM, we further confirm the validity of the relation between Eq.\ref{a eq}/\ref{b eq} and nodal coordinates.

\textit{Correlation states}--The presence of multiple-fold FBs provides an excellent platform for exploring correlated quantum states. Besides the valley and sublattice degrees of freedom in double FBs~\cite{Bultinck2020}, the quadruple FBs introduce an emergent orbital degree of freedom, which can significantly enrich the correlated states. Within the quadruple FBs, two of them (No.2/3) that constitute Dirac points possess exhibit Chern numbers of $\pm1$ labeled by $o$-orbital while the Chern numbers of the remaining two bands (No.1/4 $o'$-orbital) vanish. Therefore, in the spinless case with the two valleys, the total Chern numbers of $\pm 2$ can be realized
in the Quantum Hall (QH) states at integer filling\cite{Zhang2019_Nearly,Zhang2019_Twisted,Bultinck2020}. With
a fractional filling of the high-Chern-number FBs\cite{Ledwith2020},
fractional quantum Hall states at a variety of filling factors
can be realized. We use the half-filling scenario (four out of the eight FBs are filled) as an primary example. Assuming the intrasublattice-intraorbital interaction scale is much larger than other scales, the ground state of this interaction is the one where each of the eight bands is completely filled or empty~\cite{Bultinck2020}. There are $C^8_4=70$ degenerate many-body ground states (see SM), while only $C^4_2=6$ ground states are present in the conventional TBG. However, for intervalley-coherent (IVC) states, the types of ground states increase significantly, and an exotic combination of the QH and IVC states, such as time-reversal IVC (T-IVC) and Kramers IVC (K-IVC) states, can be realized (see SM)~\cite{Bultinck2020}. We schematically show the two representative examples: orbital related QH and  [K-IVC]$_o$[K-IVC]$_{o^{\prime}}$ states in Fig.4 (c) and (d). The former state features three fully occupied $o$-orbital FBs and one fully occupied $o^{\prime}$-orbital FB, where the $o$-orbital can contribute to the Chern number $C_T=+1$. In the latter state,  both o and o' orbital sectors feature  K-IVC states with anti-parallel pseudospin between two orbitals.



Now we introduce the effect of single-particle dispersion, intersublattice-intraorbital interaction and intrasublattice-interorbital interaction as perturbations, so the high degeneracy of the ground states can be lifted. The single-particle dispersion introduces a hopping between the horizontal pair states (A$\leftrightarrow$B) (denoted by red curves in Fig.\ref{fig6}(c)), yielding a energy reduction of the order of $J$. For the latter two interaction terms, the corresponding energy increments are proportional to $\lambda_{A/S}\sum_{i=1,2}\hat{n}_i(1-\hat{n}_{\bar{i}})$, where $\hat{n}_i$ is the particle number operator for one of horizontal (A$\leftrightarrow$B) or vertical ($o\leftrightarrow o'$) pair states. They are denoted by green and blue wavy lines in Fig.\ref{fig6} (c) and (d), respectively. It is evident that three types of energy corrections are nonzero when only one of the pair states is occupied. Therefore, the energy correction can be formulated as $\Delta E = N_A \lambda_A + N_S \lambda_S - N_J J$\cite{explain}. From the energy, we identify the anti-orbital K-IVC state in Fig.\ref{fig6}(d) as the ground state at charge neutrality. Varying doping and twisted angles may induce transitions between different correlated states, necessitating detailed Hartree Frock calculations, which we defer to future investigations.

\textit{Discussion and conclusion}--In recent experiments, FBs have been successfully achieved in twisted-bilayer optical lattice systems\cite{Meng2023}. Our proposed model can be implemented within this tunable optical lattice platform, thanks to its large spatial scale, making it possible to achieve alternative magnetic field. In summary, introducing alternative fluxes opens a venue for undiscovered multiple magic angles and another generation of absolutely FBs with higher degeneracy. Specifically, we report that the double and quadruple FBs appear recursively in TBG with alternative magnetic fluxes, and adjusting the twisted angle and magnetic phase can control the transition between double and quadruple FBs. The quadruple FBs share the same origin with the conventional TBG, and by using holomorphic functions, we can generate analytical WFs to show the absolute flatness of the quadruple bands. Moreover, additional orbitals in the quadruple bands lead to more diversified strongly correlated physics. Our work demonstrates that the magnetic phase is a unique tuning parameter to tailor the electronic structure in moir\'{e} twistronics, which can significantly enrich topological and correlated phenomena.

\textit{Acknowledgments}-- We thank Zhida Song, Simon Becker and Xu Zhang for the helpful discussions and comments. C.-K.C. was supported by JST Presto Grant No. JPMJPR2357.

%

%

\clearpage

\appendix
\begin{widetext}

\tableofcontents

\section{Effective model, symmetry analysis, topological property and condition for multi-fold degenerate flat bands}

In this section, we provide a comprehensive derivation of the chirally symmetric continuum model that describes twisted bilayer graphene (TBG) under a periodic alternating magnetic field. We engage in an in-depth analysis of the symmetry associated with this system, including approximate symmetries in the low energy and correspondence of symmetries between momentum space and real space. Furthmore, we elucidate the conditions for the emergence of multi-fold degenerate flat bands, and topological properties for the flat bands. In addition, we analyze the impact of intra-sublattice interlayer coupling on these flat bands.

\subsection{Chirally symmetric continuum model}

\begin{figure}[b]
\centerline{\includegraphics[width=0.7\textwidth]{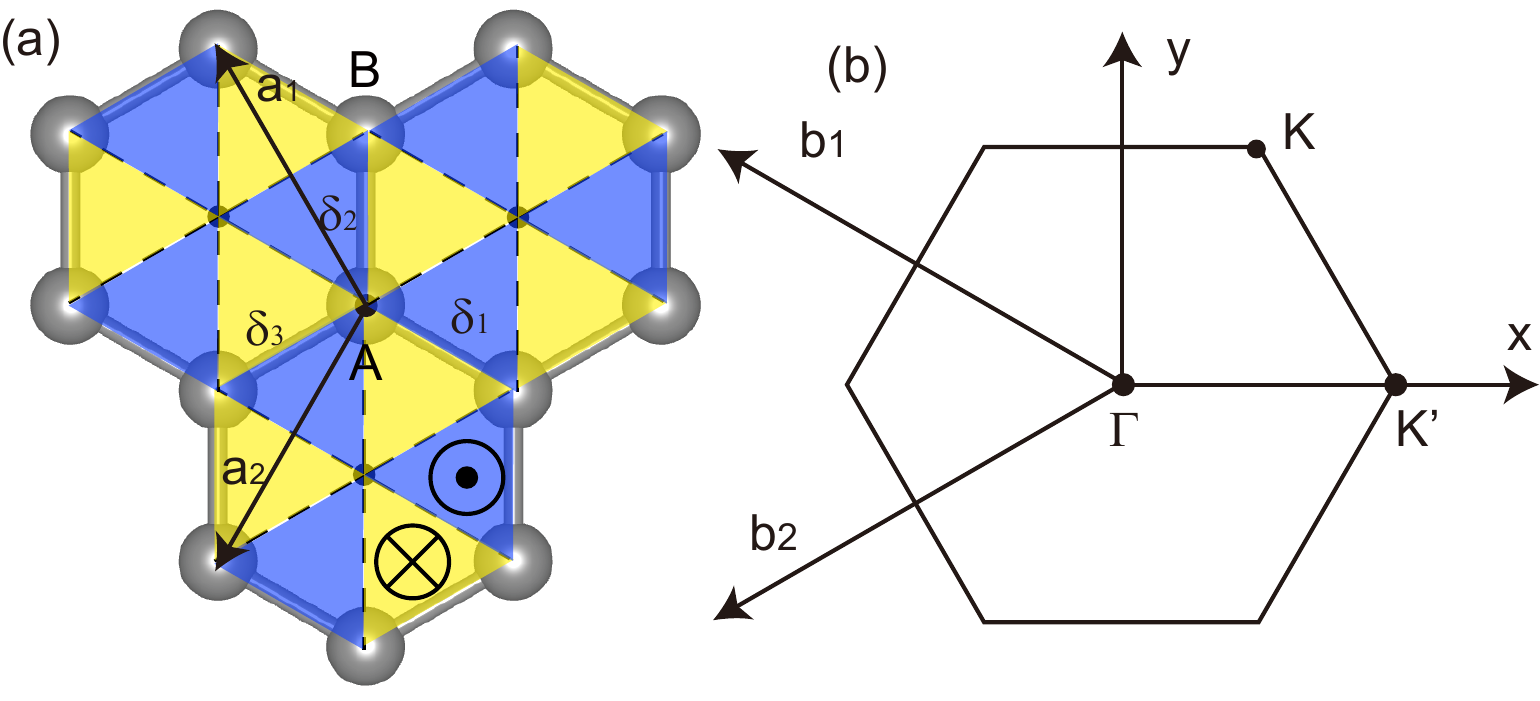}}
\captionsetup{justification=raggedright}
\caption{(color online) (a) Distribution of periodic local magnetic fluxes in hexagonal plaquette of monolayer graphene with sublattices A and B. The blue region ($\bigodot$) and yellow region ($\bigotimes$) denote outward-to-the-plane and inward-to-the-plane magnetic fluxes, respectively.  (b) The monolayer Brillouin Zone.
\label{graphene} }
\end{figure}

The monolayer graphene, composed of sublattices A and B as depicted in Fig.\ref{graphene}(a), is characterized by translation vectors $\mathbf{a}_{1}=a(-\frac{1}{2},\frac{\sqrt{3}}{2})$ and $\mathbf{a}_{2}=a(-\frac{1}{2},-\frac{\sqrt{3}}{2})$, where $a$ denotes the lattice constant. The corresponding reciprocal vectors, illustrated in Fig.\ref{graphene}(b), are $\mathbf{b}_{1}=\frac{4\pi}{a\sqrt{3}}(-\frac{\sqrt{3}}{2},\frac{1}{2})$ and $\mathbf{b}_{2}=\frac{4\pi}{a\sqrt{3}}(-\frac{\sqrt{3}}{2},-\frac{1}{2})$.
It is a well-established fact that two Dirac points are located at $\mathbf{K}$ and $\mathbf{K}^{\prime}$ points in monolayer graphene. Through the utilization of the $p_z$ orbital, a tight-binding Hamiltonian can be established to to describe Dirac points, which can be expressed as follows
\begin{eqnarray}\label{Hamiltonian}
H(\mathbf{k})=\left(\begin{array}{cc}
0 & h_{12}(\mathbf{k})  \\
h^*_{12}(\mathbf{k}) & 0
\end{array}\right),
\end{eqnarray}
where $h_{12}(\mathbf{k})=t(e^{i \mathbf{k} \cdot \mathbf{\delta}{1}}+e^{i \mathbf{k} \cdot \mathbf{\delta}{2}}+e^{i \mathbf{k} \cdot \delta_{3}})$ with $t$ representing the nearest-neighbor hopping along the $\mathbf{\delta}$ bonds. These bonds are defined by vectors $\mathbf{\delta}_{1}=a(\frac{1}{2},-\frac{\sqrt{3}}{6})$, $\mathbf{\delta}_{2}=a(0,\frac{\sqrt{3}}{3})$ and $\mathbf{\delta}_{3}=a(-\frac{1}{2},-\frac{\sqrt{3}}{6})$, as illustrated in Fig. \ref{graphene}(a). By conducting a first-order expansion of Eq.\ref{Hamiltonian} around $\mathbf{K}$ and implementing a gauge transformation denoted as $g=diag(e^{-i\frac{\pi}{3}}, e^{i\frac{\pi}{3}})$, we can get the standard Dirac Hamiltonian $h_D(\mathbf{K}+\mathbf{k})=-v_0(\sigma_x\mathbf{k}_x+\sigma_y\mathbf{k}_y)$ with Dirac velocity $v_0=-\frac{\sqrt{3}}{2}t$.

We explore the influence of an periodic alternative magnetic fluxes in monolayer graphene, as shown in Fig.\ref{graphene}(a). In this depiction, the blue region ($\bigodot$) corresponds to outward-to-the-plane magnetic fluxes, while the yellow region ($\bigotimes$) signifies inward-to-the-plane magnetic fluxes. The presence of the alternating magnetic fluxes introduces a magnetic phase $\varphi$ into the nearest-neighbor hopping process, and subsequently the tight-binding Hamiltonian Eq.\ref{Hamiltonian} becomes
\begin{eqnarray}
H^{\varphi}(\mathbf{k})=\left(\begin{array}{cc}
0 &  e^{i\varphi}h_{12}(\mathbf{k})\\
e^{-i\varphi}h^{*}_{12}(\mathbf{k}) & 0
\end{array}\right).
\end{eqnarray}

Then, the corresponding Dirac Hamiltonian around $\mathbf{K}$ becomes
\begin{eqnarray}
h_D^{\varphi}(\mathbf{K}+\mathbf{k})=-v_0\left(\begin{array}{cc}
0 &  e^{i\varphi}(\mathbf{k}_x-i\mathbf{k}_y) \\
 e^{-i\varphi}(\mathbf{k}_x+i\mathbf{k}_y) & 0
\end{array}\right).
\end{eqnarray}

In the real space, the Hamiltonian $h_D^{\varphi}(\mathbf{K}+\mathbf{k})$ can be written as
\begin{eqnarray}
h_D^{\varphi}(\mathbf{r})=-v_0\left(\begin{array}{cc}
0 &  e^{i\varphi}2{\partial} \\
 e^{-i\varphi}2\bar{\partial} & 0
\end{array}\right),
\end{eqnarray}
where ${\partial}=\frac{1}{2i}(\partial_{x}-i \partial_{y})$ and $\bar{\partial}=\frac{1}{2i}(\partial_{x}+i \partial_{y})$. Similar to conventional TBG\cite{Vishwanath2019SM}, in the basis $\Phi(\mathbf{r})=(\psi^{t}_{A}, \chi^{t}_{B}, \psi^{b}_{A},  \chi^{b}_{B})^{T}$ with layer index $t/b$, the chirally symmetric model of TBG with magnetic phase can be written as
\begin{eqnarray}\label{model_phase}
H^{\varphi}(\mathbf{r})&=&\left(\begin{array}{cccc}
0 & e^{i\varphi}2{\partial}  &  0 & e^{i\beta}\alpha_0 U^{*}(-\mathbf{r}) \\
e^{-i\varphi}2\bar{\partial} &  0   & e^{i\beta} \alpha_0U(\mathbf{r})   & 0\\
0 & e^{-i\beta} \alpha_0U^{*}(\mathbf{r}) & 0   & e^{i\varphi}2{\partial} \\
e^{-i\beta} \alpha_0U(-\mathbf{r}) &  0   & e^{-i\varphi}2\bar{\partial}   & 0\\
\end{array}\right),
\end{eqnarray}
where $U(\mathbf{r})=e^{-i \mathbf{q}_{1}\cdot \mathbf{r}}+e^{-i \mathbf{q}_{2}\cdot \mathbf{r}}e^{i\phi}+e^{-i \mathbf{q}_{3}\cdot \mathbf{r}}e^{-i\phi}$ with $\phi=2\pi/3$, $\mathbf{q}_1=k_{\theta}(-1,0)$ and $\mathbf{q}_{2,3}=k_{\theta}(1/2,\mp\sqrt{3}/2)$. The above Hamiltonian $H^{\varphi}(\mathbf{r})$ is characterized by a dimensionless real parameter $\alpha_0e^{i\beta}=\frac{\omega_{AB}}{v_0k_{\theta}}$, where $\omega_{AB}$ is the strength of the interlayer coupling with a constant phase $\beta$. Unlike the magnetic phase $\varphi$, the global phase $\beta$ can be gauged away. By reshuffling the basis to $\Phi(\mathbf{r})=(\psi^{t}_{A}, \psi^{b}_{A}, \chi^{t}_{B},  \chi^{b}_{B})^{T}$, and performing gauge transformation $g_{\varphi}=diag(e^{i\frac{\varphi}{2}}, e^{i\frac{\varphi}{2}}, e^{-i\frac{\varphi}{2}}, e^{-i\frac{\varphi}{2}})$ the Hamiltonian $H^{\varphi}(\mathbf{r})$ Eq.\ref{model_phase} reads
\begin{eqnarray}
H^{\varphi}(\mathbf{r})&=&\left(\begin{array}{cccc}
0 & 0  &  2{\partial} & e^{i(-\varphi+\beta)} \alpha_0U^{*}(-\mathbf{r}) \\
0 &  0   &e^{i(-\varphi-\beta)} \alpha_0U^{*}(\mathbf{r})   & 2{\partial}\\
2\bar{\partial} &e^{i(\varphi+\beta)} \alpha_0U(\mathbf{r}) & 0   & 0\\
e^{i(\varphi-\beta)}\alpha_0 U(-\mathbf{r}) &   2\bar{\partial}   &0   & 0\\
\end{array}\right).
\end{eqnarray}

Then, under gauge transformation $g_{\beta}=diag(e^{-i\frac{\beta}{2}}, e^{i\frac{\beta}{2}}, e^{-i\frac{\beta}{2}}, e^{i\frac{\beta}{2}})$, the global phase $\beta$ can be gauged away, and finally the Hamiltonian $H^{\varphi}(\mathbf{r})$ becomes
\begin{eqnarray}
H^{\varphi}(\mathbf{r})=\left(\begin{array}{cc}
0 & D^{\varphi*}(-\mathbf{r})   \\
D^{\varphi}(\mathbf{r}) &  0  \\
\end{array}\right),
D^{\varphi}(\mathbf{r})=\left(\begin{array}{cc}
2\bar{\partial} & \alpha(\varphi) U(\mathbf{r})     \\
\alpha(\varphi) U(-\mathbf{r}) &  2\bar{\partial}
\end{array}\right)\label{model2},
\end{eqnarray}
where we define a complex parameter $\alpha(\varphi)=\alpha_0e^{i\varphi}$. The chiral symmetric model $H(\mathbf{r})$ ($\varphi=0$) of conventional TBG can be written as
 \begin{eqnarray}
H(\mathbf{r})=\left(\begin{array}{cc}
0 & D^{*}(-\mathbf{r})   \\
D(\mathbf{r}) &  0  \\
\end{array}\right),
D(\mathbf{r})=\left(\begin{array}{cc}
2\bar{\partial} & \alpha_0 U(\mathbf{r})     \\
\alpha_0 U(-\mathbf{r}) &  2\bar{\partial}
\end{array}\right).
\end{eqnarray}

\subsection{Approximate symmetry in the low energy}

The Hamiltonian $H(\mathbf{r})$ of conventional TBG belongs to magnetic point group $6^{\prime}22^{\prime}$, including $C_{3z}=e^{i\frac{2\pi}{3}\tau_z}\sigma_0$, $C_{2z}T=\tau_x\sigma_0\mathcal{K}$, $C_{2y}=\tau_y\sigma_y$ and $C_{2x}T=i\tau_z\sigma_y\mathcal{K}$ with sublattice index $\tau$ and layer index $\sigma$, where $T$ is time reversal operator and $\mathcal{K}$ is the complex conjugation. These symmetries and the moir\'{e} translations generate the magnetic space group $P6^{\prime}2^{\prime}2$ (No.177.151 in BNS settings). In addition, we can introduce an anti-unitary particle-hole operator $\mathcal{P}=i\tau_x\sigma_y\mathcal{K}$, which satisfies $\mathcal{P}^2=-1$. It acts on the Hamiltonian $H(\mathbf{r})$ as
\begin{eqnarray}
\mathcal{P}H(\mathbf{r})\mathcal{P}^{-1}=-H(\mathbf{r}).
\end{eqnarray}

Furthermore, we define a chiral operator $S=\tau_z\sigma_0$ with $S^2=1$. Under chiral symmetry $S$, the Hamiltonian $H(\mathbf{r})$ transforms as
\begin{eqnarray}
SH(\mathbf{r})S^{-1}=-H(\mathbf{r}).
\end{eqnarray}

By combining particle-hole and chiral operators, we can define an effective time reversal operator $\tilde{T}=\tau_y\sigma_y\mathcal{K}$ with $\tilde{T}^2=1$. Subsequently, the Hamiltonian $H(\mathbf{r})$ is transformed as follows:
\begin{eqnarray}
\tilde{T}H(\mathbf{r})\tilde{T}^{-1}=H(\mathbf{r}).
\end{eqnarray}

Due to effective time reversal operator $\tilde{T}$, we can also define other crystalline symmetries, including ${C}_{2z}=i\tau_z\sigma_y$ and ${C}_{2x}=\tau_x\sigma_0$.

In the presence of the alternating magnetic fluxes, the Hamiltonian $H_{\mathbf{k}}^{\varphi}(\mathbf{r})$ exhibits a magnetic group $P6^{\prime}$(No.168.111 in BNS settings). The corresponding magnetic point group, denoted as $6^{\prime}$, encompasses symmetries such as $C_{3z}=e^{i\frac{2\pi}{3}\tau_z}\sigma_0$ and $C_{2z}T=\tau_x\sigma_0\mathcal{K}$. Notably, the crystalline symmetries ${C}_{2x}$ and $C_{2y}$ in conventional TBG are broken due to the introduction of magnetic phase $e^{i\varphi}$. However, the particle-hole operator $\mathcal{P}=i\tau_x\sigma_y\mathcal{K}$, the chiral operator $S=\tau_z\sigma_0$ and the effective time reversal operator $\tilde{T}=\tau_y\sigma_y\mathcal{K}$ are always preserved.

Despite the disruption of crystalline symmetries $C_{2x}$ and $C_{2y}$ induced by the magnetic phase, it is noteworthy that within the low-energy regime, the Hamiltonian $H^{\varphi}(\mathbf{r})$ approximately retains these symmetries. To prove this, we can define the Hamiltonian difference $\Delta H_{x/y}$ as
\begin{eqnarray}\label{difference}
\Delta H_x=({C}_{2x})H^{\varphi}(\mathbf{r})({C}_{2x})^{-1}-H^{\varphi}({C}_{2x}\mathbf{r}),~~~
\Delta H_y=(C_{2y})H^{\varphi}(\mathbf{r})(C_{2y})^{-1}-H^{\varphi}(C_{2y}\mathbf{r}).
\end{eqnarray}

Put ${C}_{2x}=\tau_x\sigma_0$ and $C_{2y}=\tau_y\sigma_y$ into Eq.\ref{difference}, and the Hamiltonian difference $\Delta H_{x/y}$ becomes
\begin{eqnarray}
\Delta H_{x}=2\sin{{\varphi}}\left(\begin{array}{cccc}
0 & 0 & {\partial} & 0 \\
0 & 0& 0 & {\partial} \\
 -\bar{\partial} & 0 & 0 & 0 \\
  0 & -\bar{\partial} & 0 & 0
\end{array}\right),~~~\Delta H_{y}=-2\sin{{\varphi}}\left(\begin{array}{cccc}
0 & 0 & {\partial} & 0 \\
0 & 0& 0 & {\partial} \\
 -\bar{\partial} & 0 & 0 & 0 \\
  0 & -\bar{\partial} & 0 & 0
\end{array}\right),
\end{eqnarray}
where $\varphi\ne\frac{\pi}{2}$. At $\varphi=\frac{\pi}{2}$, the Dirac Hamiltonian for monolayer is from $H_{Dirac}=-v\mathbf{k}\cdot\sigma$ to $H_{Dirac}=-v\mathbf{k}\times\sigma$, and the Hamiltonian $H^{\varphi}(\mathbf{r})$ has exact rotational symmetries ${C}_{2x}$ and $C_{2y}$.

 In the momentum lattice, we can arrange the basis as $\hat{\Psi}_{\mathbf{k}}=\sum_{m,n}\oplus(\hat{\psi}_{A ; \mathbf{k}+\mathbf{g}_{mn}}^{t}, \hat{\psi}_{B ; \mathbf{k}+\mathbf{g}_{mn}}^{t}, \hat{\psi}_{A; \mathbf{k}+\mathbf{g}_{mn}+\mathbf{q}_1}^{b},\hat{\psi}_{B; \mathbf{k}+\mathbf{g}_{mn}+\mathbf{q}_1}^{b})^{T}$ with $\mathbf{g}_{mn}$ being the lattice of momenta. Then, the the Hamiltonian difference $\Delta H_{x/y}$ can be rewritten as
\begin{eqnarray}
\Delta H_{x/y}=\pm2\sin{{\varphi}}\sum_{m,n}\oplus\left(\begin{array}{cc}
 (\mathbf{k}_x+\mathbf{g}_{mn,x})\sigma_x+(\mathbf{k}_y+\mathbf{g}_{mn,y})\sigma_y & 0 \\
 0 & (\mathbf{k}_x+\mathbf{g}_{mn,x}+\mathbf{q}_{1,x})\sigma_x+(\mathbf{k}_y+\mathbf{g}_{mn,y}+\mathbf{q}_{1,y})\sigma_y
\end{array}\right).
\end{eqnarray}
Consequently, the Hamiltonian difference $\Delta H_{x/y}$ is proportional to the magnitude of momentum $k$. In the low energy, the momentum $k$ is exceedingly small, and the $\Delta H_{x/y}$ approximately equal to zero. This implies that within the low-energy spectrum, the symmetries ${C}_{2x}$ and $C_{2y}$ are effectively retained. In Fig.\ref{symmetry}(b) and (c), we present the band structures corresponding to the magnetic phase $\varphi=0.254\pi$ at magic $\alpha_M=1.379$ along general lines $\Gamma\beta_1$, $\Gamma\beta_2$ and $\Gamma\beta_3$, where these lines in Fig.\ref{symmetry}(a) are interconnected by ${C}_{2x}$ and $C_{2y}$ symmetries. The general lines $\Gamma\beta_{1,2,3}$ remain consistent in the low-energy spectrum, while there are differences in the high-energy spectrum. Thus, it is evident that within the low-energy spectrum, the Hamiltonian $H^{\varphi}(\mathbf{r})$ approximately retains ${C}_{2x}$ and $C_{2y}$ symmetries.

\begin{figure}
\centerline{\includegraphics[width=0.9\textwidth]{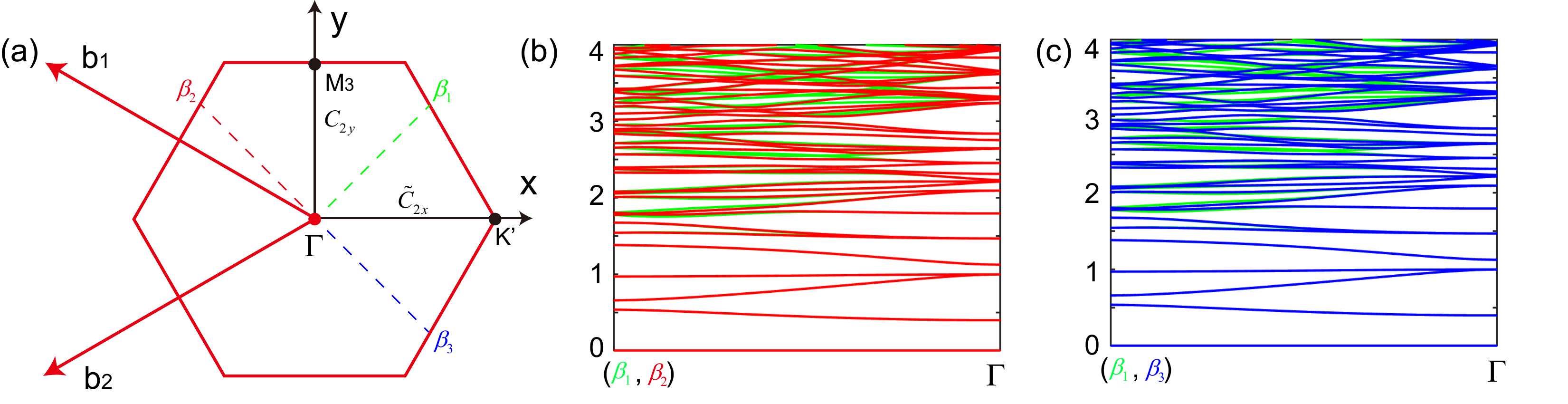}}
\captionsetup{justification=raggedright}
\caption{(color online) (a) Three general lines $\Gamma\beta_1$, $\Gamma\beta_2$ and $\Gamma\beta_3$ in moir\'{e} BZ are connected each other by ${C}_{2x}$ and $C_{2y}$ symmetries. (b) The band structure along $\Gamma\beta_1$ and $\Gamma\beta_2$. (c) The band structure along $\Gamma\beta_1$ and $\Gamma\beta_3$. The green, red and blue label the band structure along $\Gamma\beta_1$, $\Gamma\beta_2$ and $\Gamma\beta_3$, respectively.
\label{symmetry} }
\end{figure}


\subsection{ Symmetry correspondence between momentum space and real space}

Here, we discuss the symmetry correspondence between momentum space and real space in the wave function. The four-component wave function can be expressed as $\Psi_\mathbf{k}(\mathbf{r})=(\psi^{\top}_\mathbf{k}(\mathbf{r})\ \chi^{\top}_\mathbf{k}(\mathbf{r}))^{\top}$, where $\psi^{\top}_\mathbf{k}(\mathbf{r})$ and $\chi^{\top}_\mathbf{k}(\mathbf{r})$ denote the two component wave function. By acting $C_{2z}T$ symmetry on the wave function $\Psi_\mathbf{k}(\mathbf{r})$, we obtain the following result:
\begin{eqnarray}
C_{2z}T\left(\begin{array}{c}
\psi_\mathbf{k}(\mathbf{r}) \\
\chi_\mathbf{k}(\mathbf{r})
\end{array}\right)=\left(\begin{array}{c}
\chi^{*}_\mathbf{k}(\mathbf{r})\\
\psi^{*}_\mathbf{k}(\mathbf{r})
\end{array}\right)=\left(\begin{array}{c}
\psi_\mathbf{k}(-\mathbf{r}) \\
\chi_\mathbf{k}(-\mathbf{r})
\end{array}\right).
\end{eqnarray}
Hence, $\chi_\mathbf{k}(\mathbf{r})=\psi^{*}_\mathbf{k}(-\mathbf{r})$, and the wave function can be rewritten as $\Psi_\mathbf{k}(\mathbf{r})=(\psi^{\top}_\mathbf{k}(\mathbf{r})\ \psi^{\dagger}_\mathbf{k}(-\mathbf{r}))^{\top}$. Due to ${C}_{2x}$ symmetry, the wave function $\Psi_\mathbf{k}(\mathbf{r})$ becomes
\begin{eqnarray}
{C}_{2x}\left(\begin{array}{c}
\psi_\mathbf{k}(\mathbf{r}) \\
\psi^{*}_\mathbf{k}(-\mathbf{r})
\end{array}\right)=\left(\begin{array}{c}
\psi^{*}_\mathbf{k}(-\mathbf{r})\\
\psi_\mathbf{k}(\mathbf{r})
\end{array}\right)=\left(\begin{array}{c}
\psi_{{C}_{2x}\mathbf{k}}({C}_{2x}\mathbf{r}) \\
\psi^{*}_{{C}_{2x}\mathbf{k}}(-{C}_{2x}\mathbf{r})
\end{array}\right),
\end{eqnarray}
Then, $\psi_\mathbf{k}(\mathbf{r})=\psi^{*}_{{C}_{2x}\mathbf{k}}(-{C}_{2x}\mathbf{r})$. Hence, the ${C}_{2x}$ symmetry in momentum space corresponds to $C^{r}_{2y}$ symmetry in real space, indicating that the norm of wave function $\psi_\mathbf{k}(\mathbf{r})$ on the ${C}_{2x}$-symmetric $\mathbf{k}$ path exhibits $C^{r}_{2y}$ symmetry in real space. Similarly, the norm of wave function $\psi_\mathbf{k}(\mathbf{r})$ on the ${C}_{2y}$-symmetric $\mathbf{k}$ path displays $C^{r}_{2x}$ symmetry in real space.

\subsection{Topological property}

\begin{figure}
\centerline{\includegraphics[width=0.8\textwidth]{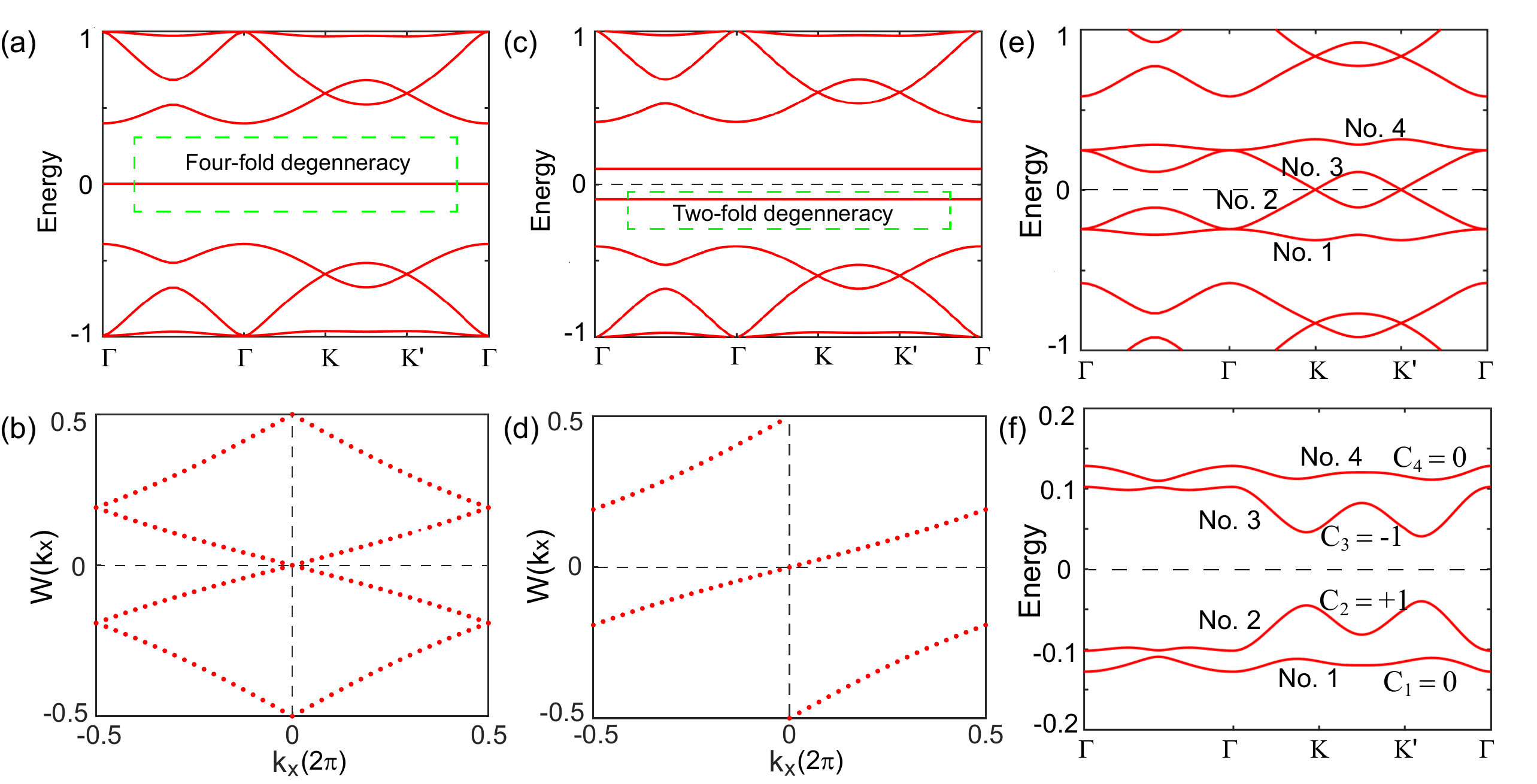}}
\captionsetup{justification=raggedright}
\caption{(color online) (a) and (b) Band structures and Wilson loops $W(\mathbf{k}_x)$ of quadruple flat bands at magic value $\alpha_M=1.379e^{i\varphi_N}$ with magnetic phase $\varphi_N=0.254\pi$. (c) and (d) Band structures with $V=0.1$ and Wilson loops $W(\mathbf{k}_x)$ of occupied double flat bands at magic value a $\alpha_M=1.379e^{i\varphi_N}$ with magnetic phase $\varphi_N=0.254\pi$. (e) and (f) Band structures at $\alpha_M=1.0e^{i\varphi_N}$  and at magic value $\alpha_M=1.379e^{i\varphi_N}$ with $V^{\prime}=0.2$. The green dotted boxes in (a) and (b) highlighted quadruple flat bands and occupied double flat bands. Mometntum $\mathbf{k}_x$ is along $\mathbf{b}_\text{I}$ direction.
\label{Wilson loops} }
\end{figure}

Here, we explore the topological properties for quadruple flat bands. Fig.\ref{Wilson loops} (b) illustrates Wilson loop $W(\mathbf{k}_x)$ corresponding to the quadruple flat bands, highlighted by the green dotted box in Fig.\ref{Wilson loops} (a), where we perform an integration procedure along the $\mathbf{b}_\text{II}$ direction and present the resulting spectra along the $\mathbf{b}_\text{I}$ direction. The partner switch in the Wannier centers from $\textbf{k}=-\textbf{b}_\text{I}/2$ to $\textbf{k}=0$ indicates that quadruple flat bands possess a nontrivial topology $\mathbb{Z}_{2}=1$, consistent with the argument in Ref.\cite{Song2021SM}. Furthermore, by introducing a staggered potential term $V\sigma_z$ on both layers, the Dirac points at $K$ and $K^{\prime}$ can be gapped while the two degeneracy at $\Gamma$ in Fig.\ref{Wilson loops} (e) will always be maintained. Under the effect of the staggered potential, the quadruple flat bands will split into two double flat bands in Fig.\ref{Wilson loops} (c), and Fig.\ref{Wilson loops} (d) shows Wilson loop $W(\mathbf{k}_x)$ of occupied double flat bands, indicating that the occupied double flat bands have Chern number $C=+1$. To investigate the topological properties of every flat band, we add an $C_3$-breaking term $V^{\prime}\sigma_x$ to the interlayer coupling along $\mathbf{q}_1$ direction, which can gap the Dirac points and two degeneracy at $\Gamma$. Fig.\ref{Wilson loops} (f) show the corresponding band structure with $V^{\prime}=0.2$ at magic value $\alpha_M=1.379e^{i\varphi_N}$, where the quadruple flat bands split into four non-degenerate bands. By calculating the Chern number, band No.2/3 origining from Dirac point possess $C_2=+1$ and $C_3=-1$, while the Chern number of band No.1/4 from band inversion vanish, as shown in Fig.\ref{Wilson loops} (f).


\subsection{Condition for the emergence of multi-fold degenerate flat bands}

In the main text, we have presented the conditions for the emergence of multi-fold degenerate flat bands, which are twice the degeneracy of magic angles. We will provide a rigorous proof of this assertion, and the eigenfunction of the Hamiltonian $H^{\varphi}(\mathbf{r})$ can be written as
\begin{eqnarray}\label{eigenfunction}
\left(\begin{array}{cc}
0 & \mathcal{D}_\mathbf{k}^{\varphi*}(-\mathbf{r}) \\
\mathcal{D}_\mathbf{k}^{\varphi}(\mathbf{r}) & 0
\end{array}\right)\left(\begin{array}{c}
\psi^n_\mathbf{k}(\mathbf{r}) \\
\chi^n_\mathbf{k}(\mathbf{r})
\end{array}\right)=E_\mathbf{k}\left(\begin{array}{c}
\psi^n_\mathbf{k}(\mathbf{r}) \\
\chi^n_\mathbf{k}(\mathbf{r})
\end{array}\right),
\end{eqnarray}
where $\psi^n_\mathbf{k}(\mathbf{r})$ and $\chi^n_\mathbf{k}(\mathbf{r})$ are two component wave functions with $n$-fold degeneracy. For zero energy flat bands($E_\mathbf{k}=0$), the eigenfunction Eq.\ref{eigenfunction} becomes
\begin{eqnarray}
\mathcal{D}_\mathbf{k}^{\varphi}(\mathbf{r})\psi^n_\mathbf{k}(\mathbf{r})=0,~~
\mathcal{D}_\mathbf{k}^{\varphi*}(-\mathbf{r})\chi^n_\mathbf{k}(\mathbf{r})=0.
\end{eqnarray}
Then, the zero energy eigenstates can be written as
\begin{eqnarray}\label{eigenstate}
\Psi^n_{\mathbf{k},1}(\mathbf{r})=\left(\begin{array}{c}
\psi^n_\mathbf{k}(\mathbf{r}) \\
0
\end{array}\right),~~\Psi^n_{\mathbf{k},2}(\mathbf{r})=\left(\begin{array}{c}
0 \\
\chi^n_\mathbf{k}(\mathbf{r})
\end{array}\right).
\end{eqnarray}
Due to $\chi^n_\mathbf{k}(\mathbf{r})=\psi^{n*}_\mathbf{k}(-\mathbf{r})$, the eigenstates Eq.\ref{eigenstate} become
\begin{eqnarray}
\Psi^n_{\mathbf{k},1}(\mathbf{r})=\left(\begin{array}{c}
\psi^n_\mathbf{k}(\mathbf{r}) \\
0
\end{array}\right),~~\Psi^n_{\mathbf{k},2}(\mathbf{r})=\left(\begin{array}{c}
0 \\
\psi^{n*}_\mathbf{k}(-\mathbf{r})
\end{array}\right).\label{n-fold}
\end{eqnarray}

The relation between $D^{\varphi}_{\mathbf{k}}(\mathbf{r})$ and $T_{\mathbf{k}}$ reads
\begin{eqnarray}
D^{\varphi}_{\mathbf{k}}(\mathbf{r})=(2\bar{\partial}-\mathbf{k})(I-\alpha({\varphi}) T_{\mathbf{k}}).
\end{eqnarray}
If $T_{\mathbf{k}}$ has  eigenvalue $\frac{1}{\alpha({\varphi})}$ independent of $\mathbf{k}$, the eigenfunction reads
\begin{eqnarray}
T_{\mathbf{k}} \phi^n_{\mathbf{k}}(\mathbf{r})=\frac{1}{\alpha({\varphi})} \phi^n_{\mathbf{k}}(\mathbf{r}),
\end{eqnarray}
where $\phi^n_{\mathbf{k}}(\mathbf{r})$ are eigenvectors of $T_{\mathbf{k}}$ with n-fold degeneracy. Then, the eigenfunction of $D^{\varphi}_{\mathbf{k}}(\mathbf{r})$ can be written as
\begin{eqnarray}
D^{\varphi}_{\mathbf{k}}(\mathbf{r}) \phi^n_{\mathbf{k}}(\mathbf{r})=(2\bar{\partial}-\mathbf{k})(I-\alpha({\varphi}) T_{\mathbf{k}}) \phi^n_{\mathbf{k}}(\mathbf{r})=0,
\end{eqnarray}
indicating that $\psi^n_\mathbf{k}(\mathbf{r})=\phi^n_{\mathbf{k}}(\mathbf{r})$ are eigenvectors of $D^{\varphi}_{\mathbf{k}}(\mathbf{r})$ with $n$-fold degeneracy. Hence, based on Eq.\ref{n-fold}, the degeneracy of flat bands are twice the degeneracy of real eigenvalues.



\subsection{Effect of the intra-sublattice interlayer coupling}

\begin{figure}
\centerline{\includegraphics[width=0.7\textwidth]{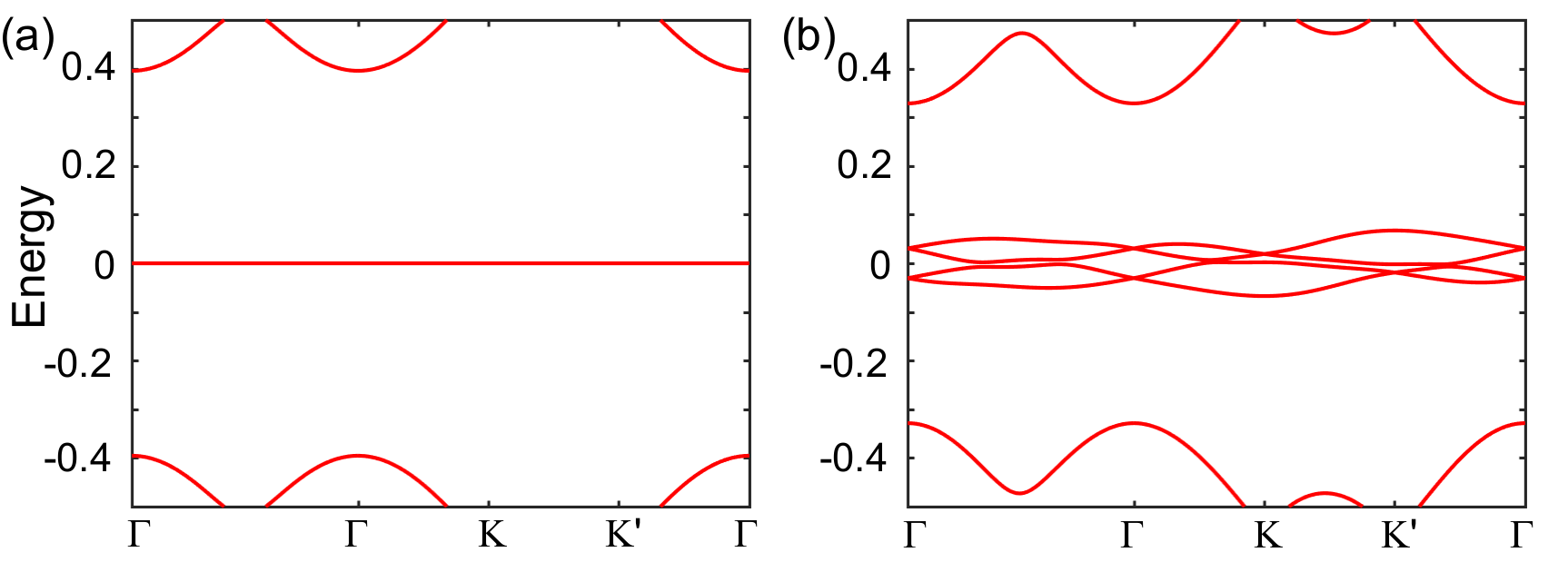}}
\captionsetup{justification=raggedright}
\caption{(color online) The band structures at magic value $\alpha_M=1.379e^{i\varphi_N}$ with magnetic phase $\varphi_N=0.254\pi$ in the absence and presence of intra-sublattice interlayer coupling: (a) $\omega_{AA}=0$, (b) $\omega_{AA}/\omega_{AB}=0.15$. The symbols $\omega_{AA}$ and $\omega_{AB}$  represent the intra-sublattice and inter-sublattice interlayer couplings, respectively. \label{fig6}}
\end{figure}

Here, we discuss the effect of the intra-sublattice interlayer coupling on the absolutely flat bands. Usually, the area of the AA stacking region undergoes a reduction due to lattice relaxation effects, and the absolutely flat bands can be obtained at magic angle when the intra-sublattice interlayer coupling $\omega_{AA}$ vanishes\cite{Vishwanath2019SM}. Then, we take the magic value $\alpha_M=1.379e^{i\varphi_N}$ as an example with magnetic phase $\varphi_N=0.254\pi$, Fig.\ref{fig6}(a) shows the the absolutely flat bands with four-fold degeneracy in the absence of intra-sublattice interlayer coupling. However, when $\omega_{AA}$ is switched on, the absolutely flat bands exhibit a slight dispersion, as shown in Fig.\ref{fig6} (b).
\\

\newpage

\section{Theta function and flat band wave function}

In this section, we introduce the definitions of the theta function and delve into its inherent properties. Subsequently, we utilize these properties of the theta function to derive analytical expressions for the flat band wave function. Furthermore, we investigate the nodal evolution of flat band wave function in real space.

\subsection{Theta function}

The theta function\cite{Mumford} can be defined as
\begin{eqnarray}
\vartheta(z | \tau)=\sum_{n=-\infty}^{\infty} e^{\pi i \tau n^{2}+2 \pi i n z}
\end{eqnarray}

Where $n \in \mathbb{Z}$, $\tau$ are complex parameters with $\operatorname{Im}(\tau)>0$, and $z$ is complex variable. It can be checked that
\begin{eqnarray}
\vartheta(z+1 | \tau)&=& \sum_{n=-\infty}^{+\infty} e^{\pi i \tau n^{2}+2 \pi i n (z+1)}=\sum_{n=-\infty}^{+\infty} e^{\pi i \tau n^{2}+2 \pi i n z}=\vartheta(z | \tau),
\\
\quad \vartheta(z+\tau | \tau)&=& \sum_{n=-\infty}^{+\infty} e^{\pi i \tau n^{2}+2 \pi i n (z+\tau)}=e^{-\pi i \tau-2 \pi i z}\sum_{n=-\infty}^{+\infty} e^{\pi i(n+1)^{2} \tau+2 \pi i (n+1) z}=e^{-\pi i \tau-2 \pi i z} \vartheta(z | \tau),
\\
\vartheta(-z | \tau)&=&\sum_{n=-\infty}^{\infty} e^{\pi i \tau n^{2}-2 \pi i n z}=\sum_{n=-\infty}^{\infty} e^{\pi i \tau (-n)^{2}+2 \pi i (-n) z}=\vartheta(z | \tau)
\end{eqnarray}

All zeros of the theta functions are simple zeros and are given as
\begin{eqnarray}
\vartheta(z | \tau)=0 \quad  \Longleftrightarrow z=m+n \tau+\frac{1}{2}+\frac{\tau}{2},~m, n \in \mathbb{Z}.
\end{eqnarray}

The theta functions with rational characteristics $a$ and $b$ are defined as
\begin{eqnarray}
\vartheta_{a, b}(z | \tau) = \sum_{n=-\infty}^{+\infty} e^{i \pi \tau(n+a)^{2}} e^{2 i \pi(n+a)(z+b)},
\end{eqnarray}
where $\tau$ are complex parameters with $\operatorname{Im}(\tau)>0$, and the range of $a$ is [0,1]. $z$ is complex variables. The relation between $\vartheta_{a, b}(z | \tau)$ and $\vartheta(z | \tau)$ is
\begin{eqnarray}
\vartheta_{a, b}(z | \tau) = e^{\pi i a^{2} \tau+2 \pi i a(z+b)} \vartheta(z+a \tau+b| \tau).
\end{eqnarray}
Then, it can be checked that
\begin{eqnarray}
\vartheta_{a, b}(z+1 | \tau)= e^{2 i \pi a} \vartheta_{a, b}(z | \tau),\vartheta_{a\pm 1, b}(z | \tau)= \vartheta_{a, b}(z | \tau), \vartheta_{a, b}(z+\tau | \tau)=e^{-2 i \pi\left(z+b+\frac{\tau}{2}\right)} \vartheta_{a, b}(z | \tau)m,\label{property}
\end{eqnarray}

The simple zeros are
\begin{eqnarray}
\vartheta_{a, b}(z | \tau)=0 \leftrightarrow z=\frac{1+\tau}{2}-(a \tau+b)+(m \tau+n),~m, n \in \mathbb{Z},\label{zero}
\end{eqnarray}

\subsection{Wave function for quadruple flat bands}

Here, we turn to construct wave functions for flat bands No.$1/2$, and the wave functions for flat bands No.$3/4$ can be generated by chiral symmetry. As discussed regarding the nodal characteristics of wave function at $\mathbf{K}$ point in the main text, the flat band No.$1$ exhibit two nodes at the AA and BA stackings with linear real-space dependence, while the flat band No.$2$ features a single node at the AB stacking with real-space quadratic dependence. Based on these nodal features, the analytical expressions for flat band wave functions can be constructed as
\begin{eqnarray}
\Psi_{\mathbf{k},i}(\mathbf{r})=f_{\mathbf{k}}(z)\Psi_{\mathbf{K},i}(\mathbf{r}),f_{\mathbf{k}}(z)=\frac{\vartheta_{a, b}(\frac{z}{{L}_1} | \frac{{L}_2}{{L}_1})\vartheta_{a^{\prime}, b^{\prime}}(\frac{z}{{L}_1} | \frac{{L}_2}{{L}_1})}{\vartheta_{e, f}(\frac{{z}}{{L}_1} | \frac{{L}_2}{{L}_1})\vartheta_{e^{\prime}, f^{\prime}}(\frac{z}{{L}_1} | \frac{{L}_2}{{L}_1})},\label{expression}
\end{eqnarray}
where $i=1,2$ is band index, and the origin of $\mathbf{k}$ is chosen at $\Gamma$ in the moir\'{e} BZ. The parameters $a$/$a^{\prime}$ and $b$/$b^{\prime}$ are functions of $\mathbf{k}$, while $e$/$e^{\prime}$ and $f$/$f^{\prime}$ are determined by nodal points of wave functions $\Psi_{\mathbf{K},i}(\mathbf{r})$. If the two nodal points of $\Psi_{\mathbf{K},i}(\mathbf{r})$ are located at  $c_{\text{I}}\mathbf{L}_{\text{I}}+c_{\text{II}}\mathbf{L}_{\text{II}}$ and $c^{\prime}_{\text{I}}\mathbf{L}_{\text{I}}+c^{\prime}_{\text{II}}\mathbf{L}_{\text{II}}$ in the moir\'{e} unit cell, by using Eq.\ref{zero}, the parameters $e$/$e^{\prime}$ and $f$/$f^{\prime}$ can be expressed as
\begin{eqnarray}
e=\frac{1}{2}-c_{\text{II}},f=\frac{1}{2}-c_{\text{I}},
e^{\prime}=\frac{1}{2}-c^{\prime}_{\text{II}},f^{\prime}=\frac{1}{2}-c^{\prime}_{\text{I}}
\end{eqnarray}

According to Bloch theorem, $H_\mathbf{k}^{\varphi}(\mathbf{r}+\mathbf{L}_{\text{I/II}})=UH_\mathbf{k}^{\varphi}(\mathbf{r})U^{-1}$ with $U=\operatorname{diag}(e^{-i2\pi/3}, e^{i2\pi/3})$, the wave functions $\Psi_{\mathbf{k},i}(\mathbf{r})$ shall obey the Bloch boundary condition
\begin{eqnarray}
\Psi_{\mathbf{k},i}(\mathbf{r}+\mathbf{L}_{\text{I/II}})=e^{i \mathbf{k}\cdot\mathbf{L}_{\text{I/II}}}U \Psi_{\mathbf{k},i}(\mathbf{r})\label{periodic}.
\end{eqnarray}
Putting Eq.\ref{expression} into Eq.\ref{periodic}, $f_{\mathbf{k}}(z)$ also satisfies Bloch boundary condition
\begin{eqnarray}
f_{\mathbf{k}}(z+\text{L}_{\text{I/II}})=e^{i(\mathbf{k}-\mathbf{K})\cdot\mathbf{L}_{\text{I/II}}}f_{\mathbf{k}}(z).
\end{eqnarray}
Then, by using $\vartheta_{a, b}(z+1 | \tau)= e^{2 i \pi a} \vartheta_{a, b}(z | \tau)$ and $\vartheta_{a, b}(z+\tau | \tau)=e^{-2 i \pi\left(z+b+\frac{\tau}{2}\right)}$ in the Eq.\ref{property}, we can get
\begin{eqnarray}
e^{2i\pi(a+a^{\prime}-e-e^{\prime})}=e^{i(\mathbf{k}-\mathbf{K}) \cdot \mathbf{L}_{\text{I}}},~~e^{-2i\pi(b+b^{\prime}-f-f^{\prime})}=e^{i(\mathbf{k}-\mathbf{K}) \cdot \mathbf{L}_{\text{II}}}.
\end{eqnarray}
Therefore,
\begin{eqnarray}
a+a^{\prime}=e+e^{\prime}+\frac{(\mathbf{k}-\mathbf{K}) \cdot \mathbf{L}_{\text{I}}}{2 \pi},~~b+b^{\prime}=f+f^{\prime}-\frac{(\mathbf{k}-\mathbf{K}) \cdot \mathbf{L}_{\text{II}}}{2 \pi}.\label{relations1}
\end{eqnarray}

By denoting the two nodal positions of $\Psi_{\mathbf{k},i}(\mathbf{r})$ as  $l\mathbf{L}_{\text{I}}+m\mathbf{L}_{\text{II}}$ and $l^{\prime}\mathbf{L}_{\text{I}}+m^{\prime}\mathbf{L}_{\text{II}}$, abbreviated as ($l$,$m$) and ($l^{\prime}$,$m^{\prime}$), the relationship between nodal coordinates and rational characteristics can be expressed as
\begin{eqnarray}
l=\frac{1}{2}-b,l^{\prime}=\frac{1}{2}-b^{\prime},m=\frac{1}{2}-a,m^{\prime}=\frac{1}{2}-a^{\prime}.
\end{eqnarray}
Combined with Eq.\ref{relations1}, the relationship between nodal coordinates reads
\begin{eqnarray}
m+m^{\prime}=1-e-e^{\prime}-\frac{(\mathbf{k}-\mathbf{K}) \cdot \mathbf{L}_{\text{I}}}{2 \pi},~~l+l^{\prime}=1-f-f^{\prime}+\frac{(\mathbf{k}-\mathbf{K}) \cdot\mathbf{L}_{\text{II}}}{2 \pi}.
\end{eqnarray}

For flat band No.$1$, the two nodal points of $\Psi_{\mathbf{K},1}(\mathbf{r})$ are located at $AA$ and $BA$ stacking points with $\mathbf{r}_{AA}=0$ and $\mathbf{r}_{BA}=\frac{-2\mathbf{L}_{\text{I}}-\mathbf{L}_{\text{II}}}{3}$, and we can get
\begin{eqnarray}
e=\frac{1}{2},~~f=\frac{1}{2},~~~~e^{\prime}=\frac{5}{6},~~ f^{\prime}=\frac{7}{6}.
\end{eqnarray}

 Hence, the wave function $\Psi_{\mathbf{k},1}(\mathbf{r})$ for flat band No.$1$ and the relationship between nodal coordinates and rational characteristics are given by
\begin{eqnarray}\label{wave function1}
\Psi_{\mathbf{k},1}(\mathbf{r})=\frac{\vartheta_{a_{1}, b_{1}}(\frac{z}{{L}_1} | \frac{{L}_2}{{L}_1})\vartheta_{a^{\prime}_{1}, b^{\prime}_{1}}(\frac{z}{{L}_1} | \frac{{L}_2}{{L}_1})}{\vartheta_{\frac{1}{2}, \frac{1}{2}}(\frac{z}{{L}_1} | \frac{{L}_2}{{L}_1})\vartheta_{\frac{5}{6}, \frac{7}{6}}(\frac{z}{{L}_1} | \frac{{L}_2}{{L}_1})} \Psi_{\mathbf{K},1}(\mathbf{r}),
\end{eqnarray}

\begin{eqnarray}
a_1+a^{\prime}_1=\frac{4}{3}+\frac{(\mathbf{k}-\mathbf{K}) \cdot \mathbf{L}_{\text{I}}}{2 \pi},~~b_1+b^{\prime}_1=\frac{5}{3}-\frac{(\mathbf{k}-\mathbf{K}) \cdot \mathbf{L}_{\text{II}}}{2 \pi}.\label{band1}
\end{eqnarray}

The corresponding relationship between nodal coordinates is
\begin{eqnarray}
l_1+l^{\prime}_1=-\frac{2}{3}+\frac{(\mathbf{k}-\mathbf{K}) \cdot \mathbf{L}_{\text{II}}}{2 \pi},~~m_1+m^{\prime}_1=-\frac{1}{3}-\frac{(\mathbf{k}-\mathbf{K}) \cdot \mathbf{L}_{\text{I}}}{2 \pi}.
\end{eqnarray}

For flat band No.$2$, the nodal point of $\psi_{\mathbf{K},2}(\mathbf{r})$ is located at AB stacking point with $\mathbf{r}_{AB}=\frac{-\mathbf{L}_{\text{I}}+\mathbf{L}_{\text{II}}}{3}$
\begin{eqnarray}
e=\frac{1}{6},~~f=\frac{5}{6},~~~~e^{\prime}=\frac{1}{6},~~ f^{\prime}=\frac{5}{6}
\end{eqnarray}

Hence, the wave function $\Psi_{\mathbf{k},2}(\mathbf{r})$ for flat band No.$2$ and relationship between nodal coordinates and rational characteristics can be expressed as
\begin{eqnarray}\label{wave function2}
\Psi_{\mathbf{k},2}(\mathbf{r})=\frac{\vartheta_{a_{2}, b_{2}}(\frac{z}{{L}_1} | \frac{{L}_2}{{L}_1})\vartheta_{a^{\prime}_{2}, b^{\prime}_{2}}(\frac{z}{{L}_1} | \frac{{L}_2}{{L}_1})}{[\vartheta_{\frac{1}{6}, \frac{5}{6}}(\frac{z}{{L}_1} | \frac{{L}_2}{{L}_1})]^2}\Psi_{\mathbf{K},2}(\mathbf{r}),
\end{eqnarray}

\begin{eqnarray}
a_2+a^{\prime}_2=\frac{1}{3}+\frac{(\mathbf{k}-\mathbf{K}) \cdot \mathbf{L}_{\text{I}}}{2 \pi},~~b_2+b^{\prime}_2=\frac{5}{3}-\frac{(\mathbf{k}-\mathbf{K}) \cdot \mathbf{L}_{\text{II}}}{2 \pi},\label{band2}
\end{eqnarray}

The corresponding relationship between nodal coordinates is
\begin{eqnarray}
l_2+l^{\prime}_2=-\frac{2}{3}+\frac{(\mathbf{k}-\mathbf{K}) \cdot \mathbf{L}_{\text{II}}}{2 \pi},~~m_2+m^{\prime}_2=\frac{2}{3}-\frac{(\mathbf{k}-\mathbf{K}) \cdot \mathbf{L}_{\text{I}}}{2 \pi}.
\end{eqnarray}

Due to moir\'{e} lattice translation symmetry, the nodal positions of wave function $\Psi_{\mathbf{K},i}(\mathbf{r})$ can be increased by any integer multiple of the translation vectors $\mathbf{L}_{\text{I}/\text{II}}$. Hence, Eq.\ref{band1} and Eq.\ref{band2} can be written in uniform form with arbitrary integers $n_{\text{I}}$ and $n_{\text{II}}$
\begin{eqnarray}
a_i+a^{\prime}_i=\frac{1}{3}+\frac{(\mathbf{k}-\mathbf{K}) \cdot \mathbf{L}_{\text{I}}}{2 \pi}+n_{\text{I}},~~b_i+b^{\prime}_i=\frac{2}{3}-\frac{(\mathbf{k}-\mathbf{K}) \cdot \mathbf{L}_{\text{II}}}{2 \pi}+n_{\text{II}}.
\end{eqnarray}
Then, the uniform relationship between nodal coordinates is
\begin{eqnarray}
l_i+l^{\prime}_i=\frac{1}{3}+\frac{(\mathbf{k}-\mathbf{K}) \cdot \mathbf{L}_{\text{II}}}{2 \pi}+n_{\text{II}},~~m_i+m^{\prime}_i=\frac{2}{3}-\frac{(\mathbf{k}-\mathbf{K}) \cdot \mathbf{L}_{\text{I}}}{2 \pi}+n_{\text{I}}.\label{nodal_relation}
\end{eqnarray}

\begin{figure}
\centerline{\includegraphics[width=0.8\textwidth]{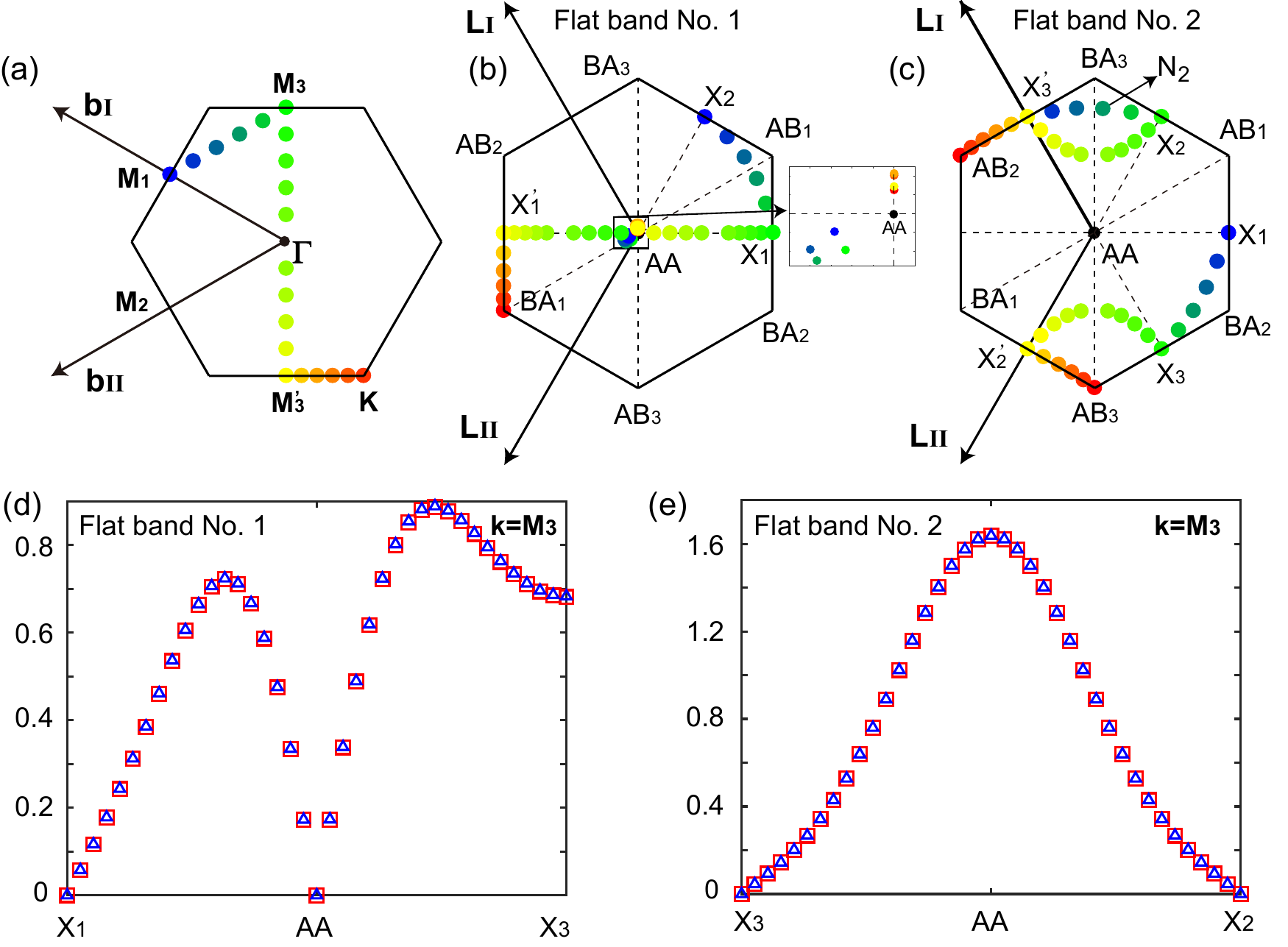}}
\captionsetup{justification=raggedright}
\caption{(color online) (a)-(c) Momentum points at three paths $\mathbf{M}_1\mathbf{M}_3$, $\mathbf{M}_3\mathbf{M^{'}_3}$  and $\mathbf{M^{'}_3}\mathbf{K}$, and a high point $\mathbf{M}_3=0.5\mathbf{b}_1-0.5\mathbf{b}_2$ . (b) and (c) Wave function nodal points for FBs No. $1/2$ in moir\'{e} unit cell indicated by corresponding colors in (a). (d) and (e) The absolute value of wave functions $|\Psi_{\mathbf{k},1/2}(\mathbf{r})|$ for flat bands No.$1/2$ in moir\'{e} unit cell at $\mathbf{k}=\mathbf{M}_3$, and the wave functions from Hamiltonian $H^{\varphi}(\mathbf{r})$ and Eq.\ref{wave function1}/\ref{wave function2} are labelled in red squares and blue triangles. \label{fig5}}
\end{figure}

\subsection{Nodal evolution in real space}

Here, we discuss the nodal evolution and flat band wave function in the moir\'{e} BZ. As the momentum $\textbf{k}$ varies, the nodal points for the FBs move in real space and their coordinates obey Eq.\ref{nodal_relation}. We take three representative paths $\mathbf{M}_1\mathbf{M}_3$, $\mathbf{M}_3\mathbf{M}'_3$ (${C}_{2y}$-symmetric) and $\mathbf{M}'_3\mathbf{K}$ (${C}_{2x}$-symmetric) to show how nodal points evolve in the moir\'{e} unit cell, which is denoted by circles with blue-green-yellow-red gradient color in Fig.\ref{fig5}(a). The corresponding evolution of nodes for bands No.$1,2$ in the moir\'{e} unit cell is displayed in Fig.\ref{fig5} (b) and (c). As previously discussed, the norm of WFs $\psi_\mathbf{k}(\mathbf{r})$ on the ${C}_{2y}$-symmetric $\mathbf{M}_3\mathbf{M}'_3$ path and ${C}_{2x}$-symmetric $\mathbf{M}'_3\mathbf{K}$ path preserves ${C}^{r}_{2x}$ and ${C}^{r}_{2y}$ symmetries in the real space, respectively. When $\mathbf{k}$ is located at $\mathbf{M}_1$ point, the FB WF ${\psi}_{\mathbf{k},1}$ (${\psi}_{\mathbf{k},2}$) has two nodal points at $\text{AA}$ ($\text{X}'_3$) and $\text{X}_2$ ($\text{X}_1$). As $\mathbf{k}$ varies from $\mathbf{M}_1$ to $\mathbf{M}_3$, for ${\psi}_{\mathbf{k},1}$ one nodal point moves continuously from $\text{X}_{2}$ to $\text{X}_{1}$, while the other first moves away from $\text{AA}$ and then returns. Correspondingly, two nodal points of the ${\psi}_{\mathbf{k},2}$ move to $\text{X}_2$ and $\text{X}_{3}$. It is apparent that the nodes of both bands are ${C}^{r}_{2x}$-symmetric at $\mathbf{M}_3$. With $\mathbf{k}$ further varying along the $C_{2y}$-symmetric  $\text{M}_3\text{M}'_3$ path, both nodes of ${\psi}_{\mathbf{k},1}$ move along the $\text{X}_{1}$--$\text{AA}$ lines, and nodes of ${\psi}_{\mathbf{k},1}$ move from $\text{X}_{2}$/$\text{X}_3$ to $\text{X}'_{3}$/$\text{X}'_{2}$, preserving ${C}^{r}_{2x}$ symmetry. Finally, as $\mathbf{k}$ varies along the the ${C}_{2x}$-symmetric $\mathbf{M^{'}_3}\mathbf{K}$ path, one nodal point of ${\psi}_{\mathbf{k},1}$ moves along $\text{X}'_{1}$--$\text{BA}_{1}$ line and the other first moves away from $\text{AA}$ along $\text{AA}$--$\text{BA}_{3}$ line and then returns. In contrast, both nodes of the ${\psi}_{\mathbf{k},2}$ move towards the AB stacking point and merge to form a second-order node at $\textbf{K}$, consistent with our previous analysis. In addition, to confirm the validity of wave function analytical expressions, we calculate the flat-band wave functions from Hamiltonian $H^{\varphi}(\mathbf{r})$ and  Eq.\ref{wave function1}/\ref{wave function2} at $\mathbf{k}=\mathbf{M}_3$, which are labelled in red squares and blue triangles in Fig.\ref{fig5} (d) and (e), and the good agreement between them verifies the validity of wave function from analytical expressions. In the following, we take $\mathbf{K}=\frac{1}{3}\mathbf{b}_\text{I}-\frac{2}{3}\mathbf{b}_\text{II}$, $\mathbf{M}_3=\frac{1}{2}\mathbf{b}_\text{I}-\frac{1}{2}\mathbf{b}_\text{II}$ and $\mathbf{P}=0.5\mathbf{b}_\text{I}-0.3\mathbf{b}_\text{II}$ points as examples to check Eq.\ref{nodal_relation}.
\\

(1) At $\mathbf{k}=\mathbf{K}$, the Eq.\ref{nodal_relation} becomes
\begin{eqnarray}
l_i+l^{\prime}_i=\frac{1}{3}+\frac{(\mathbf{K}-\mathbf{K}) \cdot \mathbf{L}_{\text{II}}}{2 \pi}+n_{\text{II}}=\frac{1}{3}+n_{\text{II}},~~m_i+m^{\prime}_i=\frac{2}{3}-\frac{(\mathbf{K}-\mathbf{K}) \cdot \mathbf{L}_{\text{I}}}{2 \pi}+n_{\text{I}}=\frac{2}{3}+n_{\text{I}}. \label{K1}
\end{eqnarray}
Since the two nodal positions of $\Psi_{\mathbf{K},1}(\mathbf{r})$ are located at $AA$ and $BA$  points with $\mathbf{r}_{AA}=0$ and $\mathbf{r}_{BA}=\frac{\mathbf{L}_{\text{I}}+2\mathbf{L}_{\text{II}}}{3}$, and the nodal position of $\Psi_{\mathbf{K},2}(\mathbf{r})$ is located at $AB$ point with $\mathbf{r}_{AB}=\frac{-\mathbf{L}_{\text{I}}+\mathbf{L}_{\text{II}}}{3}$. Then, we can get
\begin{eqnarray}
l_1=0,~m_1=0,~l^{\prime}_1=\frac{1}{3},~m^{\prime}_1=\frac{2}{3}~\text{and}~l_2=-\frac{1}{3},~m_2=\frac{1}{3},~l^{\prime}_2=-\frac{1}{3},~m^{\prime}_2=\frac{1}{3},
\end{eqnarray}
which satisfy Eq.\ref{K1} with $n_{\text{I}}=n_{\text{II}}=0$ for flat band No.1, and with $n_{\text{I}}=-1$ and $n_{\text{II}}=0$ for flat band No.2.
\\

(2) At $\mathbf{k}=\mathbf{M}_3$, the Eq.\ref{nodal_relation} becomes
\begin{eqnarray}
l_i+l^{\prime}_i=\frac{1}{3}+\frac{(\mathbf{M}_3-\mathbf{K}) \cdot \mathbf{L}_{\text{II}}}{2 \pi}+n_{\text{II}}=\frac{1}{2}+n_{\text{II}},~~m_i+m^{\prime}_i=\frac{2}{3}-\frac{(\mathbf{M}_3-\mathbf{K}) \cdot \mathbf{L}_{\text{I}}}{2 \pi}+n_{\text{I}}=\frac{1}{2}+n_{\text{I}}.\label{K2}
\end{eqnarray}
Because the two nodal positions of $\Psi_{\mathbf{M}_3,1}(\mathbf{r})$($\Psi_{\mathbf{M}_3,2}(\mathbf{r})$) are located at $AA$($X_2$) and $X_1$ ($X_3$) points with $\mathbf{r}_{AA}=0$ ($\mathbf{r}_{X_2}=\frac{\mathbf{L}_{\text{II}}}{2}$) and $\mathbf{r}_{X_1}=\frac{\mathbf{L}_{\text{I}}+\mathbf{L}_{\text{II}}}{2}$ ($\mathbf{r}_{X_3}=\frac{\mathbf{L}_{\text{I}}}{2}$), then
\begin{eqnarray}
&l_1=0,~m_1=0,~l^{\prime}_1=\frac{1}{2},~m^{\prime}_1=\frac{1}{2}~\text{and}~l_2=0,~m_2=\frac{1}{2},~l^{\prime}_2=\frac{1}{2},~m^{\prime}_2=0,
\end{eqnarray}
which satisfy Eq.\ref{K2} with $n_{\text{I}}=n_{\text{II}}=0$ for flat band No.1/2.
\\

(3) At $\mathbf{k}=\mathbf{P}$, the Eq.\ref{nodal_relation} becomes
\begin{eqnarray}
l_i+l^{\prime}_i=\frac{1}{3}+\frac{(\mathbf{P}-\mathbf{K}) \cdot \mathbf{L}_{\text{II}}}{2 \pi}+n_{\text{II}}=0.7+n_{\text{II}},~~m_i+m^{\prime}_i=\frac{2}{3}-\frac{(\mathbf{P}-\mathbf{K}) \cdot \mathbf{L}_{\text{I}}}{2 \pi}+n_{\text{I}}=0.5+n_{\text{I}}.\label{K3}
\end{eqnarray}
Since the nodal points of wave functions for flat bands No.$1/2$ are $N_1=-0.3265\mathbf{L}_1-0.5607\mathbf{L}_2$, $N^{\prime}_1=0.0266\mathbf{L}_1+0.0612\mathbf{L}_2$, $N_2=0.236\mathbf{L}_1-0.2986\mathbf{L}_2$ and $N^{\prime}_2=-0.5358\mathbf{L}_1-0.201\mathbf{L}_2$, then we can get
\begin{eqnarray}
l_1=-0.3265,m_1=-0.5607,l^{\prime}_1=0.0266,m^{\prime}_1=0.0612~\text{and}~ l_2=0.236,m_2=-0.2986,l^{\prime}_2=-0.5358,m^{\prime}_2=-0.201,
\end{eqnarray}
which satisfy Eq.\ref{K3} with $n_{\text{I}}=n_{\text{II}}=-1$ for flat band No.$1/2$.

\newpage

\section{Ground states of TBG with alternative magnetic fluxes at half-filling}

\subsection{Projected interaction and ground states}

Here, we begin with the general Coulomb interaction that describes TBG, and proceed to derive the projected interaction within the flat band basis. By utilizing this projected interaction, we deduce the intrasublattice-intraorbital interaction characterized by chiral symmetry, and its ground states with zero energy can be obtained at half-filling.

\subsubsection{Projected interaction}

The general Hamiltonian describing the Coulomb interaction in TBG can be expressed as

\begin{eqnarray}
\mathcal{\hat{H}}_{I}=\sum^{\bm{i},\bm{\delta}}_{\tau,\sigma,\tau^{\prime},\sigma^{\prime}}V_{\bm{\delta}}c^{\dag}_{\bm{i},\tau,\sigma}c_{\bm{i},\tau,\sigma}c^{\dag}_{\bm{i}+\bm{\delta},\tau^{\prime},\sigma^{\prime}}c_{\bm{i}+\bm{\delta},\tau^{\prime},\sigma^{\prime}},
\end{eqnarray}
where $c_{\bm{i},\tau,\sigma}$ and $c_{\bm{i}+\bm{\delta},\tau,\sigma}$ denote the annihilation operators for electrons in sublattice $\tau$ and layer $\sigma$ at unit cells $\bm{i}$ and $\bm{i}+\bm{\delta}$, respectively. The parameter $V_{\bm{\delta}}$ represents the strength of Coulomb interaction between two electrons separated by the vector $\bm{\delta}$. Considering the small magnitude of the twisted angle, we neglect the dependence of the Hamiltonian on this angle. By performing a Fourier transformation, the interaction Hamiltonian $\mathcal{\hat{H}}_{I}$ can be written in terms of momentum space as follows:

\begin{eqnarray}
\mathcal{\hat{H}}_{I}&=&\sum^{\bm{i},\bm{\delta}}_{\tau,\sigma,\tau^{\prime},\sigma^{\prime}}V_{\bm{\delta}}c^{\dag}_{\bm{i},\tau,\sigma}c_{\bm{i},\tau,\sigma}c^{\dag}_{\bm{i}+\bm{\delta},\tau^{\prime},\sigma^{\prime}}c_{\bm{i}+\bm{\delta},\tau^{\prime},\sigma^{\prime}},\nonumber
\\
&=&\frac{1}{N^2}\sum^{\bm{i},\bm{\delta}}_{\tau,\sigma,\tau^{\prime},\sigma^{\prime}}V_{{\bm{\delta}}}e^{i(\mathbf{k}_{1}-\mathbf{k}_{2}+\mathbf{k}_{3}-\mathbf{k}_{4})\cdot{\mathbf{R}_{i}}}c^{\dag}_{\mathbf{k}_{1},\tau,\sigma}c_{\mathbf{k}_{2},\tau,\sigma}c^{\dag}_{\mathbf{k}_{3},\tau^{\prime},\sigma^{\prime}}c_{\mathbf{k}_{4},\tau^{\prime},\sigma^{\prime}}e^{-i\mathbf{k}_{3}\cdot\bm{\delta}}e^{i\mathbf{k}_{4}\cdot\bm{\delta}},\nonumber
\\
&=&\frac{1}{N}\sum^{\bm{i},\bm{\delta}}_{\tau,\sigma,\tau^{\prime},\sigma^{\prime}}V_{\bm{\delta}}e^{-i(\mathbf{k}_{3}-\mathbf{k}_{4})\cdot{\bm{\delta}}}\delta{(\mathbf{k}_{1}-\mathbf{k}_{2}+\mathbf{k}_{3}-\mathbf{k}_{4})}c^{\dag}_{\mathbf{k}_{1},\tau,\sigma}c_{\mathbf{k}_{2},\tau,\sigma}c^{\dag}_{\mathbf{k}_{3},\tau^{\prime},\sigma^{\prime}}c_{\mathbf{k}_{4},\tau^{\prime},\sigma^{\prime}},\nonumber
\\
&=&\frac{1}{N}\sum_{\bm{\delta},\mathbf{q},\mathbf{k},\mathbf{k}^{\prime}}V_{\bm{\delta}}e^{-i\mathbf{q}\cdot{\bm{\delta}}}c^{\dag}_{\mathbf{k},\tau,\sigma}c_{\mathbf{k}+\mathbf{q},\tau,\sigma}c^{\dag}_{\mathbf{k}^{\prime},\tau^{\prime},\sigma^{\prime}}c_{\mathbf{k}^{\prime}-\mathbf{q},\tau^{\prime},\sigma^{\prime}},\nonumber
\\
&=&\frac{1}{N}\sum_{\mathbf{q}}V(\mathbf{q})\rho_{\mathbf{q}}\rho_{-\mathbf{q}},\label{general Hamiltonian}
\end{eqnarray}
where $V(\mathbf{q})=\sum_{\bm{\delta}}V_{\bm{\delta}}e^{-i\mathbf{q}\cdot{\bm{\delta}}}$,~$\rho_{\pm\mathbf{q}}=\sum_{\mathbf{k},\tau,\sigma}c^{\dag}_{\mathbf{k},\tau,\sigma}c_{\mathbf{k}\pm\mathbf{q},\tau,\sigma}$, and $\mathbf{q}$ and $\mathbf{k}$ are defined in the whole momentum space.

The low-energy physics of TBG with alternative magnetic fluxes can also be effectively described by the Bistrizer-MacDonald (BM) model, which incorporates Dirac Hamiltonians with interlayer momentum couplings between the same valley Dirac points while neglecting the interlayer coupling between different valleys. In the basis of BM model, the interaction Hamiltonian Eq.\ref{general Hamiltonian} becomes

\begin{eqnarray}
\mathcal{\hat{H}}_I=\frac{1}{2\Omega}\sum^{\mathbf{g}}_{\mathbf{q}_M \in \mathrm{MBZ}} V(\mathbf{q}_M+\mathbf{g}) \delta \rho_{\mathbf{q}_M+\mathbf{g}}\delta \rho_{-\mathbf{q}_M-\mathbf{g}},~\delta \rho_{\mathbf{q}_M+\mathbf{g}}=\sum^{\eta, \tau,\mathbf{Q}}_{\mathbf{k}_M \in \mathrm{MBZ}}c^{\dagger}_{\mathbf{k}_M+\mathbf{q}_M, \mathbf{Q}+\mathbf{g},\tau,\eta} c_{\mathbf{k}_M, \mathbf{Q},\tau,\eta},
\end{eqnarray}
where $\Omega$ denotes the total area of TBG. The symbol $\mathbf{Q}$ indicates that the creation and annihilation operators originate from either the red or green Dirac cones, which are distinguished in the momentum lattice structure depicted in Fig. 1(d) of the main text. The momentums $\mathbf{k}_M$ and $\mathbf{q}_M$ are defined within the moir\'{e} BZ, while $\mathbf{g}$ represents the moir\'{e} reciprocal vectors. Additionally, the index $\eta$ corresponds to the valley index.

In the main text, we have presented a comprehensive analysis of the twisted bilayer graphene under alternative magnetic fluxes. Our investigation reveals the existence of eight-fold degenerate flat bands at zero energy per spin, each of which possesses a non-trivial Chern number owing to the presence of chiral symmetry. Additionally, the eight-fold degenerate flat bands introduce an effective orbital degree of freedom. Thus, we can characterize the eight-fold degenerate flat bands using the sublattice index ($\kappa=A~ \text{and}~B$), the effective orbital index ($\mu=o~ \text{and}~ o^{\prime}$), and the valley index ($\eta=K~ \text{and}~ K^{\prime}$). In the eight-fold degenerate flat bands, the four bands that constitute Dirac points possess are marked by symbol $o$ while the four bands originating from band inversion are marked by symbol $o^{\prime}$. To distinguish from the unprojected interaction Hamiltonian $\mathcal{\hat{H}}_I$ with a hat, we represent the interaction Hamiltonian projected onto the eight-fold degenerate flat bands as $\mathcal{H}_I$ without the hat. Then, the projected interaction Hamiltonian\cite{Bernevig2021} is given by

\begin{eqnarray}
\mathcal{{H}}_I=\frac{1}{2 \Omega} \sum_{\mathbf{q}_M,\mathbf{g}}V(\mathbf{q}_M+\mathbf{g}) \delta \bar{\rho}_{\mathbf{q}_M+\mathbf{g}} \delta \bar{\rho}_{-\mathbf{q}_M-\mathbf{g}},~~\delta \bar{\rho}_{\mathbf{q}_M+\mathbf{g}}=\sum^{\beta,\gamma}_{\mathbf{k}_M }\{ c_{\beta,\mathbf{k}_M}^{\dagger}\left[\Lambda_{\mathbf{q}_M+\mathbf{g}}(\mathbf{k}_M)\right]_{\beta,\gamma} c_{\gamma, \mathbf{k}_M+\mathbf{q}_M}-\nu\delta_{\mathbf{q}_M,0}\delta_{\beta,\gamma}\},
\end{eqnarray}
where $\beta=(\kappa,\mu,\eta)$ and $\gamma=(\kappa^{\prime},\mu^{\prime},\eta^{\prime})$ contain the Chern number, orbital and valley indexs. $\delta \bar{\rho}_{\mathbf{q}_M+\mathbf{g}}$ denotes the Fourer components of the density operator projected into the eight-fold degenerate flat bands, while the background density at charge neutrality is denoted by $\nu$. We can explicitly express  $\delta \bar{\rho}_{\mathbf{q}_M+\mathbf{g}}$  by introducing the form factor $\left[\Lambda_{\mathbf{q}_M+\mathbf{g}}(\mathbf{k}_M)\right]_{\beta,\gamma}= \left\langle u_{\beta, \mathbf{k}_M} \mid u_{\gamma, \mathbf{k}_M+\mathbf{q}_M}\right\rangle$, where $|u_{\beta, \mathbf{k}_M}\rangle$ represents the periodic Bloch wave function associated with the flat band.

\subsubsection{Ground states of intrasublattice-intraorbital interaction}

We turn to derive the intrasublattice-intraorbital interaction Hamiltonian with chiral symmetry. We will examine how symmetries impose constraints on the structure of the form factor $\Lambda_{\mathbf{q}_M+\mathbf{g}}(\mathbf{k}_M)$, where we take chiral symmetry $S=\kappa_z\mu_0\eta_0$, $C_2T=\kappa_x\mu_0\eta_zK$ and $PT=\kappa_y\mu_0\eta_x$\cite{Patrick2021SM}. Under the influence of these symmetries, we obtain the following expression:
\begin{eqnarray}
&&\left\langle u_{A, \mu, \eta, \mathbf{k}_M} \mid u_{B, \mu^{\prime}, \eta, \mathbf{k}_M+\mathbf{q}_M}\right\rangle=\left\langle u_{A, \mu, \eta, \mathbf{k}_M}\left|S^2\right| u_{B, \mu^{\prime}, \eta,\mathbf{k}_M+\mathbf{q}_M}\right\rangle=-\left\langle u_{A, \mu, \eta, \mathbf{k}_M} \mid u_{B, \mu^{\prime}, \eta, \mathbf{k}_M+\mathbf{q}_M}\right\rangle=0,\nonumber
\\
&&\left\langle u_{\kappa, \mu, \eta, \mathbf{k}_M} \mid u_{\kappa^{\prime}, \mu^{\prime}, \eta, \mathbf{k}_M+\mathbf{q}_M}\right\rangle=\left\langle u_{\kappa, \mu, \eta, \mathbf{k}_M}\left|(C_2T)^2\right| u_{\kappa^{\prime}, \mu^{\prime}, \eta, \mathbf{k}_M+\mathbf{q}_M}\right\rangle=\left\langle u_{-\kappa, \mu, \eta, \mathbf{k}_M} \mid u_{-\kappa^{\prime}, \mu^{\prime}, \eta, \mathbf{k}_M+\mathbf{q}_M}\right\rangle^*,\nonumber
\\
&&\left\langle u_{\kappa, \mu, K, \mathbf{k}_M} \mid u_{\kappa, \mu, K, \mathbf{k}_M+\mathbf{q}_M}\right\rangle=\left\langle u_{\kappa, \mu, K, \mathbf{k}_M}\left|(PT)^2\right| u_{\kappa, \mu, K, \mathbf{k}_M+\mathbf{q}_M}\right\rangle=\left\langle u_{-\kappa, \mu, K^{\prime}, \mathbf{k}_M} \mid u_{-\kappa, \mu, K^{\prime}, \mathbf{k}_M+\mathbf{q}_M}\right\rangle,\nonumber
\\
&&\left\langle u_{\kappa, \mu, K, \mathbf{k}_M} \mid u_{-\kappa, \mu, K, \mathbf{k}_M+\mathbf{q}_M}\right\rangle=\left\langle u_{\kappa, \mu, K, \mathbf{k}_M}\left|(PT)^2\right| u_{-\kappa, \mu, K, \mathbf{k}_M+\mathbf{q}_M}\right\rangle=-\left\langle u_{-\kappa, \mu, K^{\prime}, \mathbf{k}_M} \mid u_{\kappa, \mu, K^{\prime}, \mathbf{k}_M+\mathbf{q}_M}\right\rangle.\label{symmetry constrain}
\end{eqnarray}

\begin{table}[b]
\footnotesize
\renewcommand\arraystretch{2.5}
\setlength{\tabcolsep}{0.5mm}{
    \begin{tabular}{|*{5}{c|}}
        \hline
      \multicolumn{2}{|c|}{Total Chern number} & C$_T=\pm2$ & C$_T=\pm1$ & C$_T=0$ \\ \hline
     \multirow{2}{*}{Double FBs} &   Configuration number & 2 & $\times$ & 4 \\ \cline{2-5}
       & \makecell{Half filling of 4 bands\\ each band with $C=\pm 1$}   &   Two filled bands with $C=\pm 1$ & $\times$ & \makecell{One filled band with $C=\pm 1$\\ another filled band with $C=\mp 1$} \\\hline
      \multirow{2}{*}{Quadruple FBs}  &Configuration number  & $2\times C^4_2 =12$  & $2\times C^4_3C^4_1=32$ & $4\times C^4_2+2=26$   \\ \cline{2-5}
       & \makecell{Half filling of 8 bands\\ orbital $o$: $C=\pm 1$\\ orbital $o'$: $C=0$}   & \makecell{Two filled bands with $C=\pm1$\\ two other filled bands with $C=0$}  &  \makecell{Three filled bands with $C=\pm1$ or $\mp 1$\\ another filled bands with $C=0$} & \makecell{One filled band with $C=\pm 1$\\ another filled band with $C=\mp 1$\\two other filled bands with $C=0$} \\ \hline
    \end{tabular}}
    \captionsetup{font={small}}
     \caption{Comparison of ground states between double and quadruple flat bands at half-filling.
     }
    \label{compare}
\end{table}

Based on the above constrain to the form factor $\left[\Lambda_{\pm\mathbf{q}_M\pm\mathbf{g}}(\mathbf{k}_M)\right]_{\beta,\gamma}$, the intrasublattice-intraorbital form factor in the same valley can be written as
\begin{eqnarray}
\Lambda^{\text{S},\text{intra}}_{\pm\mathbf{q}_M\pm\mathbf{g}}(\mathbf{k}_M) =\left(\begin{array}{cccccccc}
\Delta & 0 & 0 & 0 & 0 & 0 & 0 & 0   \\
0 &  \Delta^* & 0 & 0 & 0 & 0 & 0 & 0  \\
0 & 0 & \Delta& 0 & 0 & 0 & 0 & 0 \\
0 &  0 & 0 & \Delta^* & 0 & 0 & 0 & 0  \\
0 &  0 &  0 & 0 & \Delta^* & 0 & 0 & 0  \\
0 & 0 &  0 &  0 & 0 & \Delta & 0 &  0  \\
0 & 0 & 0 &  0 &  0 & 0 & \Delta^* &  0  \\
0 & 0 & 0 & 0 &  0 &  0 & 0 & \Delta   \\
\end{array}\right),\Delta=\langle u_{A, o, K, \mathbf{k}_M} \mid u_{A, o, K, \mathbf{k}_M\pm\mathbf{q}_M}\rangle=\langle u_{A, o^{\prime}, K, \mathbf{k}_M} \mid u_{A, o^{\prime}, K,\mathbf{k}_M\pm\mathbf{q}_M}\rangle,
\end{eqnarray}
where the basis is $[c^{\dagger}_{A,o,K},c^{\dagger}_{B,o,K},c^{\dagger}_{A,o^{\prime},K},c^{\dagger}_{B,o^{\prime},K},
c^{\dagger}_{A,o,K^{\prime}},c^{\dagger}_{B,o,K^{\prime}},c^{\dagger}_{A,o^{\prime},K^{\prime}},c^{\dagger}_{B,o^{\prime},K^{\prime}}]$.
By setting $\Delta=F^{\text{intra}}_{\mathbf{q}_M}(\mathbf{k}_M) e^{i \Phi^{\text{intra}}_{\mathbf{q}_M}(\mathbf{k}_M)}$, the form factor can be simplified to
\begin{eqnarray}
\Lambda^{\text{S},\text{intra}}_{\mathbf{q}_M+\mathbf{g}}(\mathbf{k}_M)=F^{\text{intra}}_{\mathbf{q}_M}(\mathbf{k}_M) e^{i \Phi^{\text{intra}}_{\mathbf{q}_M}(\mathbf{k}_M) \kappa_z \mu_0 \eta_z},
\end{eqnarray}
and the intrasublattice-intraorbital interaction Hamiltonian can be written as
\begin{eqnarray}
\mathcal{{H}}^{\text{S},\text{intra}}_I=\frac{1}{2 A} \sum_{\mathbf{q}_M,\mathbf{g}}V(\mathbf{q}_M+\mathbf{g}) \delta \bar{\rho}^{\text{S},\text{intra}}_{\mathbf{q}_M+\mathbf{g}} \delta \bar{\rho}^{\text{S},\text{intra}}_{-\mathbf{q}_M-\mathbf{g}}.
\end{eqnarray}

Then, we investigate ground states of Hamiltonian $\mathcal{{H}}^{\text{S},\text{intra}}_I$ at charge-neutrality for the spinless case, where four out of eight flat bands are occupied. As this Hamiltonian can be expressed as a sum of positive semidefinite operators for each momentum $\mathbf{q}_M$, we can establish that any state satisfying $ \delta \bar{\rho}^{\text{S},\text{intra}}_{\mathbf{q}_M+\mathbf{g}}|\Psi\rangle=0$ for all $\mathbf{q}_M$ serves as a ground state. Subsequently, we explore possible ground states for filling four bands that satisfy $ \delta \bar{\rho}^{\text{S},\text{intra}}_{\mathbf{q}_M+\mathbf{g}}|\Psi\rangle=0$ for all $\mathbf{q}_M$, along with their corresponding total Chern numbers.

\begin{figure}
\centerline{\includegraphics[width=1\textwidth]{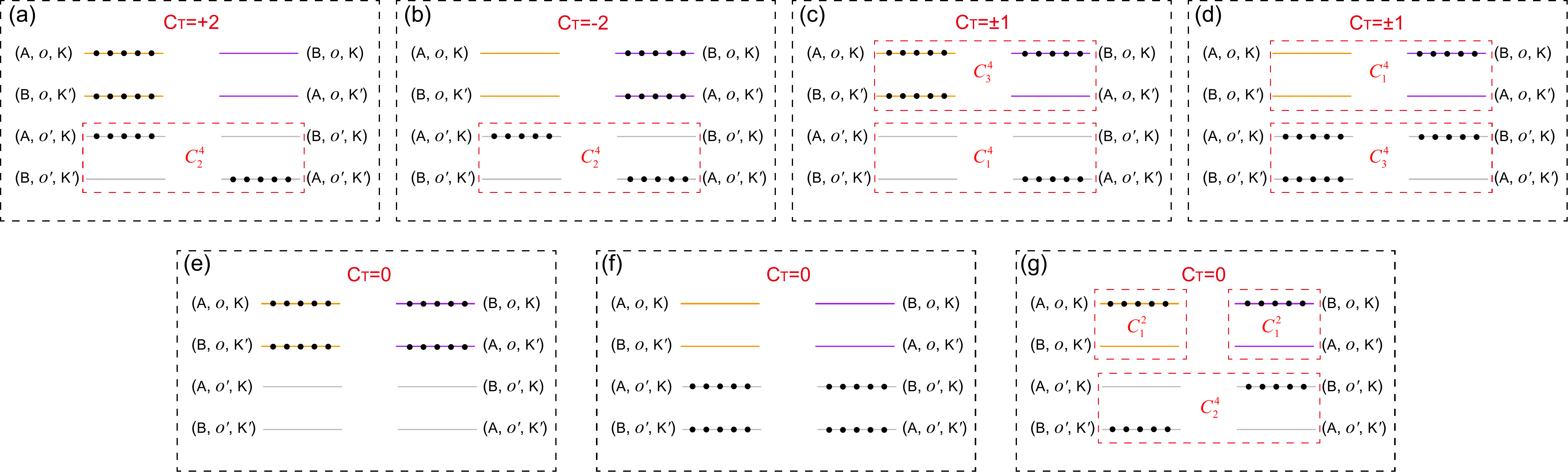}}
\captionsetup{justification=raggedright}
\caption{(color online) The configurations for ground states with total Chern number $C_T=\pm2$, $C_T=\pm1$ and $C_T=0$ in the quadratic flat bands at half-filling. The flat band labelled by gray, pink and orange lines possesses Chern number $C=0$, $C=-1$, and $C=+1$, respectively.\label{C0}}
\end{figure}

Before explore all possibel ground states, we discuss the comparison of ground states between double and quadruple flat bands at half-filling. According to established research\cite{Bultinck2020}, the conventional TBG has double degenerate flat bands in each of its two valleys. These bands possess a distinct Chern number ($C$) of $\pm 1$, due to  the topological nature of the Dirac cone.  On a half-filling scenario, taking into account both valleys, these four flat bands can be marked by sublattice index ($\kappa=A~ \text{or}~B$) and valley index ($\eta=K~ \text{or}~ K^{\prime}$). As detailed in Table \ref{compare},  the fillings with a total Chern number $C_T=\pm 2$ correspond to two possible configurations, while those with $C_T=0$ correspond to four possible configurations. Transitioning to our discussion of four-fold degeneracy, this elevated degree of degeneracy in the flat bands diversify the configurations for various ground states. Two of the flat bands for each valley exhibit a Chern number of $\pm 1$, linked to the topological nature of the Dirac cone. We have denoted these two non-trivial bands with the orbital $o$ and the remaining two flat bands with $C=0$ with orbital $o'$. Thus, considering both valleys, these eight bands can be represented with sublattice ($A,B$), valley ($K,K'$), and orbital ($o,o'$) indices. Specifically, we focus on the half filling scenario for comparative study. Even outside this scenario, the quadruple flat bands lead to a variety of ground states. In the half filling, the filling configurations can be classified to three categories based on the total Chern numbers $C_T=\pm 2, \pm 1, 0$, as shown in Fig.\ref{C0}.

\begin{table}[b]
\renewcommand\arraystretch{1.6}
\setlength{\tabcolsep}{2.5mm}{
    \begin{tabular}{|*{7}{c|}}
        \hline
      Total Chern number & Type of ground state& Q$^{|\text{C}_T|}_n$ & Energy correction & $\xi_{h}$ & $\xi_{\Lambda^{\text{S,inter}}}$ & $\xi_{\Lambda^{\text{A}}}$ \\ \hline
     \multirow{4}{*}{C$_T=\pm2$} &   Type-1 and -1$^{\prime}$ & $\text{Q}^2_1=\kappa_z\mu_0\eta_0$ & 4$\lambda_A$-4J & - & + & - \\
       & Type-2 and -2$^{\prime}$  & $\text{Q}^2_2=\kappa_z\mu_z\eta_0$  &$4\lambda_A$+4$\lambda_S$-4J & -  & - & -  \\
       & Type-(3$\sim$4) and -(3$^{\prime}\sim4^{\prime}$) & Q$^2_{(3\sim4)}$  & $4\lambda_A$+2$\lambda_S$-4J & - & 0  & - \\
       & Type-(5$\sim$6) and -(5$^{\prime}\sim6^{\prime}$) & Q$^2_{(5\sim6)}$  &2$\lambda_A$+2$\lambda_S$-2J & 0  & 0 & 0 \\ \hline
      \multirow{2}{*}{C$_T=\pm1$}  &Type-(1$\sim$3) and -(1$^{\prime}\sim3^{\prime}$) & Q$^1_{(1\sim3)}$  & $2\lambda_A$+4$\lambda_S$-2J & 0 & -  & 0 \\
       & Type-(4$\sim$16) and -(4$^{\prime}\sim16^{\prime}$) & Q$^1_{(4\sim16)}$  &2$\lambda_A$+2$\lambda_S$-2J & 0  & 0 & 0 \\ \hline
        \multirow{7}{*}{C$_T=0$}  &Type-(1$\sim$2)~{\clr[VP]} & Q$^0_{(1\sim2)}$  & 0 & + & +  & + \\
       & Type-(3$\sim$6)~{\clr[OP/VOL]} & Q$^0_{(3\sim6)}$   &4$\lambda_S$ & +  & - & + \\
        & Type-7 and -7$^{\prime}$~{\clr[VH]} & $\text{Q}^0_7=\kappa_z\mu_0\eta_z$  &4$\lambda_A$-4J& -  & + & - \\
        & Type-8 and -8$^{\prime}$~{\clr[VH]} & $\text{Q}^0_8=\kappa_z\mu_z\eta_z$  &4$\lambda_A$+4$\lambda_S$-4J& -  & - & - \\
        & Type-(9$\sim$10) and -($9^{\prime}\sim10^{\prime}$)~{\clr[VH]} & Q$^0_{(9\sim10)}$   &4$\lambda_A$+2$\lambda_S$-4J& -  & 0 & - \\
        & Type-(11$\sim$12) and -($11^{\prime}\sim12^{\prime}$)~{\clr[VH]} & Q$^0_{(11\sim12)}$    &2$\lambda_A$+2$\lambda_S$-2J& 0  & 0 & 0 \\
        & Type-(13$\sim$16) and -($13^{\prime}\sim16^{\prime}$) & Q$^0_{(13\sim16)}$   &2$\lambda_A$+2$\lambda_S$-2J& 0  & 0 & 0 \\ \hline
    \end{tabular}}
    \captionsetup{justification=raggedright}
    \caption{Energy corrections, total Chern numbers ($\text{C}_T$), and the operator Q$^{|\text{C}_T|}_n$ for the three categories of ground states. The symbols $\xi_x = +, -, 0$ indicate the different commutation relations between the operators Q$^{|\text{C}_T|}_n$ and $x$, namely commutation, anticommutation, and neither commutation nor anticommutation, where $x$ denotes the single-particle dispersion $h$ in Eq.\ref{single-particle}, the form factor $\Lambda^\text{A}$ in Eq.\ref{form factor1} and the form factor $\Lambda^\text{S,inter}$ in Eq.\ref{form factor2}. The operator Q$^{|\text{C}_T|}_n$ depends on the configuration of corresponding ground states, and Table.\ref{Q_operator} presents the corresponding operator Q$^{|\text{C}_T|}_n$ for each ground state.  }
    \label{Ground state}
\end{table}

\begin{enumerate}
    \item For $C_T=\pm 2$: When two states fill orbital $o$ with the same Chern number sign, the other states fill any two slots of the four in orbital $o'$ with $C=0$. This results in $2C^4_2=12$ possible ground states. These states are the quantum Hall (QH) states due to non-zero Chern number, and furtherly they can be labeled by type-$n$ ($\text{C}_T=+2$), as well as type-$n^{\prime}$ states ($\text{C}_T=-2$), with the index $n$ ranging from 1 to 6. Fig.\ref{QH1}(b)-(g) present the configurations for the type-$n$ QH states. Concurrently, the configurations for type-$n^{\prime}$ states serve as complementary counterparts to the type-$n$ states, achieved through a horizontal exchange of the occupied bands between the left and right bands. This category naturally breaks time-reversal symmetry $T$. In addition, type-$(1\sim2)$[Type-$(1^{\prime}\sim2^{\prime})$] QH states can host two-fold rotational symmetry ($C_{2z}$), while other states will break this symmetry.
    \item For $C_T=\pm 1$: This has two types of half filling. When one/three state(s) occupies any slot in orbital $o$ and the remaining three/one fill(s) any slot in orbital $o'$, this yields $2C^4_3C^4_1=32$ possible QH states. We also use the type-$n$ and type-$n^{\prime}$ to label these QH states, with the index $n$ ranging from 1 to 16, and Fig.\ref{QH2} presents the configurations for all QH states. It is evident that this category breaks both time-reversal symmetry $T$ and two-fold rotational symmetry ($C_{2z}$).
    \item For $C_T=0$: Three types of configurations arise. First, four filled states occupy all of orbital $o$. Second, four filled states fills in all of orbital $o'$. Third, when a state fills in one of the two slots in orbital $o$ with $C=1$ and another fills in one of the two in orbital $o$ with $C=-1$, the remaining states occupy two of the four slots in orbital $o'$. The total number of ground states is $1+1+C^2_1C^2_1C^4_2=26$. These states can be labeled by type-$n$ and type-$n^{\prime}$, as shown in Fig.\ref{QH3}. For type-$(1\sim6)$ states, under the horizontal exchange of the occupied bands between the left and right bands, these ground states are invariant, indicating that these states don't have complementary counterparts type-$(1^{\prime}\sim6^{\prime})$. Furthermore, some of states can be classified as valley-polarized (VP) states, valley orbital (OP) states, valley-orbital locking (VOL) states,and valley Hall (VH) states, which correspond to type-$(1\sim2)$, type-$(3\sim4)$, type-$(5\sim6)$, and type-$(7\sim12)$[type-$(7^{\prime}\sim12^{\prime})$], respectively. Type-$3$ and Type-$4$ states exhibit time-reversal symmetry ($T$) and two-fold rotational symmetry ($C_{2z}$), while Type-$(7\sim8)$[Type-$(7^{\prime}\sim8^{\prime})$] QH states only feature time-reversal symmetry ($T$), with all other states breaking both of these symmetries.
\end{enumerate}

\begin{table}[t]
\renewcommand\arraystretch{2.2}
\setlength{\tabcolsep}{8mm}{
    \begin{tabular}{|*{4}{c|}}
        \hline
      Total Chern number & Intervalley-coherent state & Type-n & Energy correction \\ \hline
     \multirow{2}{*}{C$_T=\pm2$} &[QH]$_{o}$[T-IVC]$_{o^{\prime}}$ &Type-(1$\sim$4)& 4$\lambda_A$+2$\lambda_S$-2J \\ \cline{2-4}
       &[QH]$_{o}$[K-IVC]$_{o^{\prime}}$ &Type-(1$\sim$4)& 2$\lambda_A$+2$\lambda_S$-4J    \\ \hline
        \multirow{8}{*}{C$_T=0$}  & \multirow{2}{*}{[T-IVC]$_{o}$[T-IVC]$_{o^{\prime}}$} & Type-(1$\sim$2)  & 4$\lambda_A$+4$\lambda_S$   \\
         &  & Type-(3$\sim$4) & 4$\lambda_A$  \\\cline{2-4}
       & {[T-IVC]$_{o}$[K-IVC]$_{o^{\prime}}$} & Type-(1$\sim$2)  &2$\lambda_A$+2$\lambda_S$-2J   \\\cline{2-4}
       & {[K-IVC]$_{o}$[T-IVC]$_{o^{\prime}}$} & Type-(1$\sim$4)   &2$\lambda_A$+2$\lambda_S$-2J  \\\cline{2-4}
        & \multirow{2}{*}{[K-IVC]$_{o}$[K-IVC]$_{o^{\prime}}$} & Type-(1$\sim$2)   &4$\lambda_S$-4J \\
            &  & Type-(3$\sim$4) & -4J  \\\hline
    \end{tabular}}
    \captionsetup{justification=raggedright}
    \caption{Total Chern numbers ($\text{C}_T$) and energy corrections  in various IVC states.   }
    \label{Ground stateIVC}
\end{table}

For ease of comparison, Table \ref{compare} tabulates the number of filling configurations. The quadruple flat bands present a richer tapestry of configurations, notably introducing systems with $C_T=\pm 1$, which is absent in the conventional TBG's double flat bands. Furthermore, Table.\ref{Ground state} presents energy corrections, total Chern numbers ($\text{C}_T$), and the operator Q$^{|\text{C}_T|}_n$ for the three categories of ground states, as will be discussed in the follow.

In addition, similar to the double flat bands\cite{Bultinck2020}, intervalley coherent (IVC) states can also be realized in the quadruple flat bands. Within the manifold of Chern zero states, composed of double flat bands, the T-IVC and K-IVC states can be described as XY aligned and anti-aligned orders. Since the flat bands labeled by $o$ and $o^{\prime}$ orbitals are topological nontrivial and trivial, we can individually analyze IVC states in these two orbitals, including both time-reversal IVC (T-IVC) or Kramers IVC (K-IVC) states\cite{Bultinck2020}. On one hand, when considering the QH state in the $o$ orbital and the IVC state in the $o^{\prime}$ orbital, a mixture of the QH and IVC state is realized, in sharp contrast to the case of double flat bands. On the other band, a combination of various types of IVC states in these two orbitals can also be achieved. Table.\ref{Ground stateIVC} summarizes possible various IVC states with total Chern numbers and energy corrections at half filling, and Fig.\ref{IVC1} and Fig.\ref{IVC2} exhibit the filling configuration for each IVC state. In Fig.\ref{IVC1} and Fig.\ref{IVC2}, we utilize IVC states as a basis to describe the flat bands\cite{Patrick2021SM,Bultinck2020}, facilitating the analysis of energy corrections, which are given by
\begin{eqnarray}
c^{\dagger}_{\pm,o,IVC}=\frac{1}{\sqrt{2}}[e^{-i\phi/2}c^{\dagger}_{A,o,K}\pm e^{i\phi/2}c^{\dagger}_{B,o,K'}],~~c^{\dagger}_{\pm,o,\overline{IVC}}=\frac{1}{\sqrt{2}}[e^{-i\phi/2}c^{\dagger}_{B,o,K}\pm e^{i\phi/2}c^{\dagger}_{A,o,K'}],\nonumber
\\
c^{\dagger}_{\pm,o^{\prime},IVC}=\frac{1}{\sqrt{2}}[e^{-i\phi/2}c^{\dagger}_{A,o^{\prime},K}\pm e^{i\phi/2}c^{\dagger}_{B,o^{\prime},K'}],~~c^{\dagger}_{\pm,o^{\prime},\overline{IVC}}=\frac{1}{\sqrt{2}}[e^{-i\phi/2}c^{\dagger}_{B,o^{\prime},K}\pm e^{i\phi/2}c^{\dagger}_{A,o^{\prime},K'}].
\end{eqnarray}
where the phase $\phi$ is related to the angle of the IVC order in the $x$-$y$ plane. Subsequently, these IVC states can be categorized into two groups based on the total Chern numbers $C_T=\pm 2, 0$, and each category encompasses distinct types of filling configurations, as illustrated in Table.\ref{Ground stateIVC}. For IVC states with $C_T=\pm 2$, we consider the QH state with $C_T=\pm 2$ in the $o$ orbital, and T-IVC or K-IVC states in the $o^{\prime}$ orbital, denoted as [QH]$_o$[T-IVC]$_{o^{\prime}}$ and [QH]$_o$[K-IVC]$_{o^{\prime}}$. Furthermore, each IVC state can accommodate four types of filling configurations, as illustrated in Fig.\ref{IVC1}. As for IVC states with $C_T=0$, we consider T-IVC or K-IVC states in the $o$ and $o^{\prime}$ orbital, resulting in four distinct IVC states denoted as [T-IVC]$_o$[T-IVC]$_{o^{\prime}}$, [T-IVC]$_o$[K-IVC]$_{o^{\prime}}$, [K-IVC]$_o$[I-IVC]$_{o^{\prime}}$ and [K-IVC]$_o$[K-IVC]$_{o^{\prime}}$. Moreover, each IVC state can also host four types of filling configurations, as presented in Fig.\ref{IVC2}.

\subsection{Energy correction of ground states}

Here, we investigate the energy correction of ground states induced by three perturbations, including single-particle dispersion, intersublattice-intraorbital and intrasublattice-interorbital interactions. We make a key assumption that the magnitude of these terms is significantly smaller compared to the interaction scale, allowing them to be treated as perturbations. Furthermore, we present an intuitive picture to understand the energy correction.

\subsubsection{Single-particle dispersion}

In the basis $\psi^{\dagger}_{\mathbf{k}_M}=[c^{\dagger}_{A,o,K,\mathbf{k}_M},c^{\dagger}_{B,o,K,\mathbf{k}_M},c^{\dagger}_{A,o^{\prime},K,\mathbf{k}_M},c^{\dagger}_{B,o^{\prime},K,\mathbf{k}_M},
c^{\dagger}_{A,o,K^{\prime},\mathbf{k}_M},c^{\dagger}_{B,o,K^{\prime},\mathbf{k}_M},c^{\dagger}_{A,o^{\prime},K^{\prime},\mathbf{k}_M},c^{\dagger}_{B,o^{\prime},K^{\prime},\mathbf{k}_M}]$, then the single-particle dispersion  can be written as $\mathcal{H}_0=\sum_{ \mathbf{k}_M }\psi^{\dagger}_{\mathbf{k}_M}h(\mathbf{k}_M)\psi_{\mathbf{k}_M}$. Under the constrain of chiral symmetry and $PT$ symmetry, the matrix $h(\mathbf{k}_M)$ can be given by
\begin{eqnarray}\label{single-particle}
h(\mathbf{k}_M)=\left(\begin{array}{cccccccc}
0 & t(\mathbf{k}_M) & 0 & 0 & 0 & 0 & 0 & 0   \\
t^*(\mathbf{k}_M) & 0 & 0 & 0 & 0 & 0 & 0 & 0  \\
0 & 0 & 0& t(\mathbf{k}_M) & 0 & 0 & 0 & 0 \\
0 &  0 & t^*(\mathbf{k}_M) & 0 & 0 & 0 & 0 & 0  \\
0 &  0 &  0 & 0 & 0 & t^*(\mathbf{k}_M) & 0 & 0  \\
0 & 0 &  0 &  0 & t(\mathbf{k}_M) & 0 & 0 & 0   \\
0 & 0 & 0 &  0 &  0 & 0 & 0 &  t^*(\mathbf{k}_M)  \\
0 & 0 & 0 & 0 &  0 &  0 & t(\mathbf{k}_M) & 0   \\
\end{array}\right)=h_x(\mathbf{k}_M)\kappa_x\mu_0\eta_0+h_y(\mathbf{k}_M)\kappa_y\mu_0\eta_z,
\end{eqnarray}
where $t(\mathbf{k}_M)=h_x(\mathbf{k}_M)-ih_y(\mathbf{k}_M)$ with real functions $h_x(\mathbf{k}_M)$ and $h_y(\mathbf{k})$, and we only consider the intra-orbital hopping $t(\mathbf{k}_M)$ in the same valley.

The analysis of the aforementioned ground states primarily focuses on the intrasublattice-intraorbital interaction, neglecting the effects of single-particle dispersion. To account for the influence of single-particle dispersion, we employ second-order perturbation theory to evaluate the perturbative correction to the energy. Then, this correction can be expressed as
\begin{eqnarray}
\Delta E=\sum_{\boldsymbol{n}}^{\prime} \frac{\langle\psi^{(0)}\left|\mathcal{H}_0\right| \psi_{\boldsymbol{n}}^{(0)}\rangle \langle\psi_{\boldsymbol{n}}^{(0)}\left|\mathcal{H}_0\right| \psi^{(0)}\rangle}{E^{(0)}-E_{\boldsymbol{n}}^{(0)}}.\label{perturbation}
\end{eqnarray}

Here, the symbol $\psi^{(0)}$ represents the unperturbed ground state, $\mathcal{H}_0$ denotes the perturbation Hamiltonian, $\psi_{\boldsymbol{n}}^{(0)}$ represents the unperturbed excited states, and $E^{(0)}$ and $E_{\boldsymbol{n}}^{(0)}$ correspond to the unperturbed ground state energy and excited state energies, respectively. Let us delve into the analysis of the Eq.\ref{perturbation}.

\begin{figure}
\centerline{\includegraphics[width=1\textwidth]{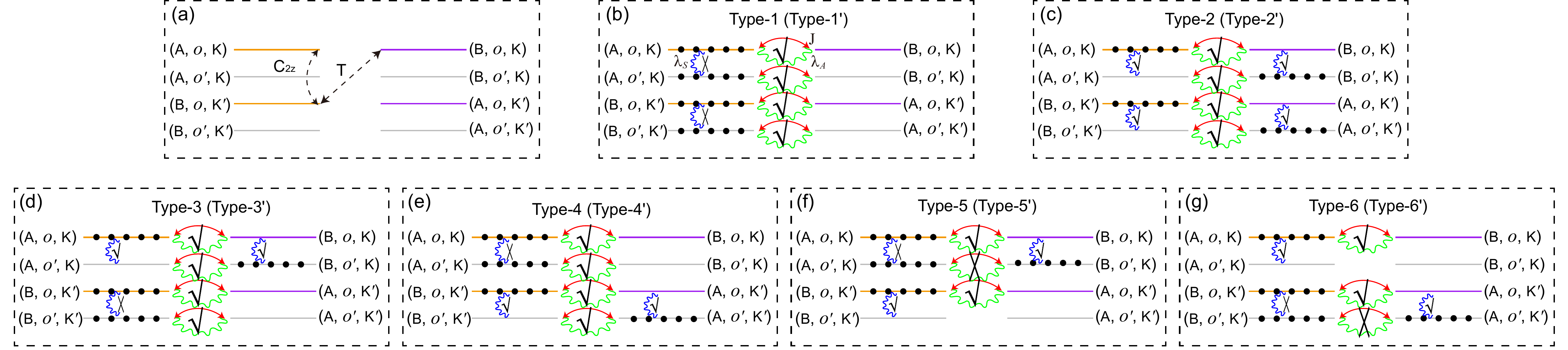}}
\captionsetup{justification=raggedright}
\caption{(color online) (a) In the spinless flat band basis labelled by sublattice ($A,B$), valley ($K,K'$), and orbital ($o,o'$) indices, the behavior of time-reversal symmetry ($T$) and two-fold rotational symmetry ($C_{2z}$) is illustrated. (b)-(g) The 12 QH states, characterized by total Chern number $C_T=\pm2$, are categorized as type-$n$ ($C_T=+2$) and type-$n^{\prime}$ ($C_T=-2$), with $n$ ranging from 1 to 6. The type-$n^{\prime}$ states correspond as complementary counterparts to the type-$n$ states, and they can be generated by horizontally exchanging the occupied bands between the left and right bands. The flat band labelled by gray, orange and pink lines possesses Chern number $C=0$, $C=+1$, and $C=-1$. The red curves represent the single-particle dispersion, while the green and blue wavy lines correspond to the intersublattice-intraorbital and intrasublattice-interorbital interactions. The parameters of $J$, $\lambda_A$ and $\lambda_S$ are the strength of the corresponding perturbation within the basis. The symbols $\surd$ or $\times$ indicate whether the processes are permitted or forbidden.   \label{QH1}}
\end{figure}

Firstly, the single-particle Hamiltonian $\mathcal{H}_0$ establishes connections between bands with different sublattice within the same orbital and valley, which can be simplified to
\begin{eqnarray}
\mathcal{H}_0=\sum_{\mathbf{k}_M}\{[h_x(\mathbf{k}_M)-i h_y(\mathbf{k}_M)]c_{A,\mathbf{k}_M}^{\dagger}c_{B,\mathbf{k}_M}+h.c\}.
\end{eqnarray}
For convenience, we have omitted the symbols $\mu$ and $\eta$.

Secondly, considering the simplified basis operator $\psi^{\dagger}_{\mathbf{k}_M}=[c^{\dagger}_{A,\mathbf{k}_M},c^{\dagger}_{B,\mathbf{k}_M}]$, there are three ground states, namely $| \Psi_{A}\rangle=\prod_{\mathbf{k}_M} c_{\mathbf{k}_M,A}^{\dagger}|0\rangle$, $| \Psi_{B}\rangle=\prod_{\mathbf{k}_M} c_{\mathbf{k}_M,B}^{\dagger}|0\rangle$ and $| \Psi_{A/B}\rangle=\prod_{\mathbf{k}_M} c_{\mathbf{k}_M,A}^{\dagger}c_{\mathbf{k}_M,B}^{\dagger}|0\rangle$.  If a state exhibits both bands connected by $\mathcal{H}_0$ being either completely filled or completely empty, it will be annihilated by the perturbation operator $\mathcal{H}_0$ due to the complete blocked of tunneling processes. Consequently, only the states $| \Psi_{A}\rangle$ and $| \Psi_{B}\rangle$ can contribute to the energy correction induced by the perturbation $\mathcal{H}_0$. For the purpose of illustration, we consider $| \psi^{(0)}\rangle=| \Psi_{A}\rangle$ as an exemplary case.

Thirdly, by applying the perturbation $\mathcal{H}_0$ to the state  $| \psi^{(0)}\rangle=| \Psi_{A}\rangle$, we can obtain
\begin{eqnarray}
H_0|\Psi_A\rangle=\sum_{\mathbf{k}_M}[h_x(\mathbf{k}_M)+i h_y(\mathbf{k}_M)]\Psi_{\mathbf{k}_M, \mathrm{eh}}\rangle,~~| \Psi_{\mathbf{k}_M, \mathrm{eh}}\rangle=c_{\mathbf{k}_M,B}^{\dagger}c_{\mathbf{k}_M,A}| \Psi_{A}\rangle,
\end{eqnarray}
where $| \Psi_{\mathbf{k}_M, \mathrm{eh}}\rangle$ represents an unperturbed excited state containing an electron-hole pair with momentum $\mathbf{k}_M$. Consequently, we can utilize $| \psi_{\boldsymbol{n}}^{(0)}\rangle=| \Psi_{\mathbf{k}_M, \mathrm{eh}}\rangle$ to calculate the energy correction described by Eq.\ref{perturbation}. Moreover, it is worth noting that $E^{(0)}$ in Eq.\ref{perturbation} should be zero, as we consider the zero-energy flat bands.

Finally, based on the preceding discussion, we can derive the expression for the energy correction resulting from  single-particle dispersion in the flat bands, which is given by
\begin{eqnarray}\label{energy correction1}
\Delta E=-J=-\frac{1}{N}\sum_{\mathbf{k}_M,\mathbf{k}_M^{\prime}} \frac{[h_x(\mathbf{k}_M)+i h_y(\mathbf{k}_M)][h_x(\mathbf{k}_M^{\prime})-i h_y(\mathbf{k}_M^{\prime})]}{\left\langle\Psi_{\mathbf{k}_M, \mathrm{eh}}\left|\mathcal{{H}}^{\text{S},\text{intra}}_I\right| \Psi_{\mathbf{k}_M^{\prime}, \mathrm{eh}}\right\rangle}.
\end{eqnarray}

Hence, by using Eq.\ref{energy correction1}, we can obtain the energy correction of all ground states induced by single-particle dispersion, as shown in Table.\ref{Ground state}.

Similarly, for various IVC states, we can also obtain the energy correction induced by single-particle dispersion, as shown in Table.\ref{Ground stateIVC}.

\begin{figure}
\centerline{\includegraphics[width=1\textwidth]{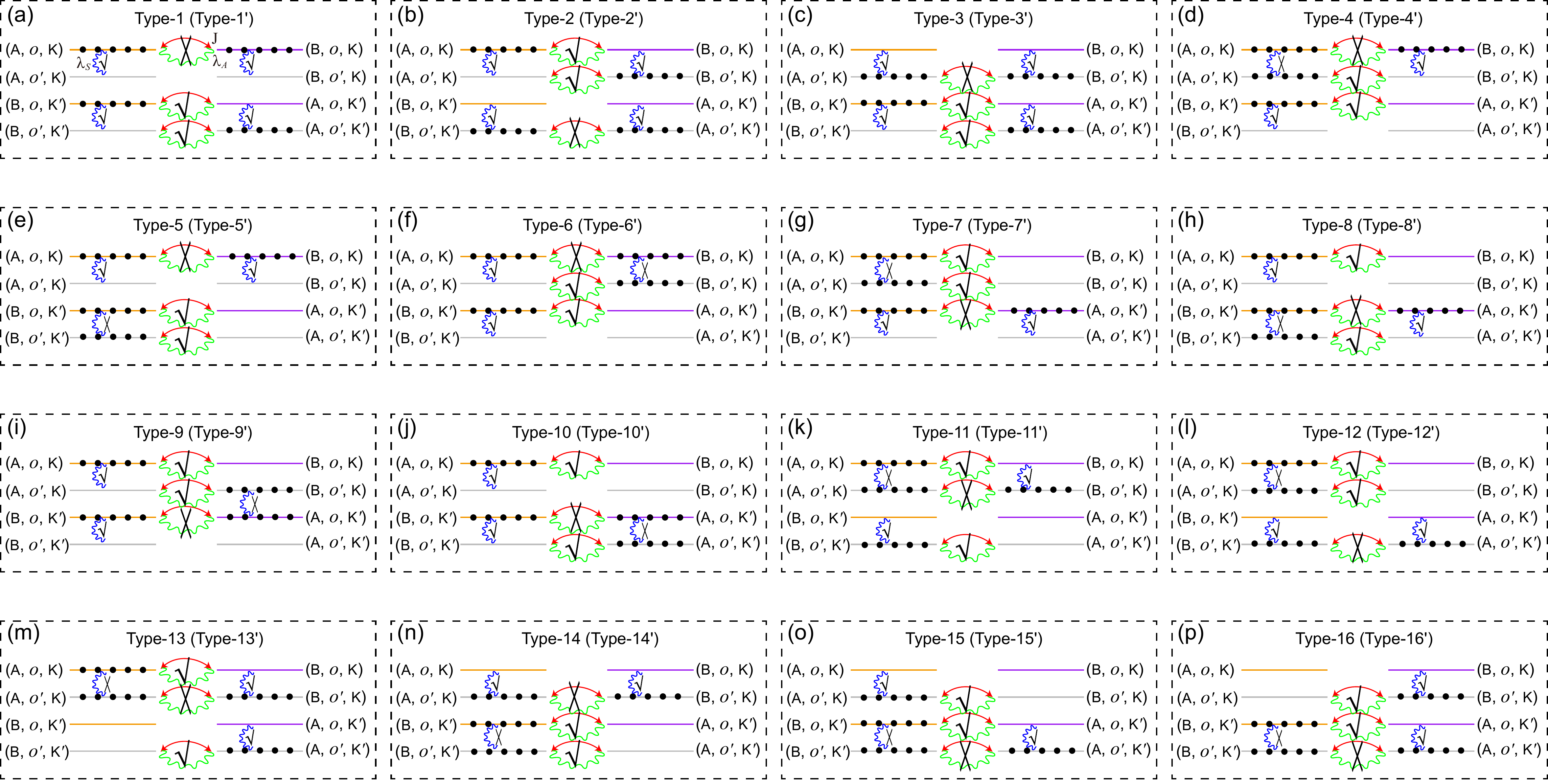}}
\captionsetup{justification=raggedright}
\caption{(color online) (a)-(p) The 32 QH states, characterized by total Chern number $C_T=\pm1$, are categorized as type-$n$ ($C_T=+1$) and type-$n^{\prime}$ ($C_T=-1$), with $n$ ranging from 1 to 16. The type-$n^{\prime}$ states correspond as complementary counterparts to the type-$n$ states, and they can be generated by horizontally shifting the occupied bands between the left and right bands.\label{QH2}}
\end{figure}

\begin{figure}
\centerline{\includegraphics[width=1\textwidth]{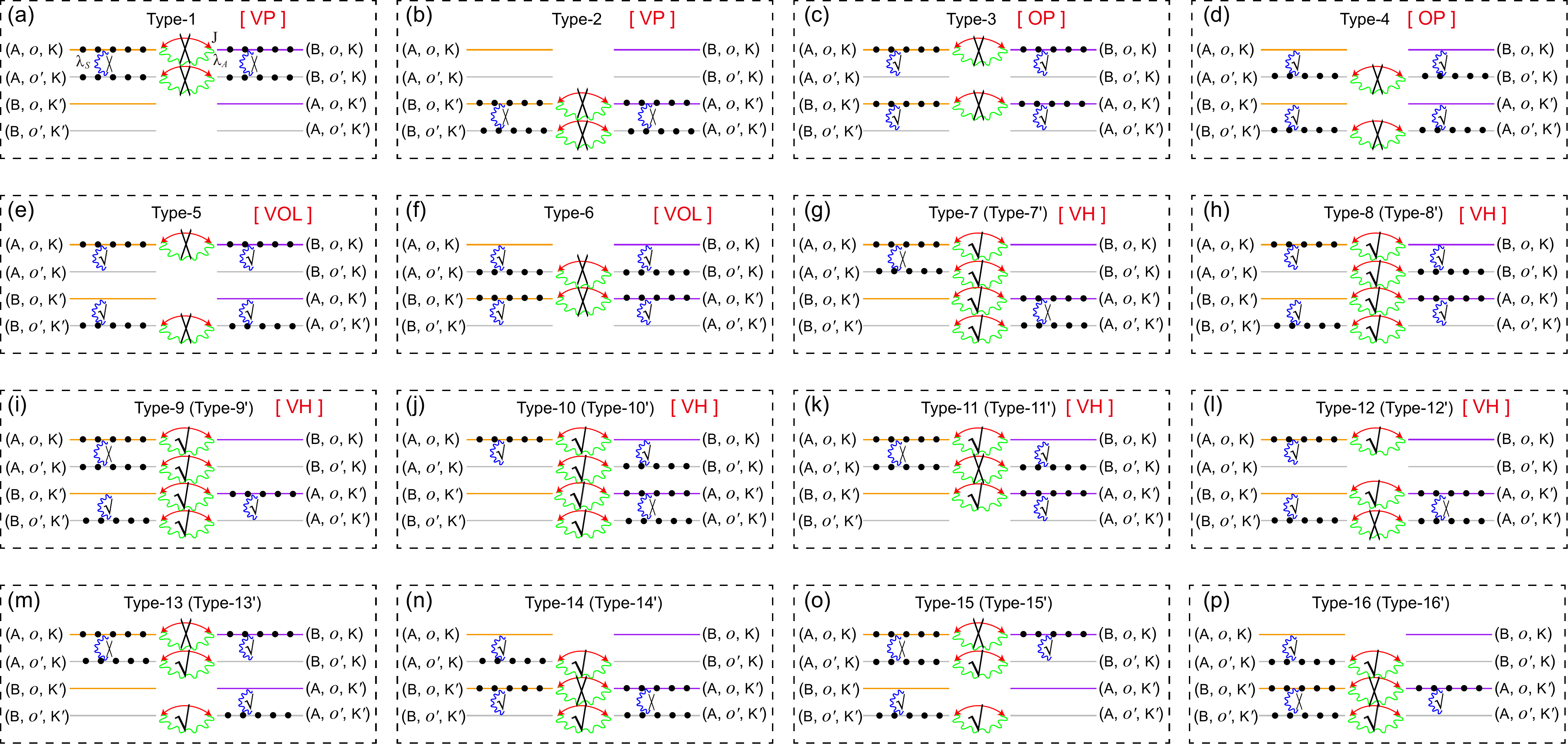}}
\captionsetup{justification=raggedright}
\caption{(color online) (a)-(p) 26 ground states with zero total Chern number. The type-$n^{\prime}$ ($n=7\sim16$) states correspond as complementary counterparts to the type-$n$ states, and they can be generated by horizontally shifting the occupied bands between the left and right bands.\label{QH3}}
\end{figure}

\subsubsection{Intersublattice-intraorbital interaction}

To account for the influence of intersublattice-intraorbital interaction, we employ first-order perturbation theory to evaluate the perturbative correction to the energy. Then, this correction can be expressed as
\begin{eqnarray}
\Delta E^A=\langle\psi^{(0)}|\mathcal{{H}}^{\text{A}}_I| \psi^{(0)}\rangle,\label{correction2}
\end{eqnarray}
where the symbol $\psi^{(0)}$ represents the unperturbed ground state, $\mathcal{{H}}^{\text{A}}_I$ denotes the intersublattice-intraorbital interaction Hamiltonian. Based on the symmetry relationships Eq.\ref{symmetry constrain}, the intersublattice-intraorbital projected interaction Hamiltonian without chiral symmetry in the same valley and orbital can be written as
\begin{eqnarray}
\mathcal{{H}}^{\text{A}}_I=\frac{1}{2 \Omega} \sum_{\mathbf{q}_M,\mathbf{g}}V(\mathbf{q}_M+\mathbf{g}) \delta \bar{\rho}^{\text{A}}_{\mathbf{q}_M+\mathbf{g}} \delta \bar{\rho}^{\text{A}}_{-\mathbf{q}_M-\mathbf{g}},~~\delta \bar{\rho}^{\text{A}}_{\pm\mathbf{q}_M\pm\mathbf{g}}=\sum^{\beta,\gamma}_{\mathbf{k}_M } c_{\beta,\mathbf{k}_M}^{\dagger}\left[\Lambda^{\text{A}}_{\pm\mathbf{q}_M\pm\mathbf{g}}(\mathbf{k}_M)\right]_{\beta,\gamma} c_{\gamma, \mathbf{k}_M\pm\mathbf{q}_M}
\end{eqnarray}

\begin{eqnarray}
\Lambda^{\text{A}}_{\pm\mathbf{q}_M\pm\mathbf{g}}(\mathbf{k}_M) =\left(\begin{array}{cccccccc}
0 & \Delta_{\pm} & 0 & 0 & 0 & 0 & 0 & 0   \\
\Delta_{\pm}^* & 0  & 0 & 0 & 0 & 0 & 0 & 0  \\
0 & 0 &0 & \Delta_{\pm}  & 0 & 0 & 0 & 0 \\
0 &  0 & \Delta_{\pm}^* & 0 & 0 & 0 & 0 & 0  \\
0 &  0 &  0 & 0 & 0 & -\Delta_{\pm}^* & 0 & 0  \\
0 & 0 &  0 &  0 & -\Delta_{\pm} & 0 & 0 &  0  \\
0 & 0 & 0 &  0 &  0 & 0 & 0 &  -\Delta_{\pm}^*  \\
0 & 0 & 0 & 0 &  0 &  0 & -\Delta_{\pm} &  0  \\
\end{array}\right),\Delta_{\pm}=\langle u_{A, o/o', K, \mathbf{k}_M} \mid u_{B, o/o', K, \mathbf{k}_M\pm\mathbf{q}_M}\rangle,
\end{eqnarray}
where the  band basis is $\psi^{\dagger}_{\mathbf{k}_M}=[c^{\dagger}_{A,o,K,\mathbf{k}_M},c^{\dagger}_{B,o,K,\mathbf{k}_M},c^{\dagger}_{A,o^{\prime},K,\mathbf{k}_M},c^{\dagger}_{B,o^{\prime},K,\mathbf{k}_M},
c^{\dagger}_{A,o,K^{\prime},\mathbf{k}_M},c^{\dagger}_{B,o,K^{\prime},\mathbf{k}_M},c^{\dagger}_{A,o^{\prime},K^{\prime},\mathbf{k}_M},c^{\dagger}_{B,o^{\prime},K^{\prime},\mathbf{k}_M}]$.
By setting $\Delta_{\pm}=F^{\text{A}}_{\pm\mathbf{q}_M}(\mathbf{k}_M) e^{-i \Phi^{\text{A}}_{\mathbf{q}_M}(\pm\mathbf{k}_M)}$, and then $\langle u_{A, o/o', K, \mathbf{k}_M-\mathbf{q}_M} \mid u_{B, o/o', K, \mathbf{k}_M}\rangle=F^{\text{A}}_{\mathbf{q}_M}(\mathbf{k}_M-\mathbf{q}_M) e^{-i \Phi^{\text{A}}_{\mathbf{q}_M}(\mathbf{k}_M-\mathbf{q}_M)}$, which can be further written as
\begin{eqnarray}
\langle u_{A, o/o', K, \mathbf{k}_M-\mathbf{q}_M} \mid u_{B, o/o', K, \mathbf{k}_M}\rangle&=&\langle u_{B, o/o', K, \mathbf{k}_M} \mid  u_{A, o/o', K, \mathbf{k}_M-\mathbf{q}_M} \rangle^*=\Delta_{-}=F^{\text{A}}_{-\mathbf{q}_M}(\mathbf{k}_M) e^{-i \Phi^{\text{A}}_{\mathbf{q}_M}(-\mathbf{k}_M)}.
\end{eqnarray}
Hence, we can get
\begin{eqnarray}\label{relationshipF}
F^{\text{A}}_{\mathbf{q}_M}(\mathbf{k}_M-\mathbf{q}_M)=F^{\text{A}}_{-\mathbf{q}_M}(\mathbf{k}_M),~~ e^{-i \Phi^{\text{A}}_{\mathbf{q}_M}(\mathbf{k}_M-\mathbf{q}_M)}=e^{-i \Phi^{\text{A}}_{-\mathbf{q}_M}(\mathbf{k}_M)}.
\end{eqnarray}
Then, the intra-orbital form factor can be simplified to
\begin{eqnarray}\label{form factor1}
\Lambda^{\text{A}}_{\pm\mathbf{q}_M\pm\mathbf{g}}(\mathbf{k}_M)=\kappa_x\mu_0 \eta_zF^{\text{A}}_{\pm\mathbf{q}_M}(\mathbf{k}_M) e^{i\Phi^{\text{A}}_{\pm\mathbf{q}_M}(\mathbf{k}_M)\kappa_z\mu_0\eta_z}.
\end{eqnarray}
Then,
\begin{eqnarray}
\delta \bar{\rho}^{\text{A}}_{\mathbf{q}_M+\mathbf{g}}&=&\sum^{\mu}_{\mathbf{k}_M }F^{\text{A}}_{\mathbf{q}_M}(\mathbf{k}_M)[ e^{-i \Phi^{\text{A}}_{\mathbf{q}_M}(\mathbf{k}_M)} (c^{\dagger}_{A,\mu,K,\mathbf{k}_M}c_{B,\mu,K,\mathbf{k}_M+\mathbf{q}_M}-c^{\dagger}_{B,\mu,K',\mathbf{k}_M}c_{A,\mu,K',\mathbf{k}_M+\mathbf{q}_M})\nonumber
\\
&+&e^{i \Phi^{\text{A}}_{\mathbf{q}_M}(\mathbf{k}_M)} (c^{\dagger}_{B,\mu,K,\mathbf{k}_M}c_{A,\mu,K,\mathbf{k}_M+\mathbf{q}_M}-c^{\dagger}_{A,\mu,K',\mathbf{k}_M}c_{B,\mu,K',\mathbf{k}_M+\mathbf{q}_M}],\nonumber
\\
\delta \bar{\rho}^{\text{A}}_{-\mathbf{q}_M-\mathbf{g}}&=&\sum^{\mu}_{\mathbf{k}_M }F^{\text{A}}_{-\mathbf{q}_M}(\mathbf{k}_M)[ e^{-i \Phi^{\text{A}}_{-\mathbf{q}_M}(\mathbf{k}_M)} (c^{\dagger}_{A,\mu,K,\mathbf{k}_M}c_{B,\mu,K,\mathbf{k}_M-\mathbf{q}_M}-c^{\dagger}_{B,\mu,K',\mathbf{k}_M}c_{A,\mu,K',\mathbf{k}_M-\mathbf{q}_M})\nonumber
\\
&+& e^{i \Phi^{\text{A}}_{-\mathbf{q}_M}(\mathbf{k}_M)}  (c^{\dagger}_{B,\mu,K,\mathbf{k}_M}c_{A,\mu,K,\mathbf{k}_M-\mathbf{q}_M}-c^{\dagger}_{A,\mu,K',\mathbf{k}_M}c_{B,\mu,K',\mathbf{k}_M-\mathbf{q}_M}]
\end{eqnarray}

Here, we turn to calculate $\Delta E^A$ equation Eq.\ref{correction2}. We take type-1 QH state with total Chern number $C_T=1$ as example, and the wave funtion can be written as $| \Psi^{C=1}_{\text{type-1}}\rangle=\prod_{\mathbf{k}_M} c_{A,o,K, \mathbf{k}_M}^{\dagger}c_{A,o', K^{\prime}, \mathbf{k}_M}^{\dagger}c_{B,o,K^{\prime}, \mathbf{k}_M}^{\dagger}c_{B,o, K, \mathbf{k}_M}^{\dagger}|0\rangle$. Then, the derivation process is as follows
\begin{eqnarray}
&&\delta \bar{\rho}^{\text{A}}_{\mathbf{q}_M+\mathbf{g}}| \Psi^{C=1}_{\text{type-1}}\rangle,\nonumber
\\
&=& \sum_{\mathbf{k}_M }F^{\text{A}}_{\mathbf{q}_M}(\mathbf{k}_M)[ -e^{-i \Phi^{\text{A}}_{\mathbf{q}_M}(\mathbf{k}_M)}c^{\dagger}_{B,o',K',\mathbf{k}_M}c_{A,o',K',\mathbf{k}_M+\mathbf{q}_M}-e^{i \Phi^{\text{A}}_{\mathbf{q}_M}(\mathbf{k}_M)}c^{\dagger}_{A,o,K',\mathbf{k}_M}c_{B,o,K',\mathbf{k}_M+\mathbf{q}_M}]
 | \Psi^{C=1}_{\text{type-1}}\rangle,\nonumber
\\
&&\delta \bar{\rho}^{\text{A}}_{-\mathbf{q}_M-\mathbf{g}}| \Psi^{C=1}_{\text{type-1}}\rangle,\nonumber
\\
&=& \sum_{\mathbf{k}_M }F^{\text{A}}_{-\mathbf{q}_M}(\mathbf{k}_M)[ -e^{-i \Phi^{\text{A}}_{-\mathbf{q}_M}(\mathbf{k}_M)}c^{\dagger}_{B,o',K',\mathbf{k}_M}c_{A,o',K',\mathbf{k}_M-\mathbf{q}_M}-e^{i \Phi^{\text{A}}_{-\mathbf{q}_M}(\mathbf{k}_M)}c^{\dagger}_{A,o,K',\mathbf{k}_M}c_{B,o,K',\mathbf{k}_M-\mathbf{q}_M}]
| \Psi^{C=1}_{\text{type-1}}\rangle,
\end{eqnarray}
where we can omit other terms due to the properties of fermions($c^{\dagger}c^{\dagger}|0\rangle=0$ and $c|0\rangle=0$). Therefore, we obtain
\begin{eqnarray}
&&\langle\Psi^{C=1}_{\text{type-1}}|\delta \bar{\rho}^{\text{A}}_{\mathbf{q}_M+\mathbf{g}} \delta \bar{\rho}^{\text{A}}_{-\mathbf{q}_M-\mathbf{g}}| \Psi^{C=1}_{\text{type-1}}\rangle,\nonumber
\\
&=&\langle\Psi^{C=1}_{\text{type-1}}|\{\sum_{\mathbf{k}'_M }F^{\text{A}}_{\mathbf{q}_M}(\mathbf{k}'_M)[ -e^{i \Phi^{\text{A}}_{\mathbf{q}_M}(\mathbf{k}'_M)}c^{\dagger}_{A,o',K',\mathbf{k}'_M}c_{B,o',K',\mathbf{k}'_M+\mathbf{q}_M}-e^{-i \Phi^{\text{A}}_{\mathbf{q}_M}(\mathbf{k}'_M)}c^{\dagger}_{B,o,K',\mathbf{k}'_M}c_{A,o,K',\mathbf{k}'_M+\mathbf{q}_M}]
\}\nonumber
\\
&&\{\sum_{\mathbf{k}_M }F^{\text{A}}_{-\mathbf{q}_M}(\mathbf{k}_M)[ -e^{-i \Phi^{\text{A}}_{-\mathbf{q}_M}(\mathbf{k}_M)}c^{\dagger}_{B,o',K',\mathbf{k}_M}c_{A,o',K',\mathbf{k}_M-\mathbf{q}_M}-e^{i \Phi^{\text{A}}_{-\mathbf{q}_M}(\mathbf{k}_M)}c^{\dagger}_{A,o,K',\mathbf{k}_M}c_{B,o,K',\mathbf{k}_M-\mathbf{q}_M}]
\} | \Psi^{C=1}_{\text{type-1}}\rangle,\nonumber
\\
&=&\langle\Psi^{C=1}_{\text{type-1}}|\{\sum_{\mathbf{k}'_M }F^{\text{A}}_{\mathbf{q}_M}(\mathbf{k}'_M)[e^{i \Phi^{\text{A}}_{\mathbf{q}_M}(\mathbf{k}'_M)}c^{\dagger}_{A,o',K',\mathbf{k}'_M}c_{B,o',K',\mathbf{k}'_M+\mathbf{q}_M}\}\nonumber
\{\sum_{\mathbf{k}_M }F^{\text{A}}_{-\mathbf{q}_M}(\mathbf{k}_M)[ e^{-i\Phi^{\text{A}}_{-\mathbf{q}_M}(\mathbf{k}_M)}c^{\dagger}_{B,o',K',\mathbf{k}_M}c_{A,o',K',\mathbf{k}_M-\mathbf{q}_M}\}\\
&+&\{\sum_{\mathbf{k}'_M }F^{\text{A}}_{\mathbf{q}_M}(\mathbf{k}'_M)e^{-i \Phi^{\text{A}}_{\mathbf{q}_M}(\mathbf{k}'_M)}c^{\dagger}_{B,o,K',\mathbf{k}'_M}c_{A,o,K',\mathbf{k}'_M+\mathbf{q}_M}
\}\{\sum_{\mathbf{k}_M }F^{\text{A}}_{-\mathbf{q}_M}(\mathbf{k}_M)[e^{i\Phi^{\text{A}}_{-\mathbf{q}_M}(\mathbf{k}_M)}c^{\dagger}_{A,o,K',\mathbf{k}_M}c_{B,o,K',\mathbf{k}_M-\mathbf{q}_M}]
\} | \Psi^{C=1}_{\text{type-1}}\rangle,\nonumber
\\
&=&\sum_{\mathbf{k}_M }\langle\Psi^{C=1}_{\text{type-1}}|\{F^{\text{A}}_{\mathbf{q}_M}(\mathbf{k}_M-\mathbf{q}_M)[e^{i \Phi^{\text{A}}_{\mathbf{q}_M}(\mathbf{k}_M-\mathbf{q}_M)}c^{\dagger}_{A,o',K',\mathbf{k}_M-\mathbf{q}_M}c_{B,o',K',\mathbf{k}_M}\}\nonumber
\{F^{\text{A}}_{-\mathbf{q}_M}(\mathbf{k}_M)[ e^{-i\Phi^{\text{A}}_{-\mathbf{q}_M}(\mathbf{k}_M)}c^{\dagger}_{B,o',K',\mathbf{k}_M}c_{A,o',K',\mathbf{k}_M-\mathbf{q}_M}\}\\
&+&\{F^{\text{A}}_{\mathbf{q}_M}(\mathbf{k}_M-\mathbf{q}_M)e^{-i \Phi^{\text{A}}_{\mathbf{q}_M}(\mathbf{k}_M-\mathbf{q}_M)}c^{\dagger}_{B,o,K',\mathbf{k}_M-\mathbf{q}_M}c_{A,o,K',\mathbf{k}_M}
\}\{F^{\text{A}}_{-\mathbf{q}_M}(\mathbf{k}_M)[e^{i\Phi^{\text{A}}_{-\mathbf{q}_M}(\mathbf{k}_M)}c^{\dagger}_{A,o,K',\mathbf{k}_M}c_{B,o,K',\mathbf{k}_M-\mathbf{q}_M}]
\} | \Psi^{C=1}_{\text{type-1}}\rangle,\nonumber
\\
&=&\sum_{\mathbf{k}_M }[F^{\text{A}}_{-\mathbf{q}_M}(\mathbf{k}_M)]^2\langle\Psi^{C=1}_{\text{type-1}}|n_{A,o',K',\mathbf{k}_M-\mathbf{q}_M}(1-n_{B,o',K',\mathbf{k}_M})+
n_{B,o,K',\mathbf{k}_M-\mathbf{q}_M}(1-n_{A,o,K',\mathbf{k}_M}) | \Psi^{C=1}_{\text{type-1}}\rangle.
\end{eqnarray}
In the second step, we have excluded terms that are inherently zero. Subsequently, in the third step, we substitute $\mathbf{k}'_M$ with $\mathbf{k}_M-\mathbf{q}_M$. Finally, in the fourth step, we make use of the relationships as defined in Eq. \ref{relationshipF}.

\begin{figure}
\centerline{\includegraphics[width=1\textwidth]{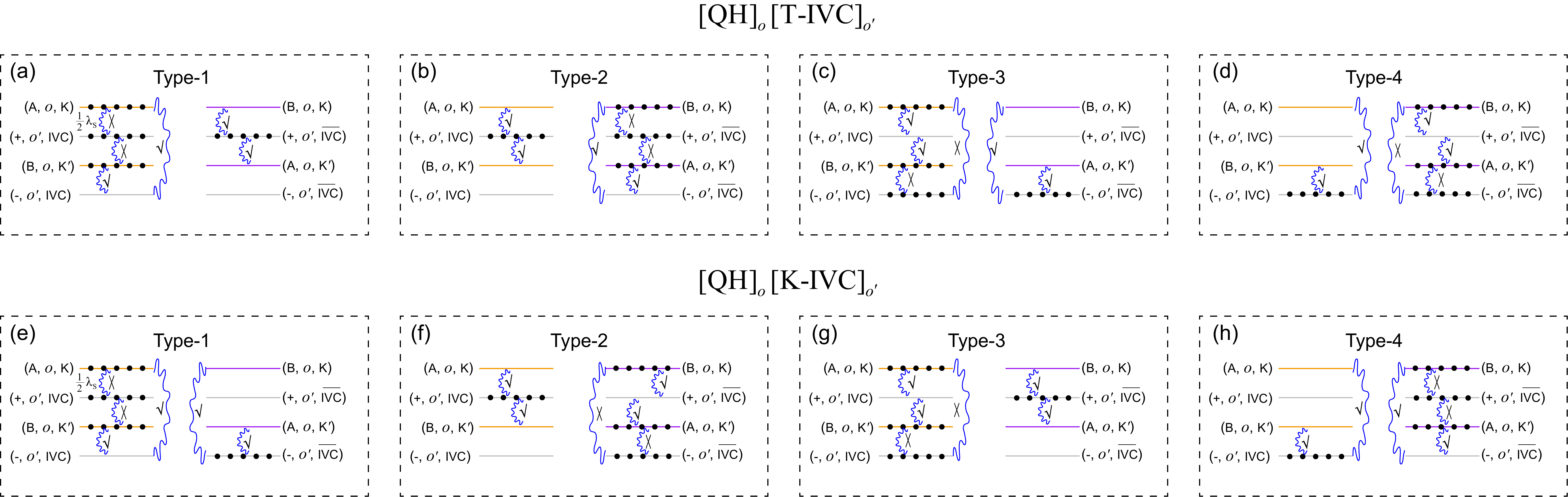}}
\captionsetup{justification=raggedright}
\caption{(color online) The filling configurations for IVC states with $C_T=\pm 2$, and we use IVC states as basis to describe $o^{\prime}$-orbital flat bands. The blue wavy line corresponds to the intrasublattice-interorbital interactions, and the symbols $\surd$ or $\times$ indicate whether the processes are permitted or forbidden.\label{IVC1}}
\end{figure}

Hence, the energy correction Eq.\ref{correction2} for the type-1 QH state with total Chern number $C_T=1$ becomes
\begin{eqnarray}\label{type-1}
&&\Delta E^A_{\text{type-1}}=\frac{1}{2 \Omega} \sum_{\mathbf{q}_M,\mathbf{g}} V_{\mathbf{q}_M+\mathbf{g}}\left\langle\Psi^{C=1}_{\text{type-1}}\left| \delta \bar{\rho}^{\text{A}}_{\mathbf{q}_M+\mathbf{g}} \delta \bar{\rho}^{\text{A}}_{-\mathbf{q}_M-\mathbf{g}}\right| \Psi^{C=1}_{\text{type-1}}\right\rangle,\nonumber
\\
&&=\frac{1}{2 \Omega} \sum_{\mathbf{q}_M,\mathbf{k}_M,\mathbf{g}}V_{\mathbf{q}_M+\mathbf{g}} [F^{\text{A}}_{-\mathbf{q}_M}(\mathbf{k}_M)]^2\left\langle\Psi^{C=1}_{\text{type-1}}\left| n_{A,o',K',\mathbf{k}_M-\mathbf{q}_M}(1-n_{B,o',K',\mathbf{k}_M})+
n_{B,o,K',\mathbf{k}_M-\mathbf{q}_M}(1-n_{A,o,K',\mathbf{k}_M})\right| \Psi^{C=1}_{\text{type-1}}\right\rangle,\nonumber
\\
&&=2\lambda_A,
\end{eqnarray}
where $\lambda_A=\frac{1}{ 2\Omega} \sum_{\mathbf{q}_M,\mathbf{k}_M,\mathbf{g}} V_{\mathbf{q}_M+\mathbf{g}} [{F}^{A}_{-\mathbf{q}_M}(\boldsymbol{k}_M)]^2$. Then, we can obtain the energy correction of other ground states induced by intersublattice-intraorbital interaction, as shown in Table.\ref{Ground state}.

Similarly, for various IVC states, we can also obtain the energy correction induced by intersublattice-intraorbital interaction, as presented in Table.\ref{Ground stateIVC}.

\begin{figure}
\centerline{\includegraphics[width=1\textwidth]{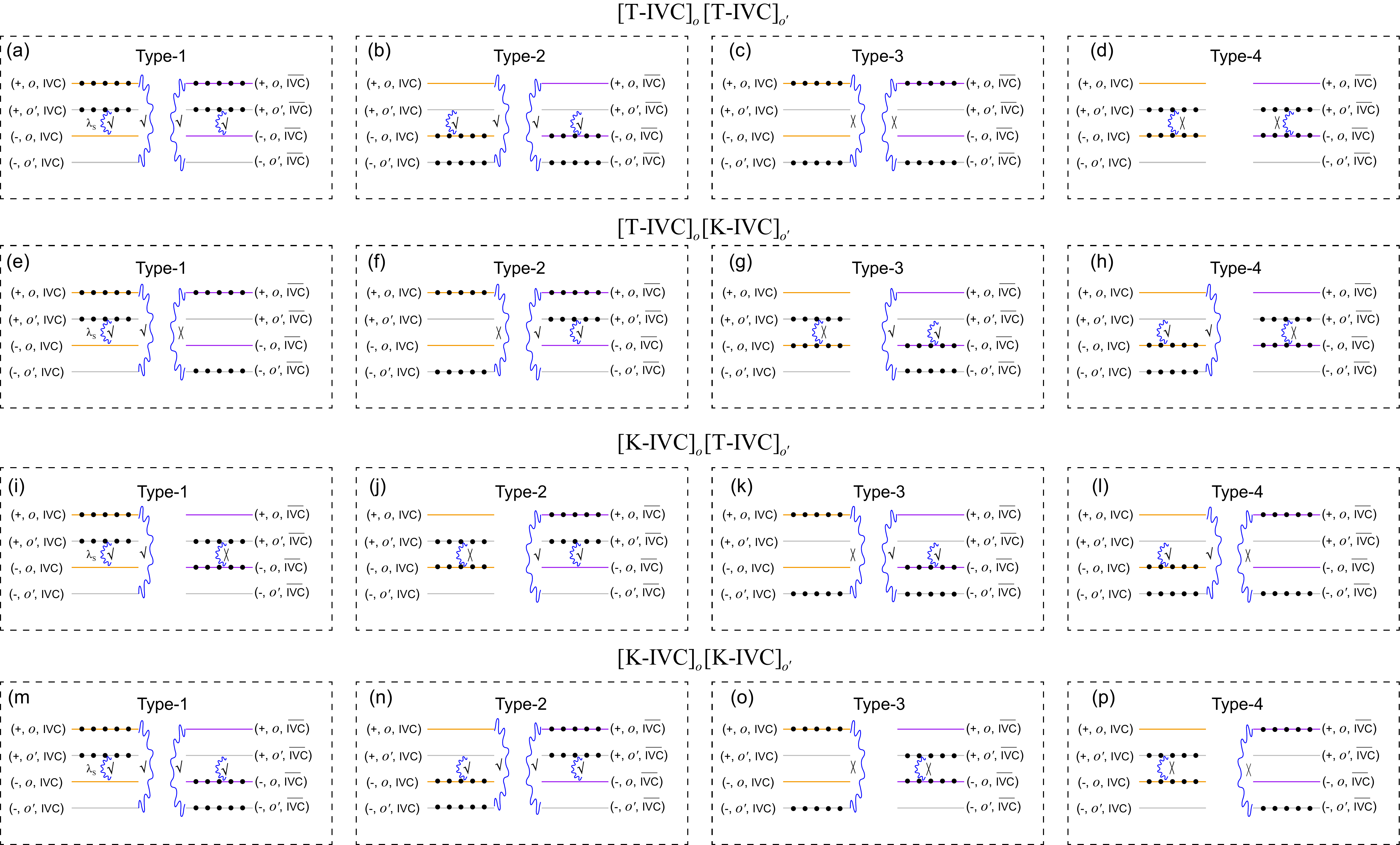}}
\captionsetup{justification=raggedright}
\caption{(color online) The filling configurations for IVC states with $C_T=0$, and we use IVC states as basis to describe all flat bands. The blue wavy line corresponds to the intrasublattice-interorbital interactions, and the symbols $\surd$ or $\times$ indicate whether the processes are permitted or forbidden.\label{IVC2}}
\end{figure}

\subsubsection{Intrasublattice-interorbital interaction}

Similar to the above perturbations, by using the symmetry relationships Eq.\ref{symmetry constrain}, the intrasublattice-interorbital projected interaction Hamiltonian with chiral symmetry in the same valley can be written as
\begin{eqnarray}
\mathcal{{H}}^{\text{S},\text{inter}}_I=\frac{1}{2 \Omega} \sum_{\mathbf{q}_M,\mathbf{g}}V(\mathbf{q}_M+\mathbf{g}) \delta \bar{\rho}^{\text{S},\text{inter}}_{\mathbf{q}_M+\mathbf{g}} \delta \bar{\rho}^{\text{S},\text{inter}}_{-\mathbf{q}_M-\mathbf{g}},\delta \bar{\rho}^{\text{S},\text{inter}}_{\pm\mathbf{q}_M\pm\mathbf{g}}=\sum^{\beta,\gamma}_{\mathbf{k}_M } c_{\beta,\mathbf{k}_M}^{\dagger}\left[\Lambda^{\text{S,inter}}_{\pm\mathbf{q}_M\pm\mathbf{g}}(\mathbf{k}_M)\right]_{\beta,\gamma} c_{\gamma, \mathbf{k}_M\pm\mathbf{q}_M},
\end{eqnarray}

\begin{eqnarray}
\Lambda^{\text{S,inter}}_{\pm\mathbf{q}_M\pm\mathbf{g}}(\mathbf{k}_M) =\left(\begin{array}{cccccccc}
0 & \Delta_{\pm,1} & 0 & 0 & 0 & 0 & 0 & 0   \\
\Delta_{\pm,2} & 0  & 0 & 0 & 0 & 0 & 0 & 0  \\
0 & 0 &0 &  \Delta_{\pm,1}^* & 0 & 0 & 0 & 0 \\
0 &  0 & \Delta^*_{\pm,2} & 0 & 0 & 0 & 0 & 0  \\
0 &  0 &  0 & 0 & 0 & -\Delta_{\pm,1}^* & 0 & 0  \\
0 & 0 &  0 &  0 & -\Delta^*_{\pm,2} & 0 & 0 &  0  \\
0 & 0 & 0 &  0 &  0 & 0 & 0 &  -\Delta_{\pm,1} \\
0 & 0 & 0 & 0 &  0 &  0 & -\Delta_{\pm,2}&  0  \\
\end{array}\right),
\end{eqnarray}
where $\Delta_{\pm,1}=\langle u_{A, o, K, \mathbf{k}_M} \mid u_{A, o^{\prime}, K, \mathbf{k}_M\pm\mathbf{q}_M}\rangle$, $\Delta_{\pm,2}=\langle u_{A, o^{\prime}, K, \mathbf{k}_M} \mid u_{A, o, K, \mathbf{k}_M\pm\mathbf{q}_M}\rangle$, and the  band basis is $\psi^{\dagger}_{\mathbf{k}_M}=[c^{\dagger}_{A,o,K},c^{\dagger}_{A,o^{\prime},K},c^{\dagger}_{B,o,K},c^{\dagger}_{B,o^{\prime},K},
c^{\dagger}_{A,o,K^{\prime}},c^{\dagger}_{A,o^{\prime},K^{\prime}},c^{\dagger}_{B,o,K^{\prime}},c^{\dagger}_{B,o^{\prime},K^{\prime}}]$. By setting $\Delta_{\pm,1}=\Delta^*_{\pm,2}=F^{S}_{\pm\mathbf{q}_M}(\mathbf{k}_M) e^{i \Phi^{S}_{\pm\mathbf{q}_M}(\mathbf{k}_M)}$, which can be further written as
\begin{eqnarray}
F^{S}_{\mathbf{q}_M}(\mathbf{k}_M-\mathbf{q}_M) e^{i \Phi^{S}_{\mathbf{q}_M}(\mathbf{k}_M-\mathbf{q}_M)}&=&\langle u_{A, o, K, \mathbf{k}_M-\mathbf{q}_M} \mid u_{A, o', K, \mathbf{k}_M}\rangle,\nonumber
\\
&=&\langle u_{A, o', K, \mathbf{k}_M} \mid  u_{A, o, K, \mathbf{k}_M-\mathbf{q}_M} \rangle^*=F^{\text{S}}_{-\mathbf{q}_M}(\mathbf{k}_M) e^{i \Phi^{\text{S}}_{-\mathbf{q}_M}(\mathbf{k}_M)}.
\end{eqnarray}
Hence, we can get
\begin{eqnarray}\label{relationshipS}
F^{\text{S}}_{\mathbf{q}_M}(\mathbf{k}_M-\mathbf{q}_M)=F^{\text{S}}_{-\mathbf{q}_M}(\mathbf{k}_M),~~ e^{i \Phi^{\text{S}}_{\mathbf{q}_M}(\mathbf{k}_M-\mathbf{q}_M)}=e^{i \Phi^{\text{S}}_{\mathbf{q}_M}(\mathbf{k}_M)}.
\end{eqnarray}

Then, the inter-orbital form factor can be simplified to
\begin{eqnarray}\label{form factor2}
\Lambda^{\text{S,inter}}_{\pm\mathbf{q}_M\pm\mathbf{g}}(\mathbf{k}_M)=\mu_x \kappa_0 \eta_zF^{\text{S}}_{\pm\mathbf{q}_M}(\mathbf{k}_M) e^{i\Phi^{\text{S}}_{\pm\mathbf{q}_M}(\mathbf{k}_M)\mu_z \kappa_z \eta_z},
\end{eqnarray}

Then,
\begin{eqnarray}
\delta \bar{\rho}^{\text{S},\text{inter}}_{\mathbf{q}_M+\mathbf{g}}&=&\sum_{\mathbf{k}_M }F^{\text{S}}_{\mathbf{q}_M}(\mathbf{k}_M)[ e^{i \Phi^{\text{S}}_{\mathbf{q}_M}(\mathbf{k}_M)} (c^{\dagger}_{A,o,K,\mathbf{k}_M}c_{A,o^{\prime},K,\mathbf{k}_M+\mathbf{q}_M}+c^{\dagger}_{B,o^{\prime},K,\mathbf{k}_M}c_{B,o,K,\mathbf{k}_M+\mathbf{q}_M}-c^{\dagger}_{A,o^{\prime},K',\mathbf{k}_M}c_{A,o,K',\mathbf{k}_M+\mathbf{q}_M},\nonumber
\\
&-&c^{\dagger}_{B,o,K',\mathbf{k}_M}c_{B,o^{\prime},K',\mathbf{k}_M+\mathbf{q}_M})+e^{-i \Phi^{\text{S}}_{\mathbf{q}_M}(\mathbf{k}_M)} (c^{\dagger}_{A,o^{\prime},K,\mathbf{k}_M}c_{A,o,K,\mathbf{k}_M+\mathbf{q}_M}+c^{\dagger}_{B,o,K,\mathbf{k}_M}c_{B,o^{\prime},K,\mathbf{k}_M+\mathbf{q}_M}\nonumber
\\
&-&c^{\dagger}_{A,o,K',\mathbf{k}_M}c_{A,o^{\prime},K',\mathbf{k}_M+\mathbf{q}_M}-c^{\dagger}_{B,o^{\prime},K',\mathbf{k}_M}c_{B,o,K',\mathbf{k}_M+\mathbf{q}_M})],\nonumber
\\
\delta \bar{\rho}^{\text{S},\text{inter}}_{-\mathbf{q}_M-\mathbf{g}}&=&\sum_{\mathbf{k}_M }F^{\text{S}}_{-\mathbf{q}_M}(\mathbf{k}_M)[ e^{i \Phi^{\text{S}}_{-\mathbf{q}_M}(\mathbf{k}_M)} (c^{\dagger}_{A,o,K,\mathbf{k}_M}c_{A,o^{\prime},K,\mathbf{k}_M-\mathbf{q}_M}+c^{\dagger}_{B,o^{\prime},K,\mathbf{k}_M}c_{B,o,K,\mathbf{k}_M-\mathbf{q}_M}-c^{\dagger}_{A,o^{\prime},K',\mathbf{k}_M}c_{A,o,K',\mathbf{k}_M-\mathbf{q}_M},\nonumber
\\
&-&c^{\dagger}_{B,o,K',\mathbf{k}_M}c_{B,o^{\prime},K',\mathbf{k}_M-\mathbf{q}_M})+e^{-i \Phi^{\text{S}}_{-\mathbf{q}_M}(\mathbf{k}_M)} (c^{\dagger}_{A,o^{\prime},K,\mathbf{k}_M}c_{A,o,K,\mathbf{k}_M-\mathbf{q}_M}+c^{\dagger}_{B,o,K,\mathbf{k}_M}c_{B,o^{\prime},K,\mathbf{k}_M-\mathbf{q}_M}\nonumber
\\
&-&c^{\dagger}_{A,o,K',\mathbf{k}_M}c_{A,o^{\prime},K',\mathbf{k}_M-\mathbf{q}_M}-c^{\dagger}_{B,o^{\prime},K',\mathbf{k}_M}c_{B,o,K',\mathbf{k}_M-\mathbf{q}_M})],\nonumber
\end{eqnarray}

Similar to the derivation of energy correction $\Delta E^A_{\text{type-1}}$. We take type-4 QH state with total Chern number $C_T=1$ as example to calculate energy correction $\Delta E^S=\langle\psi^{(0)}|\mathcal{{H}}^{\text{S,inter}}_I| \psi^{(0)}\rangle$, and the wave funtion can be written as $| \Psi^{C=1}_{\text{type-4}}\rangle=\prod_{\mathbf{k}_M} c_{A,o,K, \mathbf{k}_M}^{\dagger}c_{A,o', K, \mathbf{k}_M}^{\dagger} c_{B,o, K, \mathbf{k}_M}^{\dagger}c_{B,o,K^{\prime}, \mathbf{k}_M}^{\dagger}|0\rangle$.
Then, the derivation process is as follows
\begin{eqnarray}
&&\delta \bar{\rho}^{\text{S},\text{inter}}_{\mathbf{q}_M+\mathbf{g}}| \Psi^{C=1}_{\text{type-4}}\rangle,\nonumber
\\
&=& \sum_{\mathbf{k}_M }F^{\text{S}}_{\mathbf{q}_M}(\mathbf{k}_M)[e^{i \Phi^{\text{S}}_{\mathbf{q}_M}(\mathbf{k}_M)}c^{\dagger}_{B,o',K,\mathbf{k}_M}c_{B,o,K,\mathbf{k}_M+\mathbf{q}_M}-e^{-i \Phi^{\text{S}}_{\mathbf{q}_M}(\mathbf{k}_M)}c^{\dagger}_{B,o',K',\mathbf{k}_M}c_{B,o,K',\mathbf{k}_M+\mathbf{q}_M}]
 | \Psi^{C=1}_{\text{type-4}}\rangle,\nonumber
\\
&&\delta \bar{\rho}^{\text{S},\text{inter}}_{-\mathbf{q}_M-\mathbf{g}}| \Psi^{C=1}_{\text{type-4}}\rangle,\nonumber
\\
&=& \sum_{\mathbf{k}_M }F^{\text{S}}_{-\mathbf{q}_M}(\mathbf{k}_M)[e^{i \Phi^{\text{S}}_{-\mathbf{q}_M}(\mathbf{k}_M)}c^{\dagger}_{B,o',K,\mathbf{k}_M}c_{B,o,K,\mathbf{k}_M-\mathbf{q}_M}-e^{-i \Phi^{\text{S}}_{-\mathbf{q}_M}(\mathbf{k}_M)}c^{\dagger}_{B,o',K',\mathbf{k}_M}c_{B,o,K',\mathbf{k}_M-\mathbf{q}_M}]
 | \Psi^{C=1}_{\text{type-4}}\rangle,
\end{eqnarray}
where we can omit other terms due to the properties of fermions($c^{\dagger}c^{\dagger}|0\rangle=0$ and $c|0\rangle=0$). Therefore, we obtain

\begin{eqnarray}
&&\langle\Psi^{C=1}_{\text{type-4}}|\delta \bar{\rho}^{\text{S},\text{inter}}_{\mathbf{q}_M+\mathbf{g}} \delta \bar{\rho}^{\text{S},\text{inter}}_{-\mathbf{q}_M-\mathbf{g}}| \Psi^{C=1}_{\text{type-4}}\rangle,\nonumber
\\
&=&\langle\Psi^{C=1}_{\text{type-4}}|\{\sum_{\mathbf{k}'_M }F^{\text{S}}_{\mathbf{q}_M}(\mathbf{k}'_M)[ e^{-i\Phi^{\text{S}}_{\mathbf{q}_M}(\mathbf{k}'_M)}c^{\dagger}_{B,o,K,\mathbf{k}'_M}c_{B,o',K,\mathbf{k}'_M+\mathbf{q}_M}-e^{i \Phi^{\text{S}}_{\mathbf{q}_M}(\mathbf{k}'_M)}c^{\dagger}_{B,o,K',\mathbf{k}'_M}c_{B,o',K',\mathbf{k}'_M+\mathbf{q}_M}]
\}\nonumber
\\
&&\{\sum_{\mathbf{k}_M }F^{\text{S}}_{-\mathbf{q}_M}(\mathbf{k}_M)[e^{i \Phi^{\text{S}}_{-\mathbf{q}_M}(\mathbf{k}_M)}c^{\dagger}_{B,o',K,\mathbf{k}_M}c_{B,o,K,\mathbf{k}_M-\mathbf{q}_M}-e^{-i \Phi^{\text{S}}_{-\mathbf{q}_M}(\mathbf{k}_M)}c^{\dagger}_{B,o',K',\mathbf{k}_M}c_{B,o,K',\mathbf{k}_M-\mathbf{q}_M}]
\} | \Psi^{C=1}_{\text{type-4}}\rangle,\nonumber
\\
&=&\langle\Psi^{C=1}_{\text{type-4}}|\{\sum_{\mathbf{k}'_M }F^{\text{S}}_{\mathbf{q}_M}(\mathbf{k}'_M)[e^{-i \Phi^{\text{S}}_{\mathbf{q}_M}(\mathbf{k}'_M)}c^{\dagger}_{B,o,K,\mathbf{k}'_M}c_{B,o',K,\mathbf{k}'_M+\mathbf{q}_M}\}
\{\sum_{\mathbf{k}_M }F^{\text{S}}_{-\mathbf{q}_M}(\mathbf{k}_M)[ e^{i\Phi^{\text{S}}_{-\mathbf{q}_M}(\mathbf{k}_M)}c^{\dagger}_{B,o',K,\mathbf{k}_M}c_{B,o,K,\mathbf{k}_M-\mathbf{q}_M}\} \nonumber\\
&+&\{\sum_{\mathbf{k}'_M }F^{\text{S}}_{\mathbf{q}_M}(\mathbf{k}'_M)e^{i \Phi^{\text{S}}_{\mathbf{q}_M}(\mathbf{k}'_M)}c^{\dagger}_{B,o,K',\mathbf{k}'_M}c_{B,o',K',\mathbf{k}'_M+\mathbf{q}_M}
\}\{\sum_{\mathbf{k}_M }F^{\text{S}}_{-\mathbf{q}_M}(\mathbf{k}_M)[e^{-i\Phi^{\text{S}}_{-\mathbf{q}_M}(\mathbf{k}_M)}c^{\dagger}_{B,o',K',\mathbf{k}_M}c_{B,o,K',\mathbf{k}_M-\mathbf{q}_M}]
\} | \Psi^{C=1}_{\text{type-4}}\rangle,\nonumber
\\
&=&\sum_{\mathbf{k}_M }\langle\Psi^{C=1}_{\text{type-4}}|\{F^{\text{S}}_{\mathbf{q}_M}(\mathbf{k}_M-\mathbf{q}_M)[e^{-i \Phi^{\text{S}}_{\mathbf{q}_M}(\mathbf{k}_M-\mathbf{q}_M)}c^{\dagger}_{B,o,K,\mathbf{k}_M-\mathbf{q}_M}c_{B,o',K,\mathbf{k}_M}\}
\{F^{\text{S}}_{-\mathbf{q}_M}(\mathbf{k}_M)[ e^{i\Phi^{\text{S}}_{-\mathbf{q}_M}(\mathbf{k}_M)}c^{\dagger}_{B,o',K,\mathbf{k}_M}c_{B,o,K,\mathbf{k}_M-\mathbf{q}_M}\}\nonumber \\
&+&\{F^{\text{S}}_{\mathbf{q}_M}(\mathbf{k}_M-\mathbf{q}_M)e^{i \Phi^{\text{S}}_{\mathbf{q}_M}(\mathbf{k}_M-\mathbf{q}_M)}c^{\dagger}_{B,o,K',\mathbf{k}_M-\mathbf{q}_M}c_{B,o',K',\mathbf{k}_M}
\}\{F^{\text{S}}_{-\mathbf{q}_M}(\mathbf{k}_M)[e^{-i\Phi^{\text{S}}_{-\mathbf{q}_M}(\mathbf{k}_M)}c^{\dagger}_{B,o',K',\mathbf{k}_M}c_{B,o,K',\mathbf{k}_M-\mathbf{q}_M}]
\} | \Psi^{C=1}_{\text{type-4}}\rangle,\nonumber
\\
&=&\sum_{\mathbf{k}_M }[F^{\text{S}}_{-\mathbf{q}_M}(\mathbf{k}_M)]^2\langle\Psi^{C=1}_{\text{type-4}}|n_{B,o,K,\mathbf{k}_M-\mathbf{q}_M}(1-n_{B,o',K,\mathbf{k}_M})+
n_{B,o,K',\mathbf{k}_M-\mathbf{q}_M}(1-n_{B,o',K',\mathbf{k}_M}) | \Psi^{C=1}_{\text{type-4}}\rangle.
\end{eqnarray}

In the second step, we have excluded terms that are inherently zero. Subsequently, in the third step, we substitute $\mathbf{k}'_M$ with $\mathbf{k}_M-\mathbf{q}_M$. Finally, in the fourth step, we make use of the relationships as defined in Eq. \ref{relationshipS}.

Then, we can get
\begin{eqnarray}\label{88}
\Delta E^S_{\text{type-4}}=\frac{1}{2 \Omega} \sum_{\mathbf{q}_M,\mathbf{k}_M,\mathbf{g}} V_{\mathbf{q}_M+\mathbf{g}}\left\langle\Psi^{C=1}_{\text{type-4}}\left| \delta \bar{\rho}^{\text{S}}_{\mathbf{q}_M+\mathbf{g}} \delta \bar{\rho}^{\text{S}}_{-\mathbf{q}_M-\mathbf{g}}\right| \Psi^{C=1}_{\text{type-4}}\right\rangle=2\lambda_S
\end{eqnarray}
where $\lambda_S=\frac{1}{ 2\Omega} \sum_{\mathbf{q}_M,\mathbf{k}_M,\mathbf{g}} V_{\mathbf{q}_M+\mathbf{g}} [{F}^{S}_{-\mathbf{q}_M}(\mathbf{k}_M)]^2$. Similarly, we can obtain the energy correction of other ground states induced by intrasublattice-interorbital interaction, shown in Table.\ref{Ground state}.

We turn to consider energy correction induced by intrasublattice-interorbital interaction in IVC states. For IVC states with $C_T=\pm 2$, the flat band basis is transformed to $\psi^{\dagger}_{\mathbf{k}_M}=[c^{\dagger}_{A,o,K},\frac{1}{\sqrt{2}}(e^{-i\phi/2}c^{\dagger}_{A,o^{\prime},K}+e^{i\phi/2}c^{\dagger}_{B,o^{\prime},K'}),c^{\dagger}_{B,o,K},\frac{1}{\sqrt{2}}(e^{-i\phi/2}c^{\dagger}_{B,o^{\prime},K}+e^{i\phi/2}c^{\dagger}_{A,o^{\prime},K'}), c^{\dagger}_{A,o,K^{\prime}},\frac{1}{\sqrt{2}}(e^{-i\phi/2}c^{\dagger}_{A,o^{\prime},K}-e^{i\phi/2}c^{\dagger}_{B,o^{\prime},K'}),c^{\dagger}_{B,o,K^{\prime}},\frac{1}{\sqrt{2}}(e^{-i\phi/2}c^{\dagger}_{B,o^{\prime},K}-e^{i\phi/2}c^{\dagger}_{A,o^{\prime},K'})]$. Then, the inter-orbital form factor becomes
\begin{eqnarray}
\Lambda^{\text{IVC},1}_{\pm\mathbf{q}_M\pm\mathbf{g}}(\mathbf{k}_M)=\left(\begin{array}{cccccccc}
0 & \frac{e^{-i\phi/2}}{\sqrt{2}}\Delta_{\pm,1} & 0 & 0 & 0 & \frac{e^{-i\phi/2}}{\sqrt{2}}\Delta_{\pm,1} & 0 & 0   \\
\frac{e^{i\phi/2}}{\sqrt{2}}\Delta_{\pm,2} & 0  & 0 & 0 & 0 & 0 & -\frac{e^{-i\phi/2}}{\sqrt{2}}\Delta_{\pm,2} & 0  \\
0 & 0 &0 &  \frac{e^{-i\phi/2}}{\sqrt{2}}\Delta_{\pm,1}^* & 0 & 0 & 0& \frac{e^{-i\phi/2}}{\sqrt{2}}\Delta_{\pm,1}^* \\
0 &  0 & \frac{e^{i\phi/2}}{\sqrt{2}}\Delta^*_{\pm,2} & 0 & -\frac{e^{-i\phi/2}}{\sqrt{2}}\Delta^*_{\pm,2} & 0 & 0 & 0  \\
0 &  0 &  0 & -\frac{e^{i\phi/2}}{\sqrt{2}}\Delta_{\pm,1}^* & 0 & 0 & 0 & \frac{e^{i\phi/2}}{\sqrt{2}}\Delta_{\pm,1}^*  \\
\frac{e^{i\phi/2}}{\sqrt{2}}\Delta_{\pm,2} & 0 &  0 &  0 & 0 & 0 & \frac{e^{-i\phi/2}}{\sqrt{2}}\Delta_{\pm,2} &  0  \\
0 & -\frac{e^{i\phi/2}}{\sqrt{2}}\Delta_{\pm,1} & 0 &  0 &  0 & \frac{e^{i\phi/2}}{\sqrt{2}}\Delta_{\pm,1} & 0&  0 \\
0 & 0 &  \frac{e^{i\phi/2}}{\sqrt{2}}\Delta^*_{\pm,2} & 0 &  \frac{e^{-i\phi/2}}{\sqrt{2}}\Delta^*_{\pm,2} &  0 &0&  0  \\
\end{array}\right),
\end{eqnarray}

Similar to the derivation process of Eq.\ref{88}, for [QH]$_o$[T-IVC]$_{o^{\prime}}$ and [QH]$_o$[K-IVC]$_{o^{\prime}}$ states, we can also derive their energy corrections induced by intrasublattice-interorbital interaction, as presented in Table.\ref{Ground stateIVC}.

For IVC states with $C_T=0$, the flat band basis can be transformed to
$\psi^{\dagger}_{\mathbf{k}_M}=[\frac{1}{\sqrt{2}}(e^{-i\phi/2}c^{\dagger}_{A,o,K}+e^{i\phi/2}c^{\dagger}_{B,o,K'}),\frac{1}{\sqrt{2}}(e^{-i\phi/2}c^{\dagger}_{A,o^{\prime},K}+e^{i\phi/2}c^{\dagger}_{B,o^{\prime},K'}),\frac{1}{\sqrt{2}}(e^{-i\phi/2}c^{\dagger}_{B,o,K}+e^{i\phi/2}c^{\dagger}_{A,o,K'}),\frac{1}{\sqrt{2}}(e^{-i\phi/2}c^{\dagger}_{B,o^{\prime},K}+e^{i\phi/2}c^{\dagger}_{A,o^{\prime},K'}),
\frac{1}{\sqrt{2}}(e^{-i\phi/2}c^{\dagger}_{A,o,K}-e^{i\phi/2}c^{\dagger}_{B,o,K'}),\frac{1}{\sqrt{2}}(e^{-i\phi/2}c^{\dagger}_{A,o^{\prime},K}-e^{i\phi/2}c^{\dagger}_{B,o^{\prime},K'}),\frac{1}{\sqrt{2}}(e^{-i\phi/2}c^{\dagger}_{B,o,K}-e^{i\phi/2}c^{\dagger}_{A,o,K'}),\frac{1}{\sqrt{2}}(e^{-i\phi/2}c^{\dagger}_{B,o^{\prime},K}-e^{i\phi/2}c^{\dagger}_{A,o^{\prime},K'})$]. Then, the inter-orbital form factor becomes
\begin{eqnarray}
\Lambda^{\text{IVC},2}_{\pm\mathbf{q}_M\pm\mathbf{g}}(\mathbf{k}_M) =\left(\begin{array}{cccccccc}
0 & 0 & 0 & 0 & 0 & \Delta_{\pm,1} & 0 & 0   \\
0& 0  & 0 & 0 & \Delta_{\pm,2} & 0 & 0 & 0  \\
0 & 0 &0 &  0 & 0 & 0 & 0& \Delta_{\pm,1}^* \\
0 &  0 & 0 & 0 & 0 & 0 & \Delta^*_{\pm,2} & 0  \\
0 &  \Delta_{\pm,1} &  0 & 0 & 0 & 0 & 0 & 0  \\
\Delta_{\pm,2} & 0 &  0 &  0 & 0 & 0 & 0 &  0  \\
0 & 0 & 0 &  \Delta^*_{\pm,1} &  0 & 0 & 0&  0 \\
0 & 0 & \Delta^*_{\pm,2} & 0 &  0 &  0 &0&  0  \\
\end{array}\right),
\end{eqnarray}

Similarly, for [T-IVC]$_o$[T-IVC]$_{o^{\prime}}$, [T-IVC]$_o$[K-IVC]$_{o^{\prime}}$, [K-IVC]$_o$[I-IVC]$_{o^{\prime}}$ and [K-IVC]$_o$[K-IVC]$_{o^{\prime}}$ states, we can also obtain their energy corrections induced by intrasublattice-interorbital interaction, as shown in Table.\ref{Ground stateIVC}.

\begin{table}[t]
\renewcommand\arraystretch{2}
\setlength{\tabcolsep}{1mm}{
    \begin{tabular}{|*{5}{c|}}
     \hline
      \multirow{2}{*}{C$_T=\pm2$}&  $\text{Q}^2_1$ ~~~~~~ \vline ~~~~~~ $\text{Q}^2_2$ & $\text{Q}^2_3$ ~~~~~~~~~~ \vline ~~~~~~~~~~ $\text{Q}^2_4$ & $\text{Q}^2_5$ & $\text{Q}^2_6$ \\
        &$\kappa_z\mu_0\eta_0$ ~~~ \vline ~~~ $\kappa_z\mu_z\eta_0$ &$[\kappa_z\mu_z]\oplus[\kappa_z\mu_0]$~~\vline~$[\kappa_z\mu_0]\oplus[\kappa_z\mu_z]$ & $[\kappa_z\oplus\mu_z]\oplus[\kappa_z\oplus(-\mu_0)]$ & $[\kappa_z\oplus(-\mu_0)]\oplus[\kappa_z\oplus\mu_0]$ \\
        \hline
        \hline

       \multirow{8}{*}{C$_T=\pm1$}&{$\text{Q}^1_1$ } & $\text{Q}^1_2$ & $\text{Q}^1_3$ & $\text{Q}^1_4$  \\
        & $[\kappa_0\mu_z]\oplus[\kappa_z\mu_z]$  & $[\kappa_z\mu_z]\oplus[-\kappa_0\mu_z]$ & $[-\kappa_0\mu_z]\oplus[\kappa_z\mu_z]$  & $[\kappa_0\oplus\mu_z]\oplus[\kappa_z\oplus(-\mu_0)]$ \\ \cline{2-5}
         & $\text{Q}^1_5$ & $\text{Q}^1_6$& $\text{Q}^1_7$ & $\text{Q}^1_8$ \\
        & $[\kappa_0\mu_z]\oplus[\kappa_z\mu_0]$ & $[\kappa_0\oplus(-\mu_z)]\oplus[\kappa_z\oplus(-\mu_0)]$ & $[\kappa_z\mu_0]\oplus[\kappa_0\mu_z]$ & $[\kappa_z\oplus(-\mu_0)]\oplus[\kappa_0\oplus\mu_z]$ \\ \cline{2-5}
        & $\text{Q}^1_9$ & $\text{Q}^1_{10}$ & $\text{Q}^1_{11}$ & $\text{Q}^1_{12}$  \\
        & $[\kappa_z\mu_z]\oplus[\kappa_0\mu_z]$  & $[\kappa_z\oplus(-\mu_0)]\oplus[\kappa_0\oplus(-\mu_z)]$ & $[\kappa_z\oplus\mu_0]\oplus[(-\kappa_0)\oplus\mu_z]$  & $[\kappa_z\mu_0]\oplus[-\kappa_0\mu_z]$ \\ \cline{2-5}
         & $\text{Q}^1_{13}$ & $\text{Q}^1_{14}$& $\text{Q}^1_{15}$ & $\text{Q}^1_{16}$ \\
         & $[\kappa_z\oplus\mu_0]\oplus[-\kappa_0\oplus\mu_z]$ & $[-\kappa_0\mu_z]\oplus[\kappa_z\mu_0]$ &$[(-\kappa_0)\oplus\mu_z]\oplus[\kappa_z\oplus\mu_0]$ & $[-\kappa_0\oplus\mu_z]\oplus[\kappa_z\oplus\mu_0]$ \\
         \hline\hline

        \multirow{8}{*}{C$_T=0$}&  $\text{Q}^0_1$ & $\text{Q}^0_2$ & $\text{Q}^0_3$ & $\text{Q}^0_4$  \\
       & $\kappa_0\mu_0\eta_z$  & $-\kappa_0\mu_0\eta_z$ & $\kappa_0\mu_z\eta_0$  & $-\kappa_0\mu_z\eta_0$ \\ \cline{2-5}
        &  $\text{Q}^0_5$ & $\text{Q}^0_6$& $\text{Q}^0_7$ & $\text{Q}^0_8$ \\
         & $\kappa_0\mu_z\eta_z$ & $-\kappa_0\mu_z\eta_z$ &$\kappa_z\mu_0\eta_z$ & $\kappa_z\mu_z\eta_z$ \\ \cline{2-5}
        & $\text{Q}^0_9$ & $\text{Q}^0_{10}$ & $\text{Q}^0_{11}$ & $\text{Q}^0_{12}$  \\
       & $[\kappa_z\mu_0]\oplus[-\kappa_z\mu_z]$  & $[\kappa_z\mu_z]\oplus[-\kappa_z\mu_0]$ & $[\kappa_z\oplus\mu_0]\oplus[-\kappa_z\oplus\mu_0]$  & $[\kappa_z\oplus(-\mu_0)]\oplus[(-\kappa_z)\oplus\mu_0]$ \\ \cline{2-5}
         & $\text{Q}^0_{13}$ & $\text{Q}^0_{14}$& $\text{Q}^0_{15}$ & $\text{Q}^0_{16}$ \\
         &$[\kappa_0\oplus\mu_z]\oplus[-\kappa_0\oplus\mu_z]$ & $[(-\kappa_0)\oplus\mu_z]\oplus[\kappa_0\oplus(-\mu_z)]$ &$[\kappa_0\oplus\mu_z]\oplus[(-\kappa_0)\oplus\mu_z]$ & $[(-\kappa_0)\oplus\mu_z]\oplus[\kappa_0\oplus\mu_z]$ \\ \hline

    \end{tabular}}
    \caption{Operators Q$^{|\text{C}_T|}_n$ for each ground state.}
    \label{Q_operator}
\end{table}

\subsubsection{Intuitive picture about the energy correction}

\begin{figure}
\centerline{\includegraphics[width=1\textwidth]{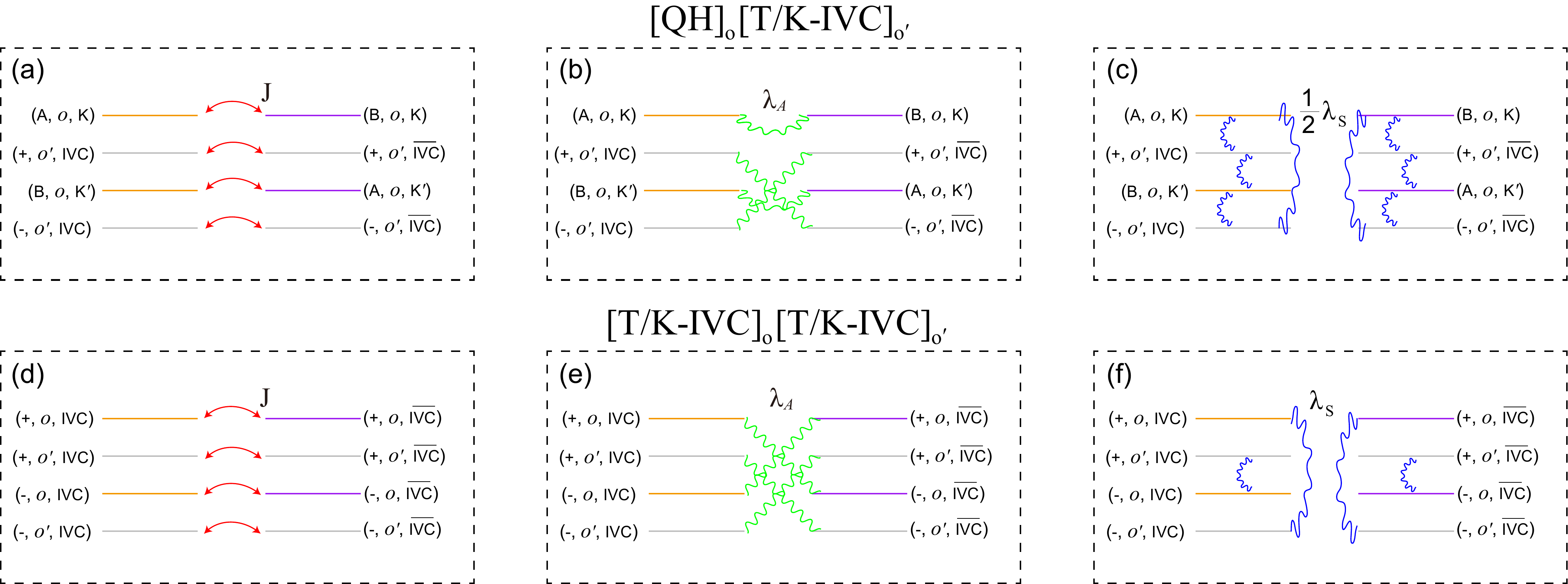}}
\captionsetup{justification=raggedright}
\caption{(color online) Schematic illustration of single-particle dispersion (red curves), intersublattice-intraorbital (green wavy lines) and intrasublattice-interorbital (blue wavy lines) interactions in the basis of IVC states for [QH]$_o$[T/K-IVC]$_{o^{\prime}}$ and [T/K-IVC]$_o$[T/K-IVC]$_{o^{\prime}}$ states. The parameters of $J$, $\lambda_A$ and $\lambda_S$ are the strength of the corresponding perturbation within the basis of IVC states.\label{interaction}}
\end{figure}

Here, we will present an intuitive picture to understand the energy correction resulting from the above single-particle dispersion and two interactions, and the eight flat bands can be organized into four pairs of bands connected by these three terms (denoted by red curves, green and blue wavy lines in Fig.(\ref{QH1}-\ref{QH3})). The single-particle dispersion describes the hopping channels between one pair of bands connected by single-particle term. On the other hand, energy corrections arising from these two interactions are proportional to $\lambda_{A/S}\hat{n}(1-\hat{n}^{\prime})$, where $\hat{n}$ and $\hat{n}^{\prime}$ represent the particle number operators for the pair of bands linked by the corresponding interactions. Then, we can summarize three distinct scenarios regarding energy corrections: (i) In a state, when all four pairs of bands connected by these terms are fully occupied or fully empty, the energy corrections vanish  due to the absence of allowed hopping and interaction channels. In VP in Fig.\ref{QH3} (a) and (b), this applies to the four pairs of bands connected by these three terms, and the energy corrections are zero shown in Table.\ref{Ground state}.
 (ii) In a state, where each pair of bands connected by the perturbation term is half-filled, the hopping and interacting channels between each pair of bands are allowed, resulting four-fold such energy correction. In type-2 and -2$^{\prime}$ QH states in Fig.\ref{QH1} (b), all four pairs of bands belong to this category and the energy corrections are $\Delta E = 4 \lambda_A + 4 \lambda_S - 4 J$ in Table.\ref{Ground state}. (iii) In a state, where two of the four pairs connected by these terms are either occupied or empty while the other two pairs are half-filled, two-fold such energy correction is generated. In type-(5$\sim$6) and -(5$^{\prime}\sim6^{\prime}$) QH states in Fig.\ref{QH1} (e) and (f), all four pairs of bands belong to this case and the energy corrections are $\Delta E = 2 \lambda_A + 2 \lambda_S - 2 J$ in Table.\ref{Ground state}. Finally, Table.\ref{Ground state} presents energy correction for each ground state, and the energy correction can be expressed by a general formula as
\begin{eqnarray}\label{correction}
 \Delta E = N_A \lambda_A + N_S \lambda_S - N_J J,
\end{eqnarray}
where the parameters $N_{A/S/J}$ are determined by the number of allowing hopping and interacting channels. It is noteworthy that the single-particle Hamiltonian and the inter-Chern-number interaction establish connections between the same pairs of bands, resulting in $N_A=N_J$.

To characterize each ground state identified by the number $n$ and total Chern number $\text{C}_T$, we introduce a corresponding operator denoted as Q$^{|\text{C}_T|}_n$. This operator possesses the property of $Q^{|\text{C}_T|}_n|\Psi^{\text{C}_T}_n\rangle=\pm|\Psi^{\text{C}_T}_n\rangle$, where $|\Psi^{\text{C}_T}_n\rangle$ represents the wave function of the ground state. We can determine the values of the parameters $N_{A/S/J}$ by utilizing the commutation relations between Q$^{|\text{C}_T|}_n$ and $x$, where $x$ denotes the single-particle dispersion $h$ in Eq.\ref{single-particle}, the form factor $\Lambda^\text{A}$ in Eq.\ref{form factor1} and the form factor $\Lambda^\text{S,inter}$ in Eq.\ref{form factor2}. When the commutation relation is a commutator or an anticommutator, these parameters take values of either 0 or 4. However, when the commutation relation is neither a commutator nor an anticommutator, these parameters take a value of 2. Table.\ref{Q_operator} presents the corresponding operator Q$^{|\text{C}_T|}_n$ for each ground state.

In addition, the general formula Eq.\ref{correction} for the energy correction, which are determined by the number of allowed hopping and interacting channels, can also be applied to IVC states in Table.\ref{Ground stateIVC}. In analogy to the schematic depiction of the three perturbations in Fig.\ref{QH1}-\ref{QH3}, Fig.\ref{interaction} also provides the schematic illustration of single-particle dispersion (red curves), intersublattice-intraorbital (green wavy lines) and intrasublattice-interorbital (blue wavy lines) interactions within the basis of IVC states. For [T/K-IVC]$_o$[T/K-IVC]$_{o^{\prime}}$ states, the strengths of the three perturbations within the basis of IVC states remain invariant, as shown in Fig.\ref{interaction}(d)-(f). However, in [QH]$_o$[T/K-IVC]${o^{\prime}}$ states, the strength and channel of intrasublattice-interorbital interaction are halved and doubled, respectively, as depicted in Fig.\ref{interaction}(c). In type-1 state of [QH]$_o$[T-IVC]$_{o^{\prime}}$ in Fig.\ref{IVC1}(a), the number of allowing channels for single-particle dispersion, intersublattice-intraorbital and intrasublattice-interorbital interactions are two, four and two, and hence $\Delta E$=4$\lambda_A$+2$\lambda_S$-2J. In type-4 state of [K-IVC]$_o$[K-IVC]$_{o^{\prime}}$ in Fig.\ref{IVC2}(p), the number of allowing channels for corresponding perturbation are four, zero and zero, and hence $\Delta E$=-4J.

\end{widetext}


\begin{thebibliography}{35}
\bibitem{explain} The parameters $N_{A/S/J}$ are determined by the number of active interacting and hopping  channels between pair states and $\lambda_{A/S},J>0$.
\end{thebibliography}
\end{document}